\documentclass[aps,pra,showpacs,amsmath,amssymb,superscriptaddress,notitlepage,onecolumn,12pt,showkeys,nofootinbib]{revtex4-2}

\usepackage{graphicx}
\usepackage{array}
\usepackage{longtable}
\usepackage{rotating,booktabs}
\usepackage{booktabs,threeparttable}
\usepackage{amssymb}
\allowdisplaybreaks[4]
\usepackage[dvipsnames]{xcolor}
\usepackage{appendix}
\usepackage{rotating}

\usepackage{changes}

\begin{document}

\newcommand{\ket}[1]{\ensuremath{\left|#1\right\rangle}}
\newcommand{\bra}[1]{\ensuremath{\left\langle #1 \right|}}
\newcommand{\braket}[2]{\ensuremath{\left\langle #1 | #2 \right\rangle}}
\newcommand{\atom}{\text{Li}}
\newcommand{\ion}{\text{Li}^+}
\newcommand{\dimer}{\text{Li}_2}
\newcommand{\dimerion}{\text{Li}_2^+}
\newcommand{\trimer}{\text{Li}_3}
\newcommand{\trimerion}{\text{Li}_3^+}

\title{Long-range additive and nonadditive potentials in a hybrid system:
Ground state atom, excited state atom, and ion}

\author{Pei-Gen Yan}
\affiliation{Department of Physics, University of New Brunswick, Fredericton, New Brunswick, E3B 5A3, Canada}

\author{Li-Yan Tang}
\email{lytang@wipm.ac.cn}
\affiliation{State Key Laboratory of Magnetic Resonance and Atomic and Molecular Physics, Wuhan Institute of Physics and Mathematics, Innovation Academy for Precision Measurement Science and Technology, Chinese Academy of Sciences, Wuhan 430071, People's Republic of China}

\author{Zong-Chao Yan}
\affiliation{Department of Physics, University of New Brunswick, Fredericton, New Brunswick, E3B 5A3, Canada}
\affiliation{State Key Laboratory of Magnetic Resonance and Atomic and Molecular Physics, Wuhan Institute of Physics and Mathematics, Innovation Academy for Precision Measurement Science and Technology, Chinese Academy of Sciences, Wuhan 430071, People's Republic of China}

\author{James F. Babb}
\email{jbabb@cfa.harvard.edu}
\affiliation{ITAMP, Center for Astrophysics \textbar\ Harvard \& Smithsonian, MS 14, 60 Garden St., Cambridge, MA 02138, USA}

\date{\today}

\begin{abstract}
We report a theoretical study on the long-range additive and nonadditive potentials for a three-body hybrid atom-atom-ion system composed of one ground $S$ state Li atom, one excited $P$ state Li atom and one ground $S$ state Li$^+$ ion, Li($2\,^{2}S$)-Li($2\,^{2}P$)-Li$^+(1\,^{1}S$). The interaction coefficients are evaluated with highly accurate wave functions calculated variationally
in Hylleraas coordinates. For this hybrid system the three-body nonadditive  interactions (appearing in second-order) induced by the energy degeneracy and enhanced by the induction effect of the Li$^+$ ion through the internal electric field can be strong and even stronger than the two-body additive interactions at the same order. We find that for particular geometries the two-body additive interactions of the system sum to zero leaving only three-body nonadditive interactions 
thus making the present system potentially a platform to explore quantum three-body effects. We also extract by
first-principles the leading coefficients
of the long-range electrostatic,
induction, and dispersion  energies of $\dimerion$ electronic states correlating to $\ion (1\, {}^1S)$--$\atom (2\,{}^2P)$,
which until now were not available
in the literature. The results should be especially valuable for the  exploration of schemes to create trimers with ultracold atoms and ions in optical lattices.
\end{abstract}

\maketitle

\newpage
\section{Introduction}\label{Int}

This paper provides detailed results for the long-range interactions between
three atomic systems, specifically, a ground state atom, a (low-lying) excited state atom, and a ground state ion, for the particular case of lithium, specifically,  Li($2\,^{2}S$)-Li($2\,^{2}P$)-Li$^+(1\,^{1}S$).
Most studies of hybrid systems consisting of ground or low-lying state atoms have been concerned with pair-wise cases, \textit{i.e.}, an atom and an ion (reviewed in Ref.~\cite{Tomza19}), or a diatomic molecule and an ion~\cite{LepDulKok10,LepVexBou11,PerRioLep15}. Recently, properties of low-lying states of triatomic cations were systematically studied~\cite{SmiTom20}. 
And, some studies, while considering excitation of atoms, do not consider ions. For three atoms, with at least one atom in a Rydberg state, there are a number of studies, such as Refs.~\cite{CanFor12,Qia16}. Other studies have considered three-body interactions  of diatomic molecules in a trapping potential~\cite{BarDalPup12,KifLiJak13}. 

We explore another possibility---three atomic systems that are in the long-range domain (sufficiently
separated such that electron
exchange is small) with one constituent charged and one constituent \textit{electronically excited.}
There are two main results: First, we give expressions for the long-range potentials as expansions in inverse-powers of separation distances and corresponding precisely evaluated coefficients for two-body (dipole-dipole and van der Waals) and three-body (van der Waals) long-range additive
and nonadditive interactions, in a manner similar to, but extending our previous work on three atoms~\cite{yan16,yan18} and
on two atoms and a ground state ion~\cite{yan20}. 
While in the present work the derived formulas are generally applicable to the hybrid 
$A(n_0S)$-$A(n_0'L)$-$A^{Q+}(n_0''S)$  systems, even involving Rydberg states,
we choose the particular states of lithium because we can evaluate the coefficients
precisely using accurate wave functions.
We discuss applications for quantum chemical studies of $\trimerion$ and, as a consequence of our formulation, for
long-range potential energies of $\dimerion$ electronic states correlating to $\ion (1\, {}^1S)$--$\atom (2\,{}^2P)$. Second, different from previous studies on the weak nonadditive interactions for three-body systems composed
of atoms~\cite{bell70,axilrod43,yan16,yan18,XuAlkGor20,lotrich97,lotrich972,lotrich00,lotrich99} 
or of two atoms and an ion~\cite{yan20}, here we find theoretical evidence of a new pure quantum three-body effect that might
have influence on constructing accurate potential surfaces. Specifically, for the
Li($2\,^{2}S$)-Li($2\,^{2}P$)-Li$^+(1\,^{1}S$) system, we find that at particular geometries the two-body additive interactions disappear leaving only three-body nonadditive interactions. These net effects of two- and three-body interactions are quite similar to those for two- and three-body interactions in the case of polar molecules confined in lattice traps~\cite{BarDalPup12,KifLiJak13}
or three Rydberg atoms under the influence of an external electric field~\cite{Qia16}, where the same goal---removal of two-body interactions---was pursued. To provide necessary context we begin with some general contextual background from molecular (chemical) physics and from ultra-cold science.

\subsection{General aspects of triatomic systems}
\label{subsec:introtrimers}
The intrinsic complexity of triatomic 
molecules\footnote{
     A widely-known aphorism (from 1981),
    attributed to Schawlow, warns atomic physicists that,
    ``a diatomic molecule is one atom too many''~\protect\cite{HecBro81}, but 
    it may be predated by an earlier observation (from 1971)
     attributed to Herschbach: ``The trouble with triatomic molecules is, 
    they have one atom too many!''~\protect\cite{Her71}.
    Recently, Gao~\protect\cite{Gao20} emphasized the emergence of 
    chemical complexity beginning with three atoms.
    } 
produces interesting phenomena such as 
conical intersections and geometric phases~\cite{MeaTru79,Bae06} and the Renner-Teller effect~\cite{JunMer80},
while consideration of three atoms at ultra-low energies leads to Efimov~\cite{Mac86,Mac07}, Borromean~\cite{KifLiJak13}, 
and Pfaffian~\cite{ParKeiCir07,BarDalPup12} states and makes the description of collisional processes, such as 
atom-diatom collisions~\cite{TizLePSeb14,deJBesSha20} and three-body recombination loss~\cite{daley09,KraMarWal06,EisKhaLau16} challenging. The demands for understanding the spectroscopy and collisional processes of specific important triatomic molecules at thermal collisional energies 
also continue to drive progress.
For example, ozone ($\textrm{O}_3$) is a vital atmospheric constituent of the planet, with quantum-mechanical collisional cross sections recently reported (see Ref.~\cite{GuiLepHon20} and references therein), tricarbon ($\textrm{C}_3$) is prominent in comets and other astrophysical~\cite{GieMooFuc20} and laboratory realms~\cite{MatKanKaw88},
and $\textrm{H}_3$, $\textrm{H}_3^+$, and their isotopologues, 
serve as long-standing theoretical benchmark systems~\cite{MeaTru79,XieZhaWan20} and are important in astrophysical applications such as, for example, to the cooling of hydrogen gas in molecular clouds~\cite{LepBucDal95} and in the evolution of the early Universe~\cite{LepStaDal02}.

Detailed procedures for calculating
and constructing potential energy surfaces (and other properties) of triatomic systems have been developed, exemplified (for the  representative molecules discussed above) by recent works such as for tricarbon~\cite{RocVar19}, for ozone~\cite{LepBusHon12,AyoBab13,VarPauTru17}, and for $\mbox{H}_3$~\cite{BooKeoMar96}.
A successful strategy to construct three-atom potential energy surfaces using semi-empirical methods requires input calculations of atom-dimer and three-atom long-range potentials~\cite{Var88,VarPai93, AngDobJan20}. To understand the dynamics of low-energy (ultra-cold) collisions, consideration of the long-range potentials is paramount,
see, for example, for atom-molecule systems~\cite{BerLarBer10,LepDulKok10,LepVexBou11,LepDul11,OlaPerRio20}
and for atom-molecular-ion
systems~\cite{Wil17,PurMilSch17,DorYurVil20}. Next, we provide an overview of the lithium dimer and trimer cations.

\subsection{Homonuclear lithum dimer and trimer cations: Excited electronic states}\label{subsec:cations}
We provide a brief overview
of relevant work on the lithium homonuclear systems $\dimerion$ and $\trimerion$ in order to demonstrate that the present work provides data previously not available in the literature.

For the diatomic lithium cation $\dimerion$, four electronic states
(ignoring fine-structure) correlate to
the separated pair $\ion (1\, {}^1S)$--$\atom (2\,{}^2P)$,
namely, $2\,{}^2\Sigma_g$, $2\,{}^2\Sigma_u$, $1\,{}^2\Pi_g$,
and $1\,{}^2\Pi_u$. Model potential method calculations were given
by Magnier~\textit{et al.}~\cite{MagRouAll99} and
by Rabli and McCarroll~\cite{RabMcC17};
a CASSCF/MRCI calculation was reported in Ref.~\cite{JasWilSie07}
(and references therein for earlier work). Magnier \textit{et al.}~\cite{MagRouAll99}
calculated long-range potential curves as functions of internuclear distance $R$, including the exchange energies and electrostatic, induction, and dispersion terms up to $\mathcal{O}(R^{-8})$, but did not give the long-range potential coefficients.
The emphasis of the present paper is on
the three-body system, but because the two-body interactions are available from our calculations, as will be shown in Secs.~\ref{subsec:extractions}--\ref{subsec:PS+}, we will extract the values of the long-range potential coefficients of the four states of $\dimerion$.  

For the triatomic lithium cation $\trimerion$, because
we have found no previous quantum chemical studies of the excited electronic states corresponding to those reported here, we present a summary of calculations on the ground electronic state of $\trimerion$.
In a series of works, Searles, Dunne, and von Nagy-Felsobuki~\cite{DunSeavon87,SeaDunvon88a,SeaDunvon88b,SeavonNag91} calculated the ground state potential energy and dipole moment surfaces,
which were utilized to calculate ro-vibrational spectra~\cite{HenMilTen88,WanvonNag95}.
Surprisingly, we have found few subsequent studies on the 
$\trimerion$ ground electronic state~\cite{TamSza00}; although, very recently as part of a systematic study exploring alkali-metal and alkali-earth-metal hybrid ion-atom diatomic and triatomic systems, \'Smia\l{}kowski and Tomza~\cite{SmiTom20} calculated
equilibrium properties of the ground $^1A_1$ and lowest triplet $^3B_2$ states of $\trimerion$. Our previous paper~\cite{yan20} supplies the long-range interactions for the ground and lowest triplet states of $\trimerion$. In advance of awaited \textit{ab initio} quantum chemical calculations, in the present work,
we calculate the long-range interaction potentials of $\trimerion$ when one Li atom is $\atom(2\, ^2P)$. 

\subsection{Similarity to lattice studies}
\label{subsec:lattice}
B\"{u}chler \textit{et al.} derived
a Hubbard model for
cold polar molecules trapped in
an optical lattice~\cite{BucMicZol07},
with the intent
of realizing a system that
could be used to
model Hamiltonians that exhibit
exotic ground state properties~\cite{BarDalPup12,KifLiJak13}. In terms of the intermolecular
interactions within the lattice,
they write
\begin{equation}
\label{eq:latticedimer}
    U_{ij} = U_0 a^3 |\mathbf{R}_i-\mathbf{R}_j|^{-3}
    +U_1 a^6 
    |\mathbf{R}_i-\mathbf{R}_j|^{-6}
\end{equation}
and
\begin{equation}
\label{eq:latticetrimer}
    W_{ijk} = W_0 a^6 |\mathbf{R}_i-\mathbf{R}_j|^{-3}
    |\mathbf{R}_i-\mathbf{R}_k|^{-3},
\end{equation}
where $U_0$, $U_1$, and $W_0$ are 
certain energy scales, $a$ is a
length scale, ($i$, $j$, $k$)
label the particles, the indices $i,j,k$ are cyclically permuted, and $\mathbf{R}_i$ are certain position vectors of the lattice site (see Ref.~\cite{BucMicZol07} for the complete definitions).
By appropriate ``dressing'' of the cold
molecules by an external static electric field and a microwave field,
they show that the two-body interactions may be tuned ``from repulsive to attractive,
and even switched off, while the three-body terms remain repulsive and strong.'' We will derive two equations
for the present system, Eqs.~(\ref{etotadd2}) and (\ref{etotnonadd2}),
respectively, that are of the same form---but with additional terms---as 
Eqs.~(\ref{eq:latticedimer})
and (\ref{eq:latticetrimer}).
We will show that at specific geometries we recover exactly Eqs.~(\ref{eq:latticedimer})
and (\ref{eq:latticetrimer}).
The anisotropies of the interactions in the present system due to the ion charge and the excited $\atom(2\, ^2P)$ atom are similar to the anisotropies
due to the intermolecular dipole-dipole interactions in the  optical-lattice-trapped cold polar molecular system~\cite{ni10,klein16,AndBurBao21,LiTobMat21}.
Further discussion will be given in Sec.~\ref{subsec:switchoff}.

\section{Theoretical Formulation}\label{The}

\begin{figure} 
\includegraphics[width=18cm,height=9cm]{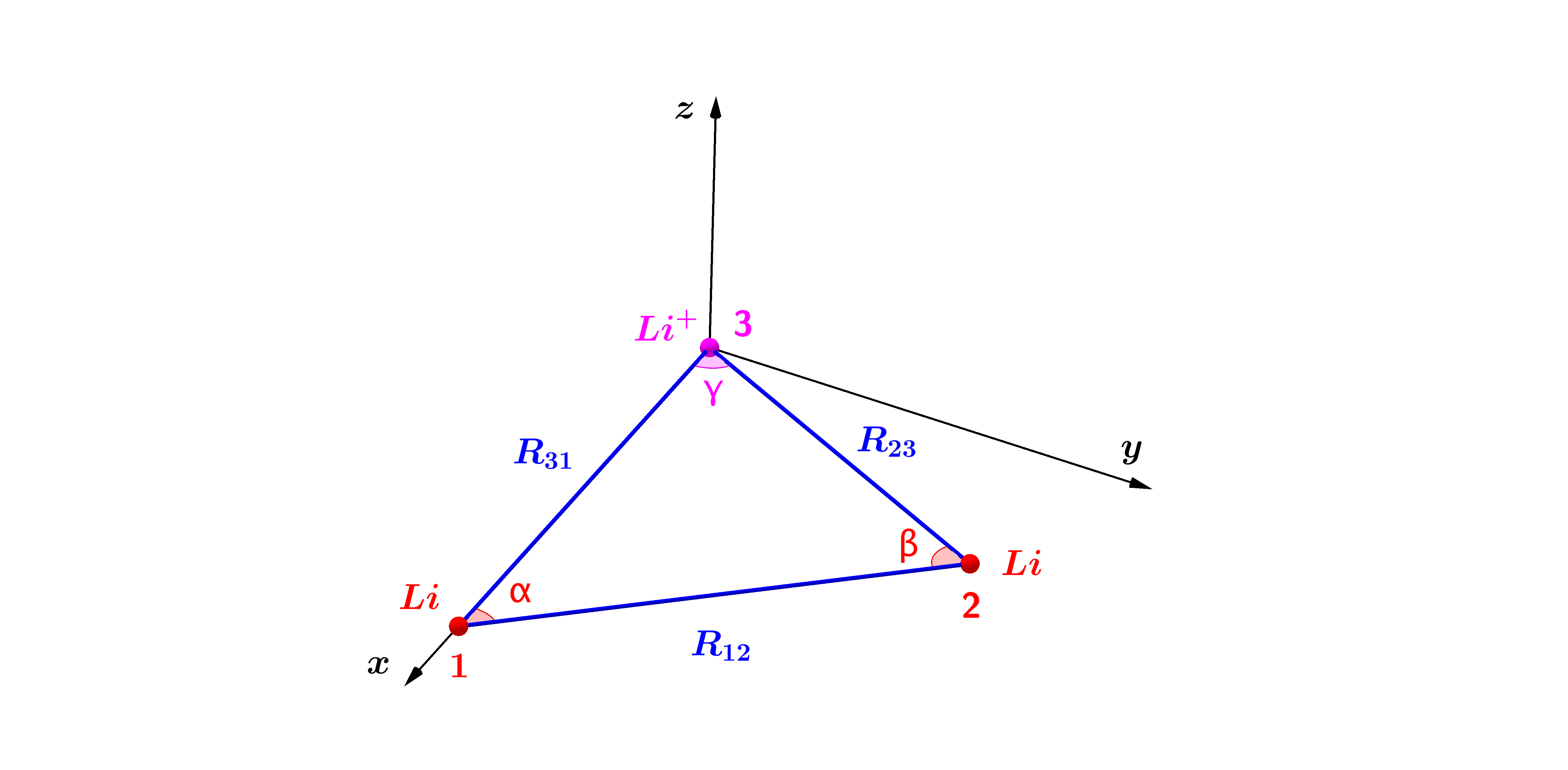}
\caption {\label{fig:coords} Configuration of the Li($2\,^2S$)-Li($2\,^2P$)-Li$^+$($1\,^{1}S$) system. The three particles define the $x\text{--}y$ plane with the two neutral atoms labeled as 1 and 2 and the ion labeled as 3, $R_{IJ}$ are the internuclear distances, and $\alpha$, $\beta$ and $\gamma$ are the interior angles. }
\end{figure}

The geometry of the three-body system is shown in Fig.~\ref{fig:coords}, in which the three particles 
define a plane with the two neutral atoms labeled as 1 and 2 
and the ion labeled as 3.  
It is important to note that due to the degeneracy of atoms 1 and 2 we can't specify
which one is in the ground state or excited state.
The interior angles of the configuration are $\alpha$, $\beta$ and $\gamma$. 

\subsection{ Coulomb potential expansion}\label{subsec:Coulomb}

In the present work, we take the electrostatic interaction $V_{123}$ between pairs of particles for the Li($n_0\,S$)-Li($n_0\,L$)-Li$^{+}$($n_0'\,S$)  system as a perturbation,
\begin{equation}\label{v123co}
H'=V_{123}=V_{12}+V_{23}+V_{31} \,,
\end{equation}
where $V_{12}$, $V_{23}$ and $V_{31}$ are the two-body mutual electrostatic interactions between atoms 1
and 2 and ion 3. For three well-separated atoms or ions, the mutual interaction energy $V_{IJ}$ can be expanded with the same method as used in Refs.~\cite{yan16, yan18, yan20},
thus, 
\begin{equation}\label{e2co}
 V_{IJ}=\sum_{l_Il_J}\sum_{m_Im_J}T_\text{$l_I-m_I$}(\boldsymbol{\sigma})T_\text{$l_Jm_J$}(\boldsymbol{\rho})W_{l_Il_J}^{m_I-m_J}(IJ) \,,
\end{equation}
where the geometry factor is
\begin{eqnarray}\label{e4co}
W_{l_Il_J}^{m_I-m_J}(IJ)&=&\frac{4\pi(-1)^{l_J}}{R_{IJ}^{l_I+l_J+1}}\frac{(l_I+l_J-m_I+m_J)!(l_I,l_J)^{-1/2}}{[(l_I+m_I)!
(l_I-m_I)!(l_J+m_J)!(l_J-m_J)!]^{1/2}}
P_{l_I+l_J}^{m_I-m_J}(\cos\theta_{IJ})\nonumber\\
&\times & \exp[{i(m_I-m_J)\Phi_{IJ}}] \,,
\end{eqnarray}
where ${\bf R}_{IJ}={\bf R}_J-{\bf R}_I$ is the relative position vector from particle $I$ to particle $J$, the notation
$(l_I,l_J,\ldots)=(2l_I+1)(2l_J+1)\ldots$, and $P_{l_I+l_J}^{m_I-m_J}(\cos\theta_{IJ})$ is the associated Legendre function with $\theta_{IJ}$ representing the angle between ${\bf R}_{IJ}$ and the $z$-axis. The 2$^{\ell}$-pole transition operator of an atom consisting of $n+1$ charged particles, in the laboratory frame
is defined as in Ref.~\cite{zhang04},
\begin{eqnarray}\label{Tl11}
T_{{\ell m}}= \sum_{i=0}^{n}q_i{\rho}_i^{\ell}Y_{\ell m}(\hat{\boldsymbol{\rho}}_i) \, ,
\end{eqnarray}
where $q_i$
is the charge of the $i$-th sub-particle of the atom. In the center of mass frame~\cite{zhang04},  $\boldsymbol{\rho}_i$ becomes
\begin{eqnarray}\label{rl11}
\boldsymbol{\rho}_i= \sum_{j=1}^{n}\epsilon_{ij}\textbf{r}_j \, ,
\end{eqnarray}
where $\textbf{r}_i=\boldsymbol{\rho}_i-\boldsymbol{\rho}_0$, $\epsilon_{ij}=\delta_{ij}-m_j/M_T$, $i=0,1,2,...,n$, $j=1,2,...,n$, and $M_T$ represents the total mass of the system. Using the formula
\begin{eqnarray}\label{yl11}
{Y}_{{\ell m}}(\hat{\textbf{r}})= \sqrt{\frac{3}{4\pi}}\prod_{i=1}^{\ell-1} \bigg{(}\sqrt{\frac{2i+3}{i+1}}\bigg{)}(\underbrace{\hat{\textbf{r}}\otimes\hat{\textbf{r}}\otimes\cdots\hat{\textbf{r}}}_{\ell})_m^{(\ell)} \, ,
\end{eqnarray}
where $\otimes$ denotes the coupling between two irreducible tensor operators, the 2$^\ell$-pole transition operator can be simplified as 
\begin{eqnarray}\label{tl222}
{T}_{{\ell}}= \sqrt{\frac{3}{4\pi}}\prod_{m=1}^{\ell-1} \bigg{(}\sqrt{\frac{2m+3}{m+1}}\bigg{)} \sum_{j_1,\cdots,j_{\ell}} \bigg{(} \sum_{i=0}^{n}q_i\epsilon_{ij_1}\epsilon_{ij_2}\cdots\epsilon_{ij_{\ell}} \bigg{)} (\underbrace{\hat{\textbf{r}}_{j_1}\otimes\hat{\textbf{r}}_{j_2}\otimes\cdots\hat{\textbf{r}}_{j_{\ell}}}_{0})_m^{(\ell)} \, .
\end{eqnarray}
For a four-body system, the explicit forms of transition operators $T_{\ell}$ with $\ell$ up to 3 can be found in Ref.~\cite{tang09}.

\subsection{The Hylleraas basis set}\label{subsec:hylleraasbasis}

The nonrelativistic Hamiltonian of the Li atom in the centre of mass frame~\cite{yan97jpb} can be written as
\begin{eqnarray}\label{ham11}
H=-\frac{1}{2\mu} \sum_{i=1}^{3}\nabla_i^2-\frac{1}{m_0} \sum_{i>j\geqslant1}^{3}\nabla_i\cdot\nabla_j+q_0 \sum_{i=1}^{3} \frac{q_i}{r_i} +  \sum_{i>j\geqslant1}^{3}\frac{q_iq_j}{r_{ij}} \, ,
\end{eqnarray}
where $\mu=m_em_0/(m_e+m_0)$ is the reduced mass between an electron $m_e$ and the nucleus $m_0$. The basis set is constructed in Hylleraas coordinates,
\begin{eqnarray}\label{basis}
\phi({\textbf{r}_1,\textbf{r}_2,\textbf{r}_3})={r}_1^{j_1}{r}_2^{j_2}{r}_3^{j_3}{r}_{12}^{j_{12}}{r}_{23}^{j_{23}}{r}_{31}^{j_{31}} e^{-\alpha r_1-\beta r_2 -\gamma r_3} \mathcal{Y}_{(\ell_1\ell_2)\ell_{12},\ell_3}^{(LM)}(\hat{\textbf{r}}_1,\hat{\textbf{r}}_2,\hat{\textbf{r}}_3) \mathcal{X}(1,2,3) \, ,
\end{eqnarray}
where 
\begin{eqnarray}
\mathcal{Y}_{(\ell_1\ell_2)\ell_{12},\ell_3}^{(LM)}(\hat{\textbf{r}}_1,\hat{\textbf{r}}_2,\hat{\textbf{r}}_3)=\sum_{m_i}&&\langle \ell_1 m_1;\ell_2 m_2 |\ell_1\ell_2; \ell_{12}m_{12}  \rangle \langle \ell_{12} m_{12};\ell_3 m_3 |\ell_{12}\ell_3; LM_{L}  \rangle  \nonumber \\
&&\times
Y_{\ell_1m_1}(\hat{\textbf{r}}_1) Y_{\ell_2m_2}(\hat{\textbf{r}}_2)
Y_{\ell_3m_3}(\hat{\textbf{r}}_3) \,  
\end{eqnarray}
is a vector-coupled product of spherical harmonics to form an eigenstate of the total angular momentum $L$ and component $M_L$, and $\mathcal{X}(1,2,3)$ is the three-electron spin-1/2 function.
The variational wave function of the Li atom is a linear combination of basis functions $\phi$ antisymmetrized. With some truncations to avoid the numerical linear dependence, all terms in Eq. (\ref{basis})  are included such that
\begin{eqnarray}\label{truncations}
j_1+j_2+j_3+j_{12}+j_{23}+j_{31} \leqslant \Omega  \, ,
\end{eqnarray}
where $\Omega$ is an integer, and the convergence for the energy eigenvalue is studied by increasing $\Omega$ progressively. The reduced matrix elements for various transition operators can be evaluated with the following basic integral
\begin{eqnarray}
\int && d{\textbf{r}}_1 d{\textbf{r}}_2 d{\textbf{r}}_3{r}_1^{j_1}{r}_2^{j_2}{r}_3^{j_3}{r}_{12}^{j_{12}}{r}_{23}^{j_{23}}{r}_{31}^{j_{31}} e^{-\alpha r_1-\beta r_2 -\gamma r_3} \nonumber \\ && \times Y^*_{\ell'_1m'_1}({\textbf{r}}_1) Y^*_{\ell'_2m'_2}({\textbf{r}}_2)
Y^*_{\ell'_3m'_3}({\textbf{r}}_3)
Y_{\ell_1m_1}({\textbf{r}}_1) Y_{\ell_2m_2}({\textbf{r}}_2)
Y_{\ell_3m_3}({\textbf{r}}_3) \, . \nonumber\\
\end{eqnarray}
The details of computational method for this integral are developed in Refs.~\cite{yan97jpb,Drake95}. Similarly, for the Li$^+$ ion, we also use the Hylleraas variational method to obtain the energies, wavefunctions and transition matrix elements. The detailed Hylleraas method for a two-electron atom is given in Ref.~\cite{yan96cpl}. 

\subsection{The zeroth-order wave function}

For the degenerate Li($n_0\,S$)-Li($n_0\,L$)-Li$^{+}$($n_0'\,S$) system with energy $E_{n_0n_0n_0'}^{(0)}=E_{n_0S}^{(0)}+E_{n_0L}^{(0)}+E_{n_0'S}^{(0)}$,
the zeroth-order wave function can be written as
\begin{eqnarray}\label{wave}
\ket{\Psi^{(0)}}=a\ket{{n_0}L;{n_0}0;{n_0'}0}+b\ket{{n_0}0;{n_0}L;{n_0'}0} \,,
\end{eqnarray}
where $a$ and $b$ are the expansion coefficients of the zeroth-order wave function in the basis set \{\ket{{n_0}L;{n_0}0;{n_0'}0}, \ket{{n_0}0;{n_0}L;{n_0'}0}\} with \ket{{n_0}0}, \ket{{n_0}L}, and \ket{{n'_0}0},
respectively, being the initial states for Li($n_0\,S$), Li($n_0\,L$) and Li$^{+}$($n_0'\,S$). The corresponding zeroth-order wave functions (or the values of $a$ and $b$) 
depend on the geometrical configuration formed by the three particles
and are determined by diagonalizing the perturbation in this basis set.
Then using the degenerate perturbation theory, we can obtain the long-range part of the interaction potential for the Li($2\,^2S$)-Li($2\,^2P$)-Li$^{+}$($1\,^{1}S$) system, which can be written as 
\begin{eqnarray}\label{etot}
\Delta E =\Delta E^{(1)}_{\text{add}}+\Delta E^{(2)}_{\text{add}}+\Delta E^{(2)}_{\text{non}} \,, 
\end{eqnarray}
where $\Delta E^{(1)}_{\text{add}}$ and $\Delta E^{(2)}_{\text{add}}$ are, respectively, the first-order and second-order additive interactions and $\Delta E^{(2)}_{\text{non}}$ is the second-order nonadditive interaction.

\subsection{The first-order additive interactions}

The first-order additive interaction $\Delta E^{(1)}_{\text{add}}$ is given by
\begin{eqnarray}\label{etot1}
&&\Delta E^{(1)}_{\text{add}}= -\frac{C_3^{(12)}(1,M)}{R_{12}^3}-\frac{C_3^{(23)}(1,M)}{R_{23}^3}-\frac{C_3^{(31)}(1,M)}{R_{31}^3} \,,
\end{eqnarray}
where $C_3^{(12)}(1,M)$ describes the dipole-dipole interaction between two neutral atoms. $C_3^{(23)}(1,M)$ and $C_3^{(31)}(1,M)$, respectively, describe the electrostatic interaction between the charge of the ion labeled as 3 and the quadrupole moments of atom~2 and atom~1; the quadrupole moment comes from the excited Li($2\,^2P$) atom, which can be atom 1 or atom 2 due to the degeneracy of the three-body system. These leading long-range interaction coefficients are given by
\begin{eqnarray}\label{c312first}
C_{3}^{(12)}(1,M)&=& (a^{*}b+b^{*}a) \frac{4\pi(-1)^{1+M}}{9(1-M)!(1+M)!}
|\langle{n_0}0\|T_1
\|{n_0}1\rangle|^2
 \,, \label{}
\end{eqnarray}
\begin{eqnarray}\label{c323first}
C_{3}^{(23)}(1,M)&=& |b|^2 Q (-1)^{1+M} \sqrt{\frac{\pi}{5}}
\left(
  \begin{array}{ccc}
    1 & 2 & 1\\
    -M & 0 & M\\
  \end{array}
\right)
\langle{n_0}1\|T_2
\|{n_0}1\rangle\,, 
\end{eqnarray}
\begin{eqnarray}\label{c331first}
C_{3}^{(31)}(1,M)&=& |a|^2 Q (-1)^{1+M} \sqrt{\frac{\pi}{5}}
\left(
  \begin{array}{ccc}
    1 & 2 & 1\\
    -M & 0 & M\\
  \end{array}
\right)
\langle{n_0}1\|T_2
\|{n_0}1\rangle \,, 
\end{eqnarray}
where $Q$ is the charge of the ion, $M$ represents the magnetic quantum number of the excited Li($2\,^2P$) atom, and $T_{\ell}$ is the 2$^{\ell}$-pole transition operator, which is defined in Sec. \ref{subsec:Coulomb}.

\subsection{The second-order additive interactions}

The second-order additive interaction $\Delta E^{(2)}_{\text{add}}$ is given by
\begin{eqnarray}\label{etotadd2}
\Delta E^{(2)}_{\text{add}}= &-& \frac{C_4^{(23)}(1,M)}{R_{23}^4}-\frac{C_4^{(31)}(1,M)}{R_{31}^4}
- \frac{C_6^{(12)}(1,M)}{R_{12}^6}\nonumber \\
&-&\frac{C_6^{(23)}(1,M)}{R_{23}^6}-\frac{C_6^{(31)}(1,M)}{R_{31}^6} - \cdots \, ,
\end{eqnarray}
where $C_4^{(23)}(1,M)$ and $C_4^{(31)}(1,M)$,
respectively, describe the induction interactions between the ion 3 and the neutral atoms 2 and 1.
The dispersion interaction coefficient between the neutral atoms 1 and 2 is given by 
$C_6^{(12)}(1,M)$,
while 
$C_6^{(23)}(1,M)$ and $C_6^{(31)}(1,M)$,
respectively, describe the interactions between the ion 3 and the two neutral atoms 2 and 1, including both the induction and the dispersion interaction coefficients. 
The corresponding expressions for
the additive coefficients are
\begin{eqnarray}
C_{4}^{(23)}(1,M)&=& |a|^2 \mathbb{T}_1 + |b|^2\mathbb{T}_3(M)
\,, \label{C423add}
\end{eqnarray}
\begin{eqnarray}
C_{4}^{(31)}(1,M)&=& |a|^2 \mathbb{T}_3(M) + |b|^2 \mathbb{T}_1
\,, \label{C431add}
\end{eqnarray}
\begin{eqnarray}
C_{6}^{(12)}(1,M)&=& |a|^2 \mathbb{T}_4(M) + |b|^2\mathbb{T}_4(M)
\,, \label{C612add}
\end{eqnarray}
\begin{eqnarray}
C_{6}^{(23)}(1,M)= |a|^2 \mathbb{T}_2
+ |b|^2 \mathbb{T}_5(M)
\,, \label{C623add}
\end{eqnarray}
\begin{eqnarray}
C_{6}^{(31)}(1,M)= |a|^2\mathbb{T}_5(M)+ |b|^2 \mathbb{T}_2
\,, \label{C631add}
\end{eqnarray}
where
\begin{eqnarray}
\mathbb{T}_1= \frac{4\pi Q^2}{9}  \sum_{\substack{n_t}}'\frac{|\langle{n_0}0\|T_{1}\|{n_{t}}1\rangle|^{2}}{E_{{n_{t}1}}-E_{n_00}^{(0)}} \, ,
\end{eqnarray}
\begin{eqnarray}
\mathbb{T}_2 &=&  \frac{4\pi Q^2}{25}  \sum_{\substack{n_t}}'\frac{|\langle{n_0}0\|T_{2}\|{n_{t}}2\rangle|^{2}}{E_{{n_{t}2}}-E_{n_00}^{(0)}} +\frac{32\pi^2}{27} \sum_{\substack{n_tn_{u}}}'\frac{|\langle{n_0}0\|T_{1}\|{n_t}1\rangle|^{2}
|\langle{n_0'}0\|T_{1}\|{n_{u}}1\rangle|^{2}}{(E_{n_t1}-E_{n_00}^{(0)})+(E_{n_{u}1}-E_{n_0'0}^{(0)})}
\, ,
\end{eqnarray}
\begin{eqnarray}
\mathbb{T}_3(M)= \frac{Q^2}{4\pi}\sum_{n_t L_t} \frac{G_1(L_t,0;1,1;1,M)|\langle{n_0}1\|T_1
\|{n_t}L_t\rangle|^2}{E_{n_tL_t}-E_{n_01}^{(0)}}  \, , 
\end{eqnarray}
\begin{eqnarray}
\mathbb{T}_4(M) = \sum_{n_tn_uL_t} \frac{G_1(L_t,1;1,1;1,M)|\langle{n_0}1\|T_1
\|{n_t}L_t\rangle|^2 \langle{ n_0' 0 }\|T_{1}\|{n_u}1\rangle|^{2}}{  (E_{n_tL_t}-E_{n_01}^{(0)})+(E_{n_u1}-E_{ n_0' 0}^{(0)}) }  \, , 
\end{eqnarray}
\begin{eqnarray}
\mathbb{T}_5(M)
&=& \sum_{n_tn_uL_t} \frac{G_1(L_t,1;1,1;1,M)|\langle{n_0}1\|T_1
\|{n_t}L_t\rangle|^2 \langle{ n_0' 0 }\|T_{1}\|{n_u}1\rangle|^{2}}{  (E_{n_tL_t}-E_{n_01}^{(0)})+(E_{n_u1}-E_{ n_0' 0}^{(0)}) } 
\nonumber \\
 &+& \frac{Q^2}{4\pi}\sum_{n_t L_t} \bigg \{ \frac{G_1(L_t,0;2,2;1,M)|\langle{n_0}1\|T_2
\|{n_t}L_t\rangle|^2}{E_{n_tL_t}-E_{n_01}^{(0)}}  \nonumber\\
&+&  \frac{G_1(L_t,0;1,3;1,M)\langle{n_0}1\|T_1
\|{n_t}L_t\rangle^* \langle{n_0}1\|T_3
\|{n_t}L_t\rangle}{E_{n_tL_t}-E_{n_01}^{(0)}}  \nonumber \\
&+& \frac{G_1(L_t,0;3,1;1,M)\langle{n_0}1\|T_3
\|{n_t}L_t\rangle^* \langle{n_0}1\|T_1
\|{n_t}L_t\rangle}{E_{n_tL_t}-E_{n_01}^{(0)}}  \bigg \} 
\, , \label{T5}
\end{eqnarray}
where $G_1$-function is defined by
\begin{eqnarray}\label{G1_main}
G_1(L_i,L_j,\ell_k,\ell_k';L,M) 
&=&  \frac{16\pi^2(\ell_k,\ell_k^{\prime})^{-1/2}}{(2L_{j}+1)^2}\sum_{M_i M_j m_k} \left(
  \begin{array}{ccc}
    L & \ell_k & L_i\\
    -M & m_k & M_i\\
  \end{array}
\right)
\left(
  \begin{array}{ccc}
    L & \ell_{k}^{\prime} & L_i\\
    -M & m_{k} & M_i\\
  \end{array}
\right)   \nonumber \\
&\times& 
\frac{(L_j+\ell_k-M_j+m_k)!(L_j+\ell_k^{\prime}-M_j+m_k)!P_{L_j+\ell_k}^{M_j-m_k}(0)P_{L_j+\ell_k^{\prime}}^{M_j-m_k}(0)}
{(L_j+M_j)!(L_j-M_j)![(\ell_k+m_k)!(\ell_k-m_k)!(\ell_k^{\prime}+m_k)!(\ell_k^{\prime}-m_k)!]^{1/2}}  \,. \nonumber \\
\end{eqnarray}

The detailed derivations are given in the Supplemental Material~\cite{Supplement}. We note that these formulas can also be used to calculate long-range interaction coefficients for other two-body or three-body systems such as the two-body Li($2\,^2S$)-Li($2\,^2P$) system, the two-body Li($2\,^2S$)-Li$^{+}$($1\,^{1}S$) system, the two-body Li($2\,^2P$)-Li$^{+}$($1\,^{1}S$) system, and the three-body Li($2\,^2S$)-Li($2\,^2S$)-Li$^{+}$($1\,^{1}S$) system.
We will provide
specific examples
below in Secs.~\ref{subsec:extractions}--\ref{subsec:PS+}.  

\subsection{The second-order nonadditive potentials}

Due to the degeneracy of the three-body system, the nonadditive potential $\Delta E^{(2)}_{\text{non}}$ starts at the second-order, 
and is given by 
\begin{eqnarray}\label{etotnonadd2}
\Delta E_{\text{non}}^{(2)}=  &-&\frac{C_{3,3}^{(12,23)}(1,M)}{R_{12}^3R_{23}^3}
-\frac{C_{3,3}^{(23,31)}(1,M)}{R_{23}^3R_{31}^3}
-\frac{C_{3,3}^{(31,12)}(1,M)}{R_{31}^3R_{12}^3} \nonumber\\
&-&\frac{C_{4,2}^{(12,23)}(1,M)}{R_{12}^4R_{23}^2}
-\frac{C_{2,4}^{(31,12)}(1,M)}{R_{31}^2R_{12}^4}- \cdots
\,, 
\end{eqnarray}
where $C_{3,3}^{(23,31)}(1,M)$ represents the dispersion nonadditive interaction coefficient. The 
remaining terms are the nonadditive induction interactions. The detailed expressions are given by 
\begin{eqnarray}
C_{3,3}^{(23,31)}(1,M)
&=& \sum_{M_u} (-1)^{M_u+M} G_4(1,M_u;1,M) \bigg\{(a^*b)\exp[i(M_u-M)\gamma] + (b^*a)\exp[-i(M_u-M) \gamma]\bigg\}   
\nonumber \\
&&\times \sum_{n_u}\bigg[
\frac{|\langle{n_0}1\|T_{1}\|{n_0}0\rangle|^2|\langle{n_0''}0\|T_{1}\|{n_u}1\rangle|^{2}}
{(E_{n_u1}-E_{n_0''0})+(E_{n_00}-E_{n_01})}
+
\frac{|\langle{n_0}0\|T_{1}\|{n_0}1\rangle|^2|\langle{n_0''}0\|T_{1}\|{n_u}1\rangle|^{2}}
{(E_{n_u1}-E_{n_0''0})+(E_{n_01}-E_{n_00})}\bigg]
\,,   \label{C62331}
\end{eqnarray}
\begin{eqnarray}
C_{4,2}^{(12,23)}(1,M)  
&=&  |a|^2 \sum_{M_t}(-1)^{M_t+M} G_5(1,M_t;2;1,M;Q)\cos(M_t \beta)  \nonumber \\
&&\times \sum_{n_t}
\frac{\langle{n_0}1\|T_{2}\|{n_0}1\rangle|\langle{n_0}0\|T_{1}\|{n_t}1\rangle|^{2}}{E_{n_s1}-E_{n_00}}
\nonumber \\
&+&  \sum_{M_t m_2'} G_6(2,M_t;1,m_{2}';1,M;Q) \bigg \{(a^*b)\exp[{-i(m_2')\beta}]+ (b^*a)\exp[{i(m_2')\beta}] \bigg \} 
 \nonumber\\ 
&& \times \sum_{n_t}
\frac{ \langle {n_0}0 \|T_1
\|{n_0}1\rangle  \langle{n_0}0\|T_{2}\|{n_t}2\rangle^*
\langle{n_0}1\|T_{1}\|{n_t}2\rangle}{E_{n_t2}-E_{n_01}} \nonumber \\
&-& \sum_{M_t} G_7(1,M_t;2;1,M;Q) \bigg \{(a^*b) \exp[{i(M_{t})\beta}]+(b^*a) \exp[{-i(M_{t})\beta}] \bigg \} \nonumber \\
&& \times \sum_{n_t}
\frac{ \langle {n_0}0 \|T_1 \|{n_0}1\rangle^*  \langle{n_0}1\|T_{2}\|{n_t}1\rangle^* \langle{n_0}0\|T_{1}\|{n_t}1\rangle}{E_{n_t1}-E_{n_00}}
\,, \label{C6421223} 
\end{eqnarray}
\begin{eqnarray}
C_{2,4}^{(31,12)}(1,M) &=& |b|^2 \sum_{M_s}(-1)^{M_s+M} G_5(1,M_s;2;1,M;Q)\cos(M_s \alpha)  \nonumber \\
&&\times  \sum_{n_s}
\frac{\langle{n_0}1\|T_{2}\|{n_0}1\rangle|\langle{n_0}0\|T_{1}\|{n_s}1\rangle|^{2}}{E_{n_s1}-E_{n_00}} 
\nonumber \\
&+&  \sum_{M_s m_1'} G_6(2,M_s;1,m_{1}';1,M;Q) \bigg \{(a^*b)\exp[{i(m_1')\alpha}]+ (b^*a)\exp[{-i(m_1')\alpha}] \bigg \}  \nonumber \\
&&  \times  \sum_{n_s}
\frac{ \langle {n_0}0 \|T_1
\|{n_0}1\rangle  \langle{n_0}0\|T_{2}\|{n_s}2\rangle^*
\langle{n_0}1\|T_{1}\|{n_s}2\rangle}{E_{n_s2}-E_{n_01}} 
\nonumber \\
&-&\sum_{M_s}G_7(1,M_s;2;1,M;Q)  \bigg \{(a^*b) \exp[{i(M_{s})\alpha}]+(b^*a) \exp[{-i(M_{s})\alpha}] \bigg \}
\nonumber \\
&& \times\sum_{n_s}
\frac{ \langle {n_0}0 \|T_1 \|{n_0}1\rangle^*  \langle{n_0}1\|T_{2}\|{n_s}1\rangle^* \langle{n_0}0\|T_{1}\|{n_s}1\rangle}{E_{n_s1}-E_{n_00}}
\,, \label{C6243112}
\end{eqnarray}
\begin{eqnarray}
C_{3,3}^{(12,23)}(1,M)
&=&  \sum_{M_t m_2'} G_6(1,M_t;2,m_{2}';1,M;Q)\bigg \{(a^*b)\exp[{-i(m_2')\beta}]+ (b^*a)\exp[{i(m_2')\beta}] \bigg \}  
\nonumber \\
&& \times \sum_{n_t}
\frac{ \langle {n_0}0 \|T_1
\|{n_0}1\rangle  \langle{n_0}0\|T_{1}\|{n_t}1\rangle^*
\langle{n_0}1\|T_{2}\|{n_t}1\rangle}{E_{n_t1}-E_{n_01}}
\nonumber \\
&-& \sum_{M_t} G_7(2,M_t;1;1,M;Q) \bigg \{(a^*b) \exp[{i(M_{t})\beta}]+(b^*a) \exp[{-i(M_{t})\beta}] \bigg \}
\nonumber \\
&& \times \sum_{n_t}
\frac{ \langle {n_0}0 \|T_1 \|{n_0}1\rangle^*  \langle{n_0}1\|T_{1}\|{n_t}2\rangle^* \langle{n_0}0\|T_{2}\|{n_t}2\rangle}{E_{n_t2}-E_{n_00}}
\,, \label{C6331223}
\end{eqnarray}
and
\begin{eqnarray}
C_{3,3}^{(31,12)}(1,M)
&=& \sum_{M_s m_1'} G_6(1,M_s;2,m_{1}';1,M;Q) \bigg \{(a^*b)\exp[{i(m_1')\alpha}]+ (b^*a)\exp[{-i(m_1')\alpha}] \bigg \} 
\nonumber \\
&&\times \sum_{n_s}
\frac{ \langle {n_0}0 \|T_1
\|{n_0}1\rangle  \langle{n_0}0\|T_{1}\|{n_s}1\rangle^*
\langle{n_0}1\|T_{2}\|{n_s}1\rangle}{E_{n_s1}-E_{n_01}}
\nonumber \\
&-& \sum_{M_s} G_7(2,M_s;1;1,M;Q)\bigg \{a^*b \exp[{i(M_{s})\alpha}]+b^*a \exp[{-i(M_{s})\alpha}] \bigg \}
\nonumber \\
&& \times \sum_{n_s}
\frac{ \langle {n_0}0 \|T_1 \|{n_0}1\rangle^*  \langle{n_0}1\|T_{1}\|{n_s}2\rangle^* \langle{n_0}0\|T_{2}\|{n_s}2\rangle}{E_{n_s2}-E_{n_00}}
\,, \label{C6333112}
\end{eqnarray}
where the functions $G_4$, $G_5$, $G_6$, and $G_7$ are defined by 
\begin{eqnarray}\label{G4}
G_4(L_i,M_i;L,M) &=&  16\pi^2\frac{[P_{L_i+L}^{M_i-M}(0)(L_i+L-M_i+M)!(L_{i},L)^{-1}]^2 }{(L_{i}+M_{i})!(L_{i}-M_{i})!(L+M)!(L-M)!}  \,,
 \\
G_5(L_i,M_i;\ell_{k};L,M;Q)&=&  \frac{8 \sqrt{\pi^3} Q  P_{\ell_{k}+L_{i}}^{M_{i}}(0)P_{L_{i}}^{M_{i}}(0) (\ell_{k}+L_{i}-M_{i})! }{(2L_i+1)^2\sqrt{2\ell_{k}+1}(l_1)!(L_{i}+M_{i})!}
\left(
  \begin{array}{ccc}
    L & \ell_{k} & L\\
    -M & 0 & M\\
  \end{array}
\right)  \,,
 \\
G_6(L_i,M_i;\ell_{k},m_{k};L,M;Q)&=& 
\frac{8\sqrt{\pi^3}Q(\ell_{k})^{-1/2}}{(2L+1)(2L_i+1)} \left(
  \begin{array}{ccc}
    L & \ell_{k} & L_{i}\\
    -M & -m_{k} & M_{i}\\
  \end{array}
\right)  \nonumber \\
&\times& \frac{P_{L+L_{i}}^{-M+M_{i}}(0)P_{\ell_{k}}^{m_{k}}(0)(L+L_{i}+M-M_{i})!(\ell_{k}-m_{k})!}{[(L+M)!
(L-M)!(L_{i}+M_{i})!(L_{i}-M_{i})!(\ell_{k}+m_{k})!
(\ell_{k}-m_{k})!]^{1/2}}  \,,
\nonumber \\ \\
G_7(L_i,M_i;\ell_{k};L,M;Q)&=&
 \frac{8\sqrt{\pi^3}Q(\ell_{k})^{-1/2}}{(2L+1)(2L_i+1)} \sum_{m_{k}} \left(
  \begin{array}{ccc}
    L & \ell_{k} & L_{i}\\
    -M & m_{k} & M_{i}\\
  \end{array}
\right)  \nonumber \\
&\times& \frac{P_{L+\ell_{k}}^{M-m_{k}}(0)P_{L_i}^{M_{i}}(0)(L+\ell_{k}-M+m_{k})!(L_{i}-M_{i})!}{[(L+M)!
(L-M)!(L_{i}+M_{i})!(L_{i}-M_{i})!(\ell_{k}+m_{k})!
(\ell_{k}-m_{k})!]^{1/2}} \,.
\nonumber\\
\end{eqnarray}

The detailed derivations are given in the Supplemental Material~\cite{Supplement}. From 
Eqs.~(\ref{c312first})--(\ref{C6333112}), we see that all of these coefficients
depend on the atomic states of the three-body system  because they include $a$ and $b$. 
In other words, these additive and nonadditive coefficients show a dependence on the configurations of the three-body system. This is clearly a kind of quantum three-body collective effect. In the following subsection, we show that these three-body nonadditive interactions 
significantly
influence the total interaction potentials. 
Because of the enhancement through the induction effect, the nonadditive interactions are large enough to be comparable to (or even stronger than) the additive interactions at the same order.

In the present paper, we only consider long-range interaction for the Li($2\,^2S$)-Li($2\,^2P$)-Li$^{+}$($1\,^{1}S$) system up to $\mathcal{O}(R^{-6})$, since the next terms are $C_7/R^7$, which come from the third-order perturbation theory. 

\subsection{Specific results extracted from the general expressions}
\label{subsec:extractions}

With the zeroth-order wave functions as shown in Eq.~(\ref{wave}), the present work can be easily related to the calculations of long-range interactions for other two-body or three-body systems. For example, if we set $a={1}$, $b=0$ and remove the terms involving the Li$^{+}$($1\,^{1}S$) ion, the formulae can be used to describe the long-range interactions for the two-body Li($2\,^2S$)-Li($2\,^2S$) system; if we set $a=\frac{1}{\sqrt{2}}$, $b=\pm\frac{1}{\sqrt{2}}$ and remove the terms involving the Li$^{+}$($1\,^{1}S$) ion, the formulae can be used to describe the long-range interactions for the two-body Li($2\,^2S$)-Li($2\,^2P$) system; if we set $a={1}$, $b=0$ and remove the terms involving the Li($2\,^2P$) atom, the formulae can be used to describe the long-range interactions for the two-body Li($2\,^2S$)-Li$^{+}$($1\,^{1}S$) system; if we set $a={1}$, $b=0$ and remove the terms involving the Li($2\,^2S$) atom, the formulae can be used to describe the long-range interactions for the two-body Li($2\,^2P$)-Li$^{+}$($1\,^{1}S$) system; and if we set $a={1}$, $b=0$ and $L=0$, the formulae can be used to describe the long-range interactions for the three-body Li($2\,^2S$)-Li($2\,^2S$)-Li$^{+}$($1\,^{1}S$) system. For these long-range \textit{additive} interaction coefficients, we have arranged the following formulae to show these connections:  
\begin{eqnarray}\label{c312new1}
 C_{3}^{(12)}(1,M) &=& C_{3,\text{dip}}^{(P-S)}
 \,, \label{C3anew1}
\end{eqnarray}
\begin{eqnarray}\label{c323new1}
C_{3}^{(23)}(1,M)&=& |b|^2 C_{3,\text{elst}}^{(P-S^+)}\,, \label{C323new1}
\end{eqnarray}
\begin{eqnarray}\label{c331new1}
C_{3}^{(31)}(1,M)&=& |a|^2 C_{3,\text{elst}}^{(P-S^+)} \,, \label{C331new1}
\end{eqnarray}
\begin{eqnarray}
C_{4}^{(23)}(1,M)= |a|^2 C_{4,\text{ind}}^{(S-S^+)} + |b|^2 C_{4,\text{ind}}^{(P-S^+)}(M)
\,, \label{C423new1}
\end{eqnarray}
\begin{eqnarray}
C_{4}^{(31)}(1,M)= |a|^2 C_{4,\text{ind}}^{(P-S^+)}(M) + |b|^2 C_{4,\text{ind}}^{(S-S^+)}
\,, \label{C431new1}
\end{eqnarray}
\begin{eqnarray}
C_{6}^{(12)}(1,M)= C_{6,\text{disp}}^{(P-S)}(M)
\,, \label{C612}
\end{eqnarray}
\begin{eqnarray}
C_{6}^{(23)}(1,M)= |a|^2 \bigg \{ C_{6,\text{ind}}^{(S-S^+)} + C_{6,\text{disp}}^{(S-S^+)} \bigg \}
+ |b|^2  \bigg \{C_{6,ind}^{(P-S^+)}(M) + C_{6,\text{disp}}^{(P-S^+)}(M)  \bigg \}
\,, \label{C623new1}
\end{eqnarray}
and
\begin{eqnarray}
C_{6}^{(31)}(1,M)= |a|^2 \bigg \{ C_{6,\text{ind}}^{(P-S^+)}(M) + C_{6,\text{disp}}^{(P-S^+)}(M)  \bigg \} + |b|^2 \bigg \{ C_{6,\text{ind}}^{(S-S^+)} + C_{6,\text{disp}}^{(S-S^+)} \bigg \}
\,, \label{C631new1}
\end{eqnarray}
where $C_{3,\text{dip}}^{(P-S)}$ and $C_{6,\text{disp}}^{(P-S)}$, respectively, represent the dipolar and dispersion interaction coefficients 
for the two-body Li($2\,^2S$)-Li($2\,^2P$) system, which have been given in the Ref.~\cite{yan16} (also see Eqs.~(51) and~(52) in the Supplemental Material~\cite{Supplement}). $C_{2n,\text{ind}}^{(S-S^+)}$ and $C_{2n,\text{disp}}^{(S-S^+)}$ represent the long-range induction and dispersion coefficients for the Li($2\,^2S$)-Li$^+$($1\,^{1}S$) system, which have been given in the Ref.~\cite{yan20} (also see Eqs.~(48)--(50) in the Supplemental Material~\cite{Supplement}).
For the two-body Li($2\,^2P$)-Li$^{+}$($1\,^{1}S$) system, we  provide more
detail below in Sec.~\ref{subsec:PS+} as no previous numerical
results for the coefficients
were found
in the literature. In short,
$C_{3,\text{elst}}^{(P-S^+)}$ represents the electrostatic interaction between the charge of the ion
and the quadrupole moment of the neutral atom; $C_{2n,\text{ind}}^{(P-S^+)}$ and $C_{2n,\text{disp}}^{(P-S^+)}$ represent the long-range induction and dispersion coefficients for the Li($2\,^2P$)-Li$^+$($1\,^{1}S$) system, where the formulae of these coefficients are given by 
Eqs.~(\ref{c312})--(\ref{c623i}) in subsection~\ref{subsec:PS+}.  Clearly, with these formulae, we can easily relate the long-range \textit{additive} interactions of the three-body Li($2\,^2S$)-Li($2\,^2P$)-Li$^{+}$($1\,^{1}S$) system to those of other two-body or three-body systems.
On the other hand,
the
\textit{nonadditive}
interactions of the three-body Li($2\,^2S$)-Li($2\,^2P$)-Li$^{+}$($1\,^{1}S$) system
are induced by the degeneracy 
and cannot be decomposed
in terms of diatomic subsystems.
This is in contrast to
the nondegenerate Li($2\,^2S$)-Li($2\,^2S$)-Li$^{+}$($1\,^{1}S$) system~\cite{yan20}, 
where the nonadditive interactions start from the third-order energy correction 
and may still be used to predict contributions
to the long-range potentials between the $\atom (2\,^2S)$ atom and the excited state dimer $\dimerion ( 2\,{}^2\Sigma_{g,u}^+,1\,{}^2\Sigma_{g,u})$
or between the $\atom (2\,^2P)$ atom and the ground state dimer $\dimerion (1\,{}^2\Sigma_{g/u}^+)$.
We note
that $C_{4,2}^{(12,23)}(1,M)$ [see Eq.~(\ref{C6421223})]
and $C_{2,4}^{(31,12)}(1,M)$ [see Eq.~(\ref{C6243112})] may be very important in the study of the interactions between the cation $\ion$($1\,^{1}S$) and the excited dimer $\dimerion ( 2\,{}^2\Sigma_{g,u}^+,1\,{}^2\Sigma_{g,u})$. 

\subsection{Orientation-dependence considerations
}\label{subsec:axis}

\begin{figure}
\includegraphics[width=16cm,height=8cm]{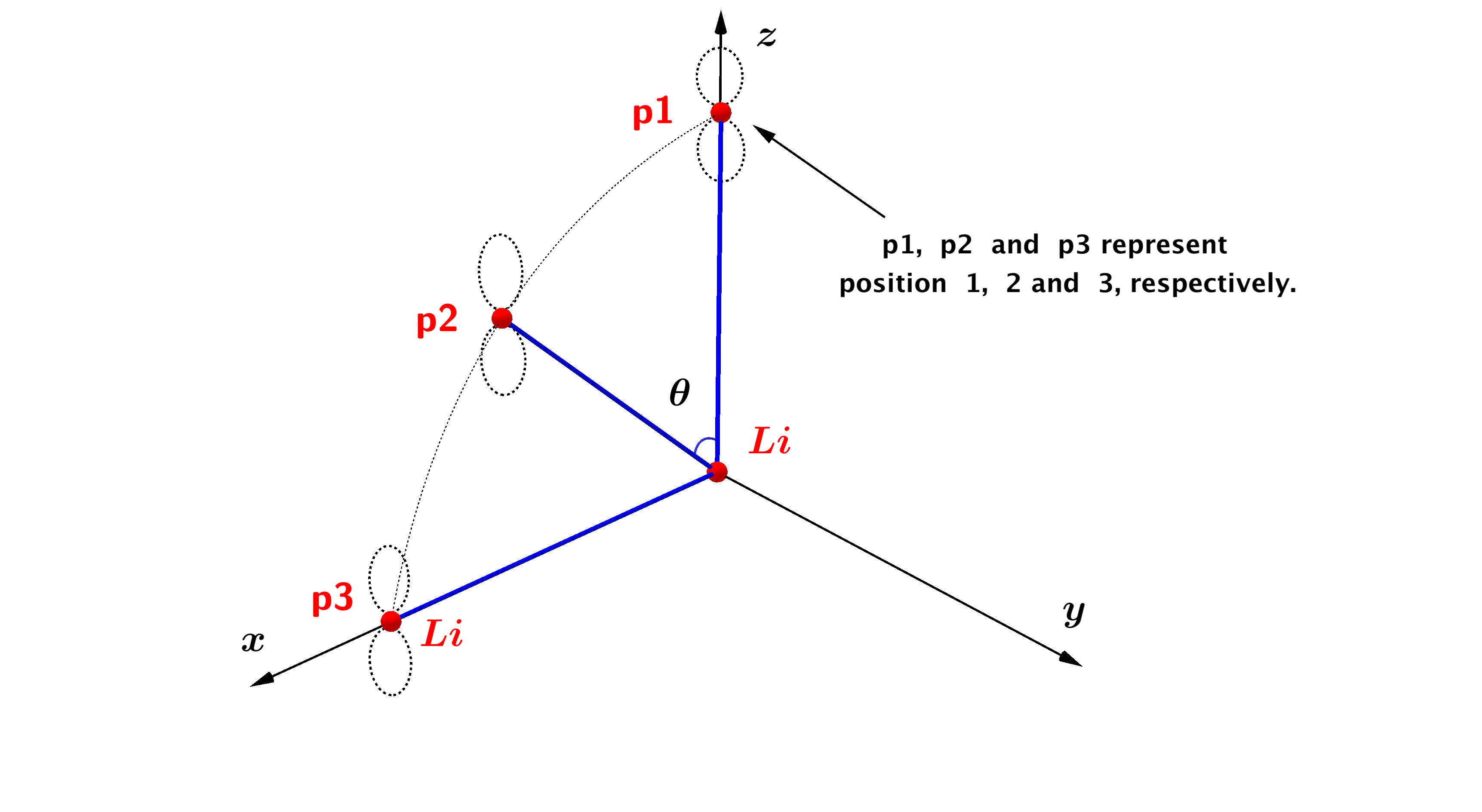}
\caption {\label{fig:axis} Simple illustration of the rotation of the two-body system from $z$-axis to $x$-axis.  }
\end{figure}

In this subsection, we describe
the orientation-dependence
of the long-range interactions
due to 
the anisotropic charge distribution of the excited Li atom.
To illustrate the orientation-dependence,
we use the two-body Li($2\,^{2}S$)-Li($2\,^{2}P$) system as an example. The rotation of the two-body system is illustrated in Fig.~\ref{fig:axis}, where the two-body system is rotated from $z$-axis to $x$-axis (or one of the two atoms is rotated from ``p1'' to ``p3''). In this process, we find that the zeroth-order wave functions of the two-body system $\Psi_{S-P}^{(0)}(M)$ and the corresponding long-range interaction coefficients change. Thus we can easily get the following inequality relation
\begin{eqnarray}
C_{n,p_1}^{(S-P)}(M) \ne C_{n,p_2}^{(S-P)}(M) \ne C_{n,p_3}^{(S-P)}(M)
\,. \label{Cnp123}
\end{eqnarray}

\begingroup
\squeezetable
\begin{table*}
\caption{The long-range interaction coefficients (in atomic units) of the $^\infty$Li($2\,^2S$)-$^\infty$Li($2\,^{2}P$) system for the two atoms lying on the $z$-axis and the $x$-axis, respectively, 
(``p1'') and (``p3'') as shown in Fig.~\ref{fig:axis}.
The numbers in parentheses represent the computational uncertainties. }\label{TabSPADD}
\begin{ruledtabular}
\begin{tabular}{lcccccc}
\multicolumn{1}{c}{``p1''}  &  \multicolumn{1}{c}{$C_{3,z}^{(S-P)}(M=0)$}  & \multicolumn{1}{c}{$C_{3,z}^{(S-P)}(M=\pm1)$}  & \multicolumn{1}{c}{$C_{6,z}^{(S-P)}(M=0)$}  & \multicolumn{1}{c}{$C_{6,z}^{(S-P)}(M=\pm1)$}
\\ \hline
\\
\multicolumn{1}{c}{$\Psi_{S-P,z}^{(0)}$($\beta$=1)}      &11.000221(2)   &$-$5.500111(1)     & 2075.40(3)  &1406.68(3)   
\\
\multicolumn{1}{c}{$\Psi_{S-P,z}^{(0)}$($\beta$=-1)}       &$-$11.000221(2) &5.500111(1)   & 2075.40(3)  &1406.68(3)  \\
\hline 
\\
\multicolumn{1}{c}{``p3''}  &  \multicolumn{1}{c}{$C_{3,x}^{(S-P)}(M=0)$}  & \multicolumn{1}{c}{$C_{3,x}^{(S-P)}(M=\pm1)$}  & \multicolumn{1}{c}{$C_{6,x}^{(S-P)}(M=0)$}  & \multicolumn{1}{c}{$C_{6,x}^{(S-P)}(M=\pm1)$}
\\ \hline
\\
\multicolumn{1}{c}{$\Psi_{S-P,x}^{(0)}$($\beta$=1)}      & $-$5.500111(1)   &2.750054(1)     &1406.68(3)  &1741.06(5)   
\\
\multicolumn{1}{c}{$\Psi_{S-P,x}^{(0)}$($\beta$=-1)}       &5.500111(1) & $-$2.750054(1) &1406.68(3) & 1741.06(5)                         \\
\end{tabular}
\end{ruledtabular}
\end{table*}
\endgroup

In our previous work, we have given the numerical values of the long-range interaction coefficients for the two atoms lying on the $z$-axis (see Table IX in Ref.~\cite{tang09}),
which would correspond
to  ``p3'' in Fig.~\ref{fig:axis}.
In the present work, we use the coordinates
of Fig.~\ref{fig:coords}, represented
as ``p1'' in Fig.~\ref{fig:axis},
which corresponds to the two atoms lying on the $x$-axis. The comparison of these long-range interaction coefficients is given in  Table~\ref{TabSPADD},
where the present values were obtained using  highly accurate variational wave functions for the Li atom in Hylleraas coordinates with finite nuclear mass effects~\cite{tang09}.
For these two specific situations
(``p1'' and ``p3''), we also find the following relations
\begin{eqnarray}
C_{n,p_3}^{(S-P)}(M=0) = C_{n,p_1}^{(S-P)}(M=\pm1)
\,, \label{Cnaxis1}
\end{eqnarray}
which are obeyed by
our numerical values of these coefficients shown in Table~\ref{TabSPADD}.

In general, the
long-range interaction coefficients are given by 
\begin{eqnarray}
C_{n,p_i}^{(S-P)}(M) = C_{n,p_1}^{(S-P)}(M,\cos\theta_i)
\,, \label{Cnaxis}
\end{eqnarray}
where $p_i$ is the position i of the atom as shown in Fig.~\ref{fig:axis} and $\theta_i$ is the corresponding polar angle. For the leading coefficients $C_3$, the formulas 
are simplified as
\begin{eqnarray}
C_{3,p_i}^{(S-P)}(M) = C_{3,p_1}^{(S-P)}(M)P_{2}(\cos\theta_i)
\,, \label{C3axis}
\end{eqnarray}
where $P_{2}(\cos\theta_i)$ is the Legendre polynomial. For the other coefficients $C_n$ with $(n>3)$, the
parts 
containing the polar angle $\theta_i$ would be coupled with the virtual states, which cannot be separated. But we can still 
utilize the present formulas 
(calculated at  the ``p1''
orientation) to give the general formulas by changing the Legendre polynomial from $P_{l_I}^{m_I}(0)$ to $P_{l_I}^{m_I}(\cos\theta_i)$. For example, we can use the formula of  Eq.~(52) in the Supplemental Material~\cite{Supplement}
to get the general leading dispersion coefficient $C_{6,p_i}^{(S-P)}(M)$
at orientation p$_i$. Similarly, for other excited Li dimer and trimers, the long-range interactions also contain such orientation dependencies.
In the next subsection,
we will apply these ideas
to derive the long-range potentials
for the $\atom (2\,{}^2P)$--$\ion (1\, {}^1S)$ system in
the ``p3'' orientation.
In Sec.~\ref{res},  we will
consider the three-body system
in detail.

\subsection{The long-range potentials for the $\atom (2\,{}^2P)$--$\ion (1\, {}^1S)$ system}\label{subsec:PS+}

In this subsection, we 
use our results
to calculate the long-range potentials
of the four states of 
$\dimerion$,
$1\,{}^2\Pi_u$,
$1\,{}^2\Pi_g$,
$2\,{}^2\Sigma_g^+$,  
and $2\,{}^2\Sigma_u^+$,
correlating to $\atom (2\,{}^2P)$--$\ion (1\, {}^1S)$ system.
We begin by writing down the long-range
potential as calculated in our coordinates, see Fig.~\ref{fig:coords},
with $V^{(P-S^+)}(R;M)$
corresponding to the two-body $\atom (2\,{}^2P)$--$\ion (1\, {}^1S)$ system
in the orientation ``p3''
of Fig.~\ref{fig:axis}, which can be written as
\begin{eqnarray}\label{etotPS+}
&&V^{(P-S^+)}(R;M)= -\frac{C_{3}(M)}{R^3}-\frac{C_{4}(M)}{R^4}-\frac{C_{6}(M)}{R^6} - \cdots \,,
\end{eqnarray}
where $C_{3}(M)$ represents the electrostatic interaction between the charge of the $\ion (1\, {}^1S)$ ion
and the quadrupole moment of the excited $\atom (2\,{}^2P)$ atom, $C_{4}(M)$ represents the leading long-range induction coefficient, which is related to the dipole polarizability of the $\atom (2\,{}^2P)$ atom,  and the $C_{6}(M)$ is the sum of long-range induction coefficients $C_{6,\text{ind}}$ and dispersion coefficients $C_{6,\text{disp}}$. The formulae of these coefficients are given by
\begin{eqnarray}\label{c312}
C_{3}(M)= C_{3,\text{elst}}^{(P-S^+)}(M)= Q (-1)^{1+M} \sqrt{\frac{\pi}{5}}
\left(
  \begin{array}{ccc}
    1 & 2 & 1\\
    -M & 0 & M\\
  \end{array}
\right)
\langle{n_0}1\|T_2
\|{n_0}1\rangle\ \, ,
\end{eqnarray}
\begin{eqnarray}\label{c423i}
C_{4}(M)= C_{4,\text{ind}}^{(P-S^+)}(M)
 &=& \frac{Q^2}{4\pi}\sum_{n_t L_t} \frac{G_1(L_t,0;1,1;1,M)|\langle{n_0}1\|T_1
\|{n_t}L_t\rangle|^2}{E_{n_tL_t}-E_{n_01}^{(0)}}
\, ,
\end{eqnarray}
\begin{eqnarray}\label{c623iall}
C_{6}(M)= C_{6,\text{disp}}^{(P-S^+)}(M)+C_{6,\text{ind}}^{(P-S^+)}(M) \, ,
\end{eqnarray}
where
\begin{eqnarray}\label{c623disp}
C_{6,\text{disp}}^{(P-S^+)}(M)
&=& \sum_{n_tn_uL_t} \frac{G_1(L_t,1;1,1;1,M)|\langle{n_0}1\|T_1
\|{n_t}L_t\rangle|^2 \langle{ n_0' 0 }\|T_{1}\|{n_u}1\rangle|^{2}}{  (E_{n_tL_t}-E_{n_01}^{(0)})+(E_{n_u1}-E_{ n_0' 0}^{(0)}) } 
\, ,
\end{eqnarray}
and
\begin{eqnarray}\label{c623i}
C_{6,\text{ind}}^{(P-S^+)}(M)
 &=& \frac{Q^2}{4\pi}\sum_{n_tn_u L_t} \bigg \{ \frac{G_1(L_t,0;2,2;1,M)|\langle{n_0}1\|T_2
\|{n_t}L_t\rangle|^2}{E_{n_tL_t}-E_{n_01}^{(0)}}  \nonumber \\
&+&  \frac{G_1(L_t,0;1,3;1,M)\langle{n_0}1\|T_1
\|{n_t}L_t\rangle^* \langle{n_0}1\|T_3
\|{n_t}L_t\rangle}{E_{n_tL_t}-E_{n_01}^{(0)}}  \nonumber \\
&+& \frac{G_1(L_t,0;3,1;1,M)\langle{n_0}1\|T_3
\|{n_t}L_t\rangle^* \langle{n_0}1\|T_1
\|{n_t}L_t\rangle}{E_{n_tL_t}-E_{n_01}^{(0)}} \bigg \} 
\, ,
\end{eqnarray}
and the $G_1$ function is defined in
Eq.~(\ref{G1_main}).

\begingroup
\squeezetable
\begin{table*}	
\caption{The long-range interaction coefficients (in atomic units) of the Li($2\,^{2}P$)-Li$^+$($1\,^2S$) system for the two particles lying on $z$-axis and $x$-axis, respectively, ``p1'' and ``p3'' as shown in Fig.~\ref{fig:axis}.
Position ``p1'' corresponds to standard molecular $\Sigma$, $\Pi$ nomenclature where the $z$-axis joins the atom and the ion. The numbers in parentheses represent the computational uncertainties.}
\label{TabP+ADD1}
\begin{ruledtabular}
\begin{tabular}{lcccccc}
\multicolumn{1}{c}{``p1''}  &  \multicolumn{1}{c}{$C_{3,z}^{(P-S^+)}(M=0)$}  & \multicolumn{1}{c}{$C_{3,z}^{(P-S^+)}(M=\pm1)$}  & \multicolumn{1}{c}{$C_{4,z}^{(P-S^+)}(M=0)$}  & \multicolumn{1}{c}{$C_{4,z}^{(P-S^+)}(M=\pm1)$} & \multicolumn{1}{c}{$C_{6,z}^{(P-S^+)}(M=0)$}  & \multicolumn{1}{c}{$C_{6,z}^{(P-S^+)}(M=\pm1)$}
\\ 
\hline
\\
\multicolumn{1}{c}{$^{\infty}$Li}   &10.8199392(2) &$-$5.4099696(1)  & 61.8515(2) &64.2838(2) &9811.485(6) &$-$1820.6261(3) 
\\
\multicolumn{1}{c}{$^{7}$Li}       &10.8192592(2)  &$-$5.4096296(1)  &61.8385(2)  &64.2911(2) &9811.274(6) &$-$1820.7205(2)                         \\
\multicolumn{1}{c}{$^{6}$Li}       &10.8191462(2)  &$-$5.4095731(1)  &61.8364(2)  &64.2925(2) &9811.239(6) &$-$1820.7362(3) 
\\
\hline
\\
\multicolumn{1}{c}{``p3''}  &  \multicolumn{1}{c}{$C_{3,x}^{(P-S^+)}(M=0)$}  & \multicolumn{1}{c}{$C_{3,x}^{(P-S^+)}(M=\pm1)$}  & \multicolumn{1}{c}{$C_{4,x}^{(P-S^+)}(M=0)$}  & \multicolumn{1}{c}{$C_{4,x}^{(P-S^+)}(M=\pm1)$}& \multicolumn{1}{c}{$C_{6,x}^{(P-S^+)}(M=0)$}  & \multicolumn{1}{c}{$C_{6,x}^{(P-S^+)}(M=\pm1)$}
\\ \hline
\\
\multicolumn{1}{c}{$^{\infty}$Li}      & $-$5.4099696(1)  &2.7049847(1)      &64.2838(2)   &63.0676(3)   &$-$1820.6261(3)   &3995.429(3) 
\\
\multicolumn{1}{c}{$^{7}$Li}       &$-$5.4096296(1) &2.7048148(1)   &64.2911(2) &63.0648(2)  &$-$1820.7205(2) &3995.276(3)                          \\
\multicolumn{1}{c}{$^{6}$Li}       &$-$5.4095731(1)  &2.7047866(1)   &64.2925(2)  &63.0643(3) &$-$1820.7362(3)    &3995.251(3)      \\
\end{tabular}
\end{ruledtabular}
\end{table*}
\endgroup

\begin{figure} 
\includegraphics[width=14cm,height=9cm]{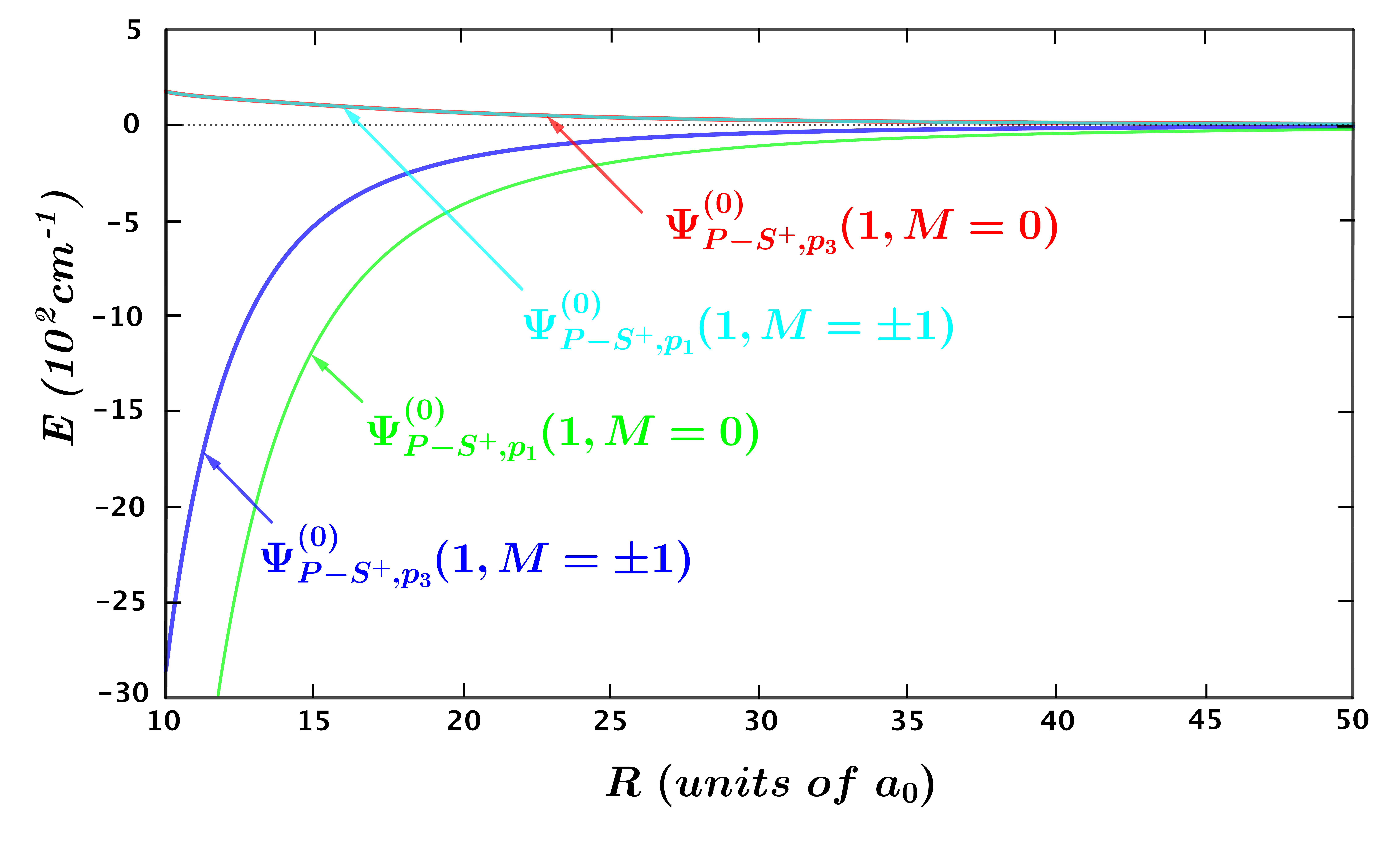}
\caption {\label{fig:dimer} Long-range potentials (in atomic units) of the Li($2\,^2P$)-Li$^+$($2\,^{1}S$) system calculated in the coordinate system of Fig.~\ref{fig:coords}, which
corresponds to ``p3'' of Fig.~\ref{fig:axis}.
}
\end{figure}

\begingroup
\begin{table*}	
\caption{Convergence of the long-range interaction coefficients $C_{3,x}^{(P-S^+)}(M)$ for the $^{\infty}$Li($2\,^{2}P$)-$^{\infty}$Li$^+$($1\,^2S$) system, where two particles lie on the $x$-axis (``p3'') as shown in Fig.~\ref{fig:axis}. $N_P$ denotes the size of the basis for the $P$ state of the $^{\infty}$Li($2\,^{2}P$) atom.}
\label{TabP+c_3con}
\begin{ruledtabular}
\begin{tabular}{lcccccc}
\multicolumn{1}{c}{$N_P$}  &  \multicolumn{1}{c}{$C_{3,x}^{(P-S^+)}(M=0)$}  & \multicolumn{1}{c}{$C_{3,x}^{(P-S^+)}(M=\pm1)$}  
\\ \hline
\multicolumn{1}{c}{1174}      & $-$5.409965844  &2.704982922    
\\
\multicolumn{1}{c}{2091}       &$-$5.409968720 &2.704984360    \\
\multicolumn{1}{c}{3543}       &$-$5.409969427  &2.704984713 
\\
\multicolumn{1}{c}{5761}       &$-$5.409969563  &2.704984781 
\\
\multicolumn{1}{c}{Extrapolated}       &$-$5.4099696(1)  &2.7049847(1) \\
\end{tabular}
\end{ruledtabular}
\end{table*}
\endgroup
\begingroup
\begin{table*}	
\caption{Convergence of the long-range interaction coefficients $C_{4,x}^{(P-S^+)}(M)$ for the $^{\infty}$Li($2\,^{2}P$)-$^{\infty}$Li$^+$($1\,^2S$) system, where two particles lie on $x$-axis (``p3'') as shown in Fig.~\ref{fig:axis}. $N_P$ denotes the size of basis for the $P$ state of the $^{\infty}$Li atom. $N_{S}$, $N_{(pp)P}$, and $N_{D}$, respectively, denote the sizes of the bases for the corresponding intermediate states of symmetries $S$, $P$, and $D$, and $(pp)P$ stands for the main configuration
of two $p$ electrons coupled to form a total angular
momentum of $P$ (since the contribution from the $(pp)P$ configuration is well converged at $N_{(pp)P}=3413$, we did not increase $N_{(pp)P}$ any further~\cite{tang09}).}
\label{TabP+c_4con}
\begin{ruledtabular}
\begin{tabular}{lcccccc}
\multicolumn{1}{c}{$(N_P,N_{S},N_{(pp)P},N_{D})$}  &  \multicolumn{1}{c}{$C_{4,x}^{(P-S^+)}(M=0)$}  & \multicolumn{1}{c}{$C_{4,x}^{(P-S^+)}(M=\pm1)$}  
\\ \hline
\multicolumn{1}{c}{(1174,1589,1106,1174)}      & 64.282596  &63.061604    
\\
\multicolumn{1}{c}{(2091,2625,2002,2091)}       &64.283174 &63.066184    \\
\multicolumn{1}{c}{(3543,4172,3413,3543)}       &64.283498  &63.067033 
\\
\multicolumn{1}{c}{(5761,6412,3413,5761)}       &64.283634  &63.067338 
\\
\multicolumn{1}{c}{Extrapolated}       &64.2838(2)  &63.0676(3) \\
\end{tabular}
\end{ruledtabular}
\end{table*}
\endgroup
\begingroup
\begin{table*}	
\caption{Convergence of the long-range interaction coefficients $C_{6,x}^{(P-S^+)}(M)$ for the $^{\infty}$Li($2\,^{2}P$)-$^{\infty}$Li$^+$($1\,^2S$) system, where two particles lie on the $x$-axis (``p3'') as shown in Fig.~\ref{fig:axis}. $N_P$ denotes the sizes of bases for the $P$ state of the $^{\infty}$Li atom. $N_{S}$, $N_{(pp)P}$, and $N_{D}$, respectively, are the sizes of basis for the corresponding intermediate states of symmetries $S$, $P$, and $D$, and $(pp)P$ stands for the main configuration
of two $p$ electrons coupled to form a total angular
momentum of $P$ (since the contribution from the $(pp)P$ configuration is well converged at $N_{(pp)P}=3413$, we did not increase $N_{(pp)P}$ any further~\cite{tang09}). $N_{S}^+$ and $N_{P}^+$,
respectively, denote the sizes of the bases for the ground state and the intermediate states of symmetry $P$ of $^{\infty}$Li$^+$.}
\label{TabP+c_6con}
\begin{ruledtabular}
\begin{tabular}{lcccccc}
\multicolumn{1}{c}{$(N_S^+,N_{P}^+;N_P,N_{S},N_{(pp)P},N_{D})$}  &  \multicolumn{1}{c}{$C_{6,x}^{(P-S^+)}(M=0)$}  & \multicolumn{1}{c}{$C_{6,x}^{(P-S^+)}(M=\pm1)$}  
\\ \hline
\multicolumn{1}{c}{(504,728,1174,1589,1106,1174)}      &$-$1820.631774  &3995.37910   
\\
\multicolumn{1}{c}{(744,1120,2091,2625,2002,2091)}       &$-$1820.627179 &3995.40911    \\
\multicolumn{1}{c}{(1050,1632,3543,4172,3413,3543)}       &$-$1820.626701  &3995.42357
\\
\multicolumn{1}{c}{(1430,2280,5761,6412,3413,5761)}       &$-$1820.626499  &3995.42669
\\
\multicolumn{1}{c}{Extrapolated}       &$-$1820.6261(3)  &3995.429(3) \\
\end{tabular}
\end{ruledtabular}
\end{table*}
\endgroup

The molecular states
for $\dimerion$
correlating to $\atom (2\,{}^2P)$--$\ion (1\, {}^1S)$ system are the $2\,{}^2\Sigma_g^+$, $2\,{}^2\Sigma_u^+$, $1\,{}^2\Pi_g$
and $1\,{}^2\Pi_u$ states
(we do not consider fine structure).
We calculated the long-range interaction
coefficients using Eqs.~(\ref{c312})--(\ref{c623i}), which include electrostatic, induction,
and dispersion energies up to $\mathcal{O}(R^{-6})$; the
corresponding numerical values
are given in Table~\ref{TabP+ADD1}
and plotted in Fig.~\ref{fig:dimer}. These
correspond to position ``p3'' of the $\atom(2\, ^2P)$ atom as indicated in Fig.~\ref{fig:axis}. Convergence studies of these long-range interaction coefficients $C_{3,x}^{(P-S^+)}(M)$, $C_{4,x}^{(P-S^+)}(M)$, and $C_{6,x}^{(P-S^+)}(M)$ are given in, respectively, Tables~\ref{TabP+c_3con}, \ref{TabP+c_4con}, and \ref{TabP+c_6con}. In these tables, $N_P$ denotes the size of the basis for the $P$ state of the $^{\infty}$Li atom, and $N_{L}$ denotes the size of the basis for the corresponding intermediate states of symmetry $L$. Similarly, $N_{S}^+$ and $N_{P}^+$ denote respectively the sizes of the bases for the ground state and the intermediate states of symmetry $P$ of the $^{\infty}$Li$^+$ ion.

\begingroup
\begin{table}
\caption{Comparison of the polarizability components ($\alpha^{p_1}_{zz}$, $\alpha^{p_1}_{xx}$) and ($\alpha^{p_3}_{zz}$, $\alpha^{p_3}_{xx}$) (in atomic units) for the excited state $2\,^2P$ of $^{\infty}\text{Li}$. For the two coordinate systems, we have $\alpha^{p_1}_{xx}$= $\alpha_1$+ $\alpha_1^T$
or $\alpha^{p_1}_{zz}$= $\alpha_1-2\alpha_1^T$(see Refs.~\cite{rerat92,rerat93,rerat97,Caffarel93}) and $\alpha^{p_3}_{zz}$= $\alpha_1$+ $\alpha_1^T$
or $\alpha^{p_3}_{xx}$= $\alpha_1-\frac{1}{2}\alpha_1^T$, respectively,
where the electric field lies in $z$ or $x$ direction,
expressed in terms of the 
principal polarizabilities $\alpha_1$ (scalar)
and $\alpha{_1^T}$ (tensor).} 
\label{6TabII}
\begin{ruledtabular}
\begin{tabular}{lllll}
\multicolumn{1}{l}{Reference} 
 & \multicolumn{1}{c}{$\alpha^{p_1}_{zz}=2C_{4,z}^{(P-S^+)}{(M=0)}$}  & \multicolumn{1}{c}{$\alpha^{p_1}_{xx}=2C_{4,z}^{(P-S^+)}{(M=\pm1)}$} \\
\hline
Pipin \& Bishop ~\cite{pipin93}(1993)
&123.634 &128.449      \\
R\'erat~\textit{et al.}~\cite{rerat97} (1997) &131 & 129     \\
Cohen \& Themelis ~\cite{cohen05} (2005)&122.94 &128.13      \\
Johnson~\textit{et al.}~\cite{johnson08} (2008) &123.81 &128.580      \\
This work &123.703(4)  &128.5676(4) \\
\hline
&
\multicolumn{1}{c}{$\alpha^{p_3}_{zz}=2C_{4,x}^{(P-S^+)}{(M=0)}$}  & \multicolumn{1}{c}{$\alpha^{p_3}_{xx}=2C_{4,x}^{(P-S^+)}{(M=\pm1)}$}\\
\hline
Pipin \& Bishop ~\cite{pipin93}(1993)
&128.449 &126.0415     \\
R\'erat~\textit{et al.}~\cite{rerat97} (1997) & 129 & 131     \\
Cohen \& Themelis ~\cite{cohen05} (2005) &128.13 &125.535     \\
Johnson~\textit{et al.}~\cite{johnson08} (2008) &128.580 &126.195     \\
This work  &128.5676(4) &126.1352(6)   \\
\end{tabular}
\end{ruledtabular}
\end{table}
\endgroup

In order to apply the results
to the $\Sigma$ and $\Pi$ molecular
states of standard molecular nomenclature, where the $z$-axis joins the atom and ion, we must first apply the considerations of the previous Sec.~\ref{subsec:axis}
to express our results in terms of position ``p1'' of Fig.~\ref{fig:axis}.
The analysis yields the coefficients, with the numerical
values given in Table~\ref{TabP+ADD1}. The corresponding general coefficients are defined by
\begin{eqnarray}
C_{3,p_i}^{(P-S^+)}(M) = C_{3,p_1}^{(P-S^+)}(M)P_{2}(\cos\theta_i)
\,, \label{C3iaxis}
\end{eqnarray}
\begin{eqnarray}
C_{4,p_i}^{(P-S^+)}(M=0)=\frac{1}{2}\alpha_{zz}^{p_i} = \frac{1}{2}[\alpha_1-2\alpha^T_{1,p_1}P_{2}(\cos\theta_i)] 
\,, \label{C4zaxis}
\end{eqnarray}
and 
\begin{eqnarray}
C_{4,p_i}^{(P-S^+)}(M=\pm1) =\frac{1}{2}\alpha_{xx}^{p_i} = \frac{1}{2}[\alpha_1+\alpha^T_{1,p_1}P_{2}(\cos\theta_i)] 
\,, \label{C4xaxis}
\end{eqnarray}
where ($\alpha^{p_i}_{zz}$, $\alpha^{p_i}_{xx}$)
are, respectively, the polarizability components, along the $z$ and $x$  directions
of an applied electric field~\cite{rerat92,rerat93,rerat97,Caffarel93}, $\alpha_1$ and $\alpha_1^T$ are the principal scalar and tensor polarizibilities of Li($2\,^{2}P$) atom~\cite{tang09},
and $P_{2}(\cos\theta_i)$ is the Legendre polynomial. The comparison of the polarizability components ($\alpha^{p_1}_{zz}$, $\alpha^{p_1}_{xx}$) and ($\alpha^{p_3}_{zz}$, $\alpha^{p_3}_{xx}$) is given in Table.~\ref{6TabII}. In the present configuration (as shown in Fig.~\ref{fig:coords}), these components can be related to the leading induction coefficients $C_{4,x}(M)$ (as shown in Table~\ref{TabP+ADD1}) by $\alpha^{p_3}_{zz}=2 C_{4,x}(M=0)$ and  $\alpha^{p_3}_{xx}=2 C_{4,x}(M=\pm1)$. According to the symmetry of the degenerate system \cite{rerat92,rerat93,rerat97,Caffarel93}, we can connect the polarizability components ($\alpha^{p_3}_{zz}$, $\alpha^{p_3}_{xx}$) with the
 principal polarizabilities,
scalar ($\alpha_1$) and tensor ($\alpha_1^T$), of the Li($2\,^2P$) atom by $\alpha_1=\frac{1}{3}(\alpha^{p_3}_{zz}+2\alpha^{p_3}_{xx})$ and $\alpha_1^T=\frac{2}{3}(\alpha^{p_3}_{zz}-\alpha^{p_3}_{xx})$. 
For example, using the
present data from Table~\ref{6TabII},
we find $\alpha_1=126.9460$
and $\alpha_1^T=1.6216$
in agreement with Ref.~\cite[Table VIII]{tang09}. Further details concerning
the polarizability components ($\alpha^{p_1}_{zz}$, $\alpha^{p_1}_{xx}$) for the two particles lying on the $z$-axis 
are given in, for example, Refs.~\cite{rerat92,rerat93,rerat97,Caffarel93}. 
For the coefficient $C_6$, we can use the formula of Eq.~(\ref{c623iall}) by changing the Legendre polynomial from $P_{l_I}^{m_I}(0)$ to $P_{l_I}^{m_I}(\cos\theta_i)$
to get the general $C_{6,p_i}^{(P-S^+)}(M)$ at orientation p$_i$.

The long-range potential energy functions 
expressed relative to ``p1''
follow from Table~\ref{TabP+ADD1}.
For example, for the $2\,{}^2\Sigma_{g,u}$ states of the $^\infty\atom_2$, 
we have
\begin{eqnarray}\label{eq:sigma}
&&V^{(P-S^+)}(R;\Sigma)= -\frac{10.8199392}{R^3}-\frac{61.8515}{R^4}-\frac{9811.485}{R^6}-\cdots\,,
\end{eqnarray}
and for the $1\,{}^1\Pi_{g,u}$
states, we have
\begin{eqnarray}\label{eq:pi}
&&V^{(P-S^+)}(R;\Pi)= \frac{5.4099696}{R^3}-\frac{64.2838}{R^4}+\frac{1820.6261}{R^6}-\cdots\,.
\end{eqnarray}
Similar expressions may be written for ${}^6\dimerion$ and ${}^7\dimerion$ using
Table~\ref{TabP+ADD1}.
To the best of our knowledge,
the expressions in Eqs.~(\ref{eq:sigma}) and (\ref{eq:pi}) are unavailable
in the literature.

Magnier~\textit{et al.}~\cite{MagRouAll99} calculated the long-range potentials with the inclusion of exchange, electrostatic, induction, and dispersion interactions up to $\mathcal{O}(R^{-8})$; the results were presented graphically.
While a direct comparison of long-range coefficients is not possible, we
can calculate the exchange energies using the expressions given by Magnier~\textit{et al.}
and add that to our long-range potentials to compare with their total potentials
for each of the four molecular states.
When the exchange energy and the long-range potential energy have opposite
signs, a long-range well or barrier results;
these singular features provide good quantitative checks between Magnier~\textit{et al.}
and the present work.
From Magnier~\textit{et al.}~\cite{MagRouAll99},
Figs.~(8)--(10), it is evident
that the exchange
energies are positive for the
$2\,{}^2\Sigma_u^+$ and $1\,{}^2\Pi_g$ states
and negative for the
$2\,{}^2\Sigma_g^+$  and  $1\,{}^2\Pi_u$
states, while the exchange splitting for the pair of $2\,{}^2\Sigma^+$ states
is larger by a factor of $R/2$
compared to the pair of $1\,{}^2\Pi$ states, where $R$ is the internuclear
distance.
Calculations show that the 
two $\Sigma$ states and the 
$1\,{}^2\Pi_u$ state form
potential wells,
while the $1\,{}^2\Pi_g$
state is purely repulsive~\cite{RabMcC17}.
It is evident from the data
in Eq.~(\ref{eq:pi})
that the net positive long-range potential and positive exchange energy completely account for the repulsive $1\,{}^2\Pi_g$ state. Of the three states
with potential wells,
the $2\,{}^2\Sigma_u^+$
state well exists at the greatest
internuclear distance, about
$25\,a_0$, with a depth
of only $127\,\text{cm}^{-1}$
according
to a recent model potential calculation~\cite{RabMcC17}.
With our long-range expansion
of Table~\ref{TabP+ADD1}
evaluated for $^\infty\atom$ as in Eq.~(\ref{eq:sigma}) 
and using Eq.~(8) of Ref.~\cite{MagRouAll99},
to estimate the contribution of the exchange energy\footnote{
      By close comparison of Eqs.~(2.13) and (3.7) of Ref. \protect\cite{ChiJan88},
      we believe that the factor ``2'' in the denominator of
      $D$ given in Eq.~(8) of Ref. \protect\cite{MagRouAll99} should be replaced by ``$m!$''.
      We evaluated the exchange energy splittings with this evident correction included.
      }, 
we find a well
of depth $119\,\text{cm}^{-1}$
at $R=25.8\,a_0$ to be compared to the depth $124\,\text{cm}^{-1}$
at $25.7\,a_0$ obtained
by Magnier~\textit{et al}.
using a long-range expansion
and the exchange energy.
We also obtain
for the the $1\,{}^2\Pi_u$ state using Eq.~(\ref{eq:pi}) a potential barrier
of $36\,\text{cm}^{-1}$ at $R=23.9
\,a_0$, compared
to $40\,\text{cm}^{-1}$
at $23.4\,a_0$ found
by  Magnier~\textit{et al}.
The agreement of the well
and the barrier
positions and energies
calculated using Eqs.~(\ref{eq:sigma}) and (\ref{eq:pi}), with the similar
calculations of Magnier~\textit{et al}. is satisfactory.
The $2\,{}^2\Sigma_u^+$ state is an example
of a long-range molecular state~\cite{JonTieLet06}.
Moreover,
we do not attempt
to reproduce the
wells of the $2\,{}^2\Sigma_u^+$ 
or $1\,{}^2\Pi_u$ states,
because
it is evident
from Ref.~\cite[Figs.~8--9]{MagRouAll99}
that
these potential wells
are fully realized
with the inclusion
of charge overlap (\textit{i.e.} in quantum-chemical
calculations~\cite{MagRouAll99,RabMcC17}).

Having thus demonstrated that two-body
long-range interaction potentials
can be extracted from our results,
and providing some coefficients
that were previously unavailable
in the literature  [\textit{viz.} Table~\ref{TabP+ADD1}
and Eqs.~(\ref{eq:sigma}) and (\ref{eq:pi})],
we turn back to the three-body system.

\section{Results and {Discussions}}\label{res}

As in Secs.~\ref{subsec:axis}--\ref{subsec:PS+},
we use highly accurate variational wave functions 
for lithium atoms and ions in Hylleraas coordinates 
with finite nuclear mass effects to evaluate
the numerical values~\cite{tang09}.
We note that in general
the zeroth-order wave functions are obtained by using degenerate perturbation theory through  Eq.~(10) in the Supplemental Material~\cite{Supplement}
and there are intrinsic geometrical dependencies
that complicate the analysis.
In particular, the zeroth-order wave functions change with the geometry (interior angles and
interatomic separations) of the three-body system.
However, when $R_{23}=R_{31}=R$,
we have the matrix elements $\Delta_{12}$=$\Delta_{21}$, see Eq.~(10) in the Supplemental Material~\cite{Supplement},
and the geometrical dependencies
don't appear in the zeroth-order wave functions,
simplifying the analysis of the three-body
system. Therefore, in this section, we consider the 
Li($2\,^{2}S$)-Li($2\,^{2}P$)-Li$^+(1\,^{1}S$) system for the configurations where $R_{23}=R_{31}=R$.

In Sec.~\ref{subsec:additive-equalR-wf} we introduce the
zeroth-order wave functions and 
in Sec.~\ref{subsec:additive-equalR-eval}
provide the numerical
values of these additive coefficients.
In Sec.~\ref{subsec:two-specific-cases}, we focus on the 
two specific arrangements of the three particles,
collinear and in an equilateral triangle,
providing the nonadditive coefficients.

\subsection{Additive coefficients: Wave functions}
\label{subsec:additive-equalR-wf}

With respect to the ``p1" orientation as shown in Fig.~\ref{fig:axis}, we calculate the long-range additive potentials for the three-body system lying collinearly
on the $z$-axis. According to degenerate perturbation theory, the corresponding zeroth-order wave functions are
\begin{eqnarray}
\Psi_{1,z}^{(0)}&=&\frac{1}{\sqrt{2}}\bigg[\ket{{n_0}1_z;{n_0}0;{n_0'}0}+\ket{{n_0}0;{n_0}1_z;{n_0'}0}\bigg] \,, \label{delta11z} \\
\Psi_{2,z}^{(0)}&=&\frac{1}{\sqrt{2}}\bigg[\ket{{n_0}1_z;{n_0}0;{n_0'}0}-\ket{{n_0}0;{n_0}1_z;{n_0'}0}\bigg] \,, \label{delta22z}
\end{eqnarray}
where the symbol $z$ indicates the three-particles lying on the $z$-axis for the configurations of $R_{23}=R_{31}=R$. 

For three particles lying in the $x$-$y$ plane as shown in Fig.~\ref{fig:coords}, the corresponding zeroth-order wave functions are
\begin{eqnarray}
\Psi_{1,\bot}^{(0)}&=&\frac{1}{\sqrt{2}}\bigg[\ket{{n_0}1;{n_0}0;{n_0'}0}+\ket{{n_0}0;{n_0}1;{n_0'}0}\bigg] \,, \label{delta11} \\
\Psi_{2,\bot}^{(0)}&=&\frac{1}{\sqrt{2}}\bigg[\ket{{n_0}1;{n_0}0;{n_0'}0}-\ket{{n_0}0;{n_0}1;{n_0'}0}\bigg] \,, \label{delta22}
\end{eqnarray}
where the symbol $\bot$ indicates specificity to the 
the $x$-$y$ planar
configuration with $R_{23}=R_{31}=R$. 
Note that Eqs.~(\ref{delta11}) and (\ref{delta22})
include the special case of the three
particles  lying collinearly on the  $x$-axis,~\textit{i.e.} the orientation ``p3''.

\subsection{Additive coefficients: Evaluation}
\label{subsec:additive-equalR-eval}

\begingroup
\squeezetable
\begin{table*}
\caption{The long-range additive interaction coefficients (in atomic units) of the Li($2\,^2S$)-Li($2\,^2P$)-Li$^+$($2\,^{1}S$) system
for two different types of the zeroth-order wave functions, where the three particles 
lie collinearly 
on the $z$-axis (similar to 
the ``p1'' orientation of the two-body system shown in Fig.~\ref{fig:axis}).
The numbers in parentheses represent the computational uncertainties.} \label{TabADDzaxis}
\begin{ruledtabular}
\begin{tabular}{lcccccc}
\multicolumn{1}{c}{Coefficients}  &  \multicolumn{1}{c}{$\Psi_{1,z}^{(0)}$}  & \multicolumn{1}{c}{$\Psi_{2,z}^{(0)}$} & \multicolumn{1}{c}{$\Psi_{1,z}^{(0)}$}
& \multicolumn{1}{c}{$\Psi_{2,z}^{(0)}$} & \multicolumn{1}{c}{$\Psi_{1,z}^{(0)}$}  & \multicolumn{1}{c}{$\Psi_{2,z}^{(0)}$}
\\
\hline
\\ 
&  \multicolumn{2}{c}{$^{\infty}$Li} & \multicolumn{2}{c}{$^{7}$Li} & \multicolumn{2}{c}{$^{6}$Li}
\\
\cline{2-3}  \cline{4-5}  \cline{6-7}
\\
\multicolumn{1}{c}{$C_{3,z}^{(12)}(1,M=0)$}   &11.000221(2)    &$-$11.000221(2) &11.001853(2) &$-$11.001853(2) &11.002125(2) &$-$11.002125(2)
\\
\multicolumn{1}{c}{$C_{3,z}^{(23)}(1,M=0)$} &5.4099696(1) &5.4099696(1) &5.4096296(1) &5.4096296(1) &5.4095731(1) &5.4095731(1)                              \\
\multicolumn{1}{c}{$C_{3,z}^{(31)}(1,M=0)$}  &5.4099696(1) &5.4099696(1) &5.4096296(1) &5.4096296(1) &5.4095731(1) &5.4095731(1)         \\
\multicolumn{1}{c}{$C_{4,z}^{(23)}(1,M=0)$} 
& 71.9539(4) &71.9539(4)   &71.9594(4)    & 71.9594(4)    &71.9604(4)    & 71.9604(4) 
\\
\multicolumn{1}{c}{$C_{4,z}^{(31)}(1,M=0)$} &71.9539(4)  &71.9539(4)   & 71.9594(4)    & 71.9594(4)   & 71.9604(4)    & 71.9604(4) 
\\
\multicolumn{1}{c}{$C_{6,z}^{(12)}(1,M=0)$} & 2075.40(3)  & 2075.40(3) & 2076.08(7) & 2076.08(7) & 2076.19(7)  & 2076.19(7) \\  
\multicolumn{1}{c}{$C_{6,z}^{(23)}(1,M=0)$} &5263.218(3)  & 5263.218(3) &5263.151(3)  &5263.151(3)  &5263.140(3)  &5263.140(3)  \\
\multicolumn{1}{c}{$C_{6,z}^{(31)}(1,M=0)$}
& 5263.218(3) &5263.218(3)  & 5263.151(3) &5263.151(3)  &5263.140(3)  & 5263.140(3)  
\\
\\
\multicolumn{1}{c}{$C_{3,z}^{(12)}(1,M=\pm1)$} &$-$5.500111(1)     &5.500111(1) &$-$5.500926(1) &5.500926(1)  &$-$5.501062(1) &5.501062(1)                                     \\
\multicolumn{1}{c}{$C_{3,z}^{(23)}(1,M=\pm1)$} & $-$2.7049847(1)& $-$2.7049847(1) &$-$2.7048148(1)&$-$2.7048148(1) 
&$-$2.7047866(1) &$-$2.7047866(1)                                   \\
\multicolumn{1}{c}{$C_{3,z}^{(31)}(1,M=\pm1)$} & $-$2.7049847(1)& $-$2.7049847(1) &$-$2.7048148(1)&$-$2.7048148(1) 
&$-$2.7047866(1) &$-$2.7047866(1)          \\
\multicolumn{1}{c}{$C_{4,z}^{(23)}(1,M=\pm1)$} & 73.1701(4) & 73.1701(4)  & 73.1859(4)   & 73.1859(4)   & 73.1885(4)   & 73.1885(4)  \\
\multicolumn{1}{c}{$C_{4,z}^{(31)}(1,M=\pm1)$} & 73.1701(4) & 73.1701(4)  & 73.1859(4)   & 73.1859(4)   & 73.1885(4)   & 73.1885(4)
    \\
\multicolumn{1}{c}{$C_6^{(12)}(1,M=\pm1)$} & 1406.68(3)  & 1406.68(3) & 1407.15(5) & 1407.15(5) & 1407.20(2)  & 1407.20(2) \\     
\multicolumn{1}{c}{$C_{6,z}^{(23)}(1,M=\pm1)$} & $-$552.8371(7) & $-$552.8371(7) & $-$552.8460(5) & $-$552.8460(5) &$-$552.8472(7)  &$-$552.8472(7)
\\
\multicolumn{1}{c}{$C_{6,z}^{(31)}(1,M=\pm1)$}
& $-$552.8371(7) & $-$552.8371(7) & $-$552.8460(5) & $-$552.8460(5) &$-$552.8472(7)  &$-$552.8472(7)
\\ 
\end{tabular}
\end{ruledtabular}
\end{table*}
\endgroup

\begingroup
\squeezetable
\begin{table*}	
\caption{The long-range additive interaction coefficients (in atomic units) of the Li($2\,^2S$)-Li($2\,^2P$)-Li$^+$($2\,^{1}S$) system
for two different types of the zeroth-order wave functions, where the three particles
lie in the $x$-$y$ plane with $R_{23}=R_{31}=R$ 
as shown in Fig.~\ref{fig:coords}. 
Note this includes the special
case of the three particles
collinear on the $x$-axis. The numbers in parentheses represent the computational uncertainties. }\label{TabADD}
\begin{ruledtabular}
\begin{tabular}{lcccccc}
\multicolumn{1}{c}{Coefficients}  &  \multicolumn{1}{c}{$\Psi_{1,\bot}^{(0)}$}  & \multicolumn{1}{c}{$\Psi_{2,\bot}^{(0)}$} & \multicolumn{1}{c}{$\Psi_{1,\bot}^{(0)}$}
& \multicolumn{1}{c}{$\Psi_{2,\bot}^{(0)}$} & \multicolumn{1}{c}{$\Psi_{1,\bot}^{(0)}$}  & \multicolumn{1}{c}{$\Psi_{2,\bot}^{(0)}$}
\\
\hline
\\
&  \multicolumn{2}{c}{$^{\infty}$Li} & \multicolumn{2}{c}{$^{7}$Li} & \multicolumn{2}{c}{$^{6}$Li}
\\
\cline{2-3}  \cline{4-5}  \cline{6-7}
\\
\multicolumn{1}{c}{$C_3^{(12)}(1,M=0)$}      &$-$5.500111(1)  & 5.500111(1)    &$-$5.500926(1)   & 5.500926(1)  &$-$5.501062(1)   & 5.501062(1)
\\
\multicolumn{1}{c}{$C_3^{(23)}(1,M=0)$}       &$-$2.7049847(1) &  $-$2.7049847(1)  &$-$2.7048148(1) &   $-$2.7048148(1)  &$-$2.7047866(1) & $-$2.7047866(1)                         \\
\multicolumn{1}{c}{$C_3^{(31)}(1,M=0)$}       & $-$2.7049847(1) & $-$2.7049847(1)            &$-$2.7048148(1)  & $-$2.7048148(1)    &$-$2.7047866(1)    & $-$2.7047866(1)     \\
\multicolumn{1}{c}{$C_4^{(23)}(1,M=0)$} & 73.1701(4) & 73.1701(4)  & 73.1859(4)   & 73.1859(4)   & 73.1885(4)   & 73.1885(4)
\\
\multicolumn{1}{c}{$C_4^{(31)}(1,M=0)$} & 73.1701(4) & 73.1701(4) & 73.1859(4)   & 73.1859(4)  & 73.1885(4)   & 73.1885(4)
\\
\multicolumn{1}{c}{$C_6^{(12)}(1,M=0)$} & 1406.68(3)  & 1406.68(3) & 1407.15(5) & 1407.15(5) & 1407.20(2)  & 1407.20(2) \\
\multicolumn{1}{c}{$C_6^{(23)}(1,M=0)$} & $-$552.8371(7) & $-$552.8371(7) & $-$552.8460(5) & $-$552.8460(5) &$-$552.8472(7)  &$-$552.8472(7)
\\
\multicolumn{1}{c}{$C_6^{(31)}(1,M=0)$}
& $-$552.8371(7) & $-$552.8371(7) & $-$552.8460(5) & $-$552.8460(5) &$-$552.8472(7) &$-$552.8472(7)   
\\
\\
\multicolumn{1}{c}{$C_3^{(12)}(1,M=\pm1)$}     & 2.750054(1)      & $-$2.750054(1) & 2.750462(1)   & $-$2.750462(1)   & 2.750530(1)   & $-$2.750530(1)                                         \\
\multicolumn{1}{c}{$C_3^{(23)}(1,M=\pm1)$}     & 1.3524924(1)  &  1.3524924(1)  & 1.3524074(1)   & 1.3524074(1)   & 1.3523932(1)    &   1.3523932(1)                                            \\
\multicolumn{1}{c}{$C_3^{(31)}(1,M=\pm1)$}    & 1.3524924(1)  & 1.3524924(1)   & 1.3524074(1)    & 1.3524074(1)   &  1.3523932(1)  &  1.3523932(1)      \\
\multicolumn{1}{c}{$C_4^{(23)}(1,M=\pm1)$}  & 72.5620(5) & 72.5620(5) & 72.5727(5)  & 72.5727(5) & 72.5745(5)  & 72.5745(5) \\
\multicolumn{1}{c}{$C_4^{(31)}(1,M=\pm1)$} & 72.5620(5) & 72.5620(5) & 72.5727(5)  & 72.5727(5) & 72.5745(5)  & 72.5745(5)
    \\
\multicolumn{1}{c}{$C_6^{(12)}(1,M=\pm1)$} & 1741.06(5) & 1741.06(5)  & 1741.59(4) & 1741.59(4)  & 1741.68(4) & 1741.68(4)
\\
\multicolumn{1}{c}{$C_6^{(23)}(1,M=\pm1)$} & 2355.190(2) & 2355.190(2)  & 2355.152(2)     & 2355.152(2) & 2355.146(2) & 2355.146(2)        \\
\multicolumn{1}{c}{$C_6^{(31)}(1,M=\pm1)$}  & 2355.190(2) & 2355.190(2)  & 2355.152(2)     & 2355.152(2)  & 2355.146(2)  & 2355.146(2)     \\
\end{tabular}
\end{ruledtabular}
\end{table*}
\endgroup

Using the degenerate perturbation theory, we find that different from the ground state Li$_3^+$ trimer (where there is no analogous quantum three-body effect for these long-range additive coefficients~\cite{yan20}), the atomic states ($a$, $b$) and the corresponding additive coefficients are changing with different geometries of the three-body system for the excited Li$_3^+$ trimer. This phenomenon is absolutely a kind of quantum three-body effect, which is caused by the degeneracy of the excited Li$_3^+$ trimer. While for the specific geometries with $R_{23}=R_{31}=R$, we find that the atomic states and the corresponding additive interaction coefficients would
remain unchanged due to the symmetry of the three-body system. This feature makes these coefficients be useful in the quantum chemistry studies. In the present paper, we give the calculations of the long-range coefficients for these specific configurations. The additive coefficients $C_n^{(IJ)}$ [to be in Eqs.~(\ref{etot1}) and~(\ref{etotadd2})] are calculated for the specific three-body Li($2\,^{2}S$)-Li($2\,^{2}P$)-Li$^+(1\,^{1}S$) system lying on the $z$-axis or in the $x$-$y$ plane. The numerical values are shown in
Tables~\ref{TabADDzaxis} and~\ref{TabADD}, where we can also find the orientation dependence (that are demonstrated in Sec.~\ref{subsec:axis} for two-body system) of the long-range interaction coefficients for the excited Li$_3^+$ trimer.

For the numerical values of these additive coefficients, we note that the leading long-range interaction coefficient between two neutral atoms $C_3^{(12)}(1,M)$ can be positive (attractive) or negative (repulsive) corresponding to the different atomic states related to the symmetry of the system.
The leading terms $C_{3,z}^{(23)}(1,M=\pm1)$, $C_{3,z}^{(31)}(1,M=\pm1)$ (see Table~\ref{TabADDzaxis}) and $C_3^{(23)}(1,M=0)$, $C_3^{(31)}(1,M=0)$ (see Table~\ref{TabADD}) are always negative, which represents the repulsive interactions between the charge of the Li$^+(1\,^{1}S$) ion and the permanent electric quadruple moments of the Li($2\,^{2}P)$ atom generated by its anisotropic charge distribution along 
the $z$-axis for $M=\pm1$ and $M=0$ states, respectively.
The leading terms $C_{3,z}^{(23)}(1,M=0)$,  $C_{3,z}^{(31)}(1,M=0)$ (see Table~\ref{TabADDzaxis}) and $C_3^{(23)}(1,M=\pm1)$, $C_3^{(31)}(1,M=\pm1)$ (see Table~\ref{TabADD}) are always positive (attractive), which is caused by induction effect of the Li$^+$($1\,^{1}S)$ atom. 
Similarly, the inductive terms $C_{4,z}^{(23)}(1,M)$, $C_{4,z}^{(31)}(1,M)$ (see Table~\ref{TabADDzaxis}) and $C_4^{(23)}(1,M)$, $C_4^{(31)}(1,M)$ (see Table~\ref{TabADD}) are also always positive (attractive). 
Their numerical values are the linear combinations of 
the inductive interactions of the Li($2\,^{2}S$)-Li$^+(1\,^{1}S$) system and the Li($2\,^{2}P$)-Li$^+(1\,^{1}S$) system [Eqs.~(\ref{C423new1}) and~(\ref{C431new1})]. The
$C_{6,z}^{(23)}(1,M)$, $C_{6,z}^{(31)}(1,M)$ (see Table~\ref{TabADDzaxis}) and
$C_6^{(23)}(1,M)$, $C_6^{(31)}(1,M)$ (see Table~\ref{TabADD}) are also the linear combinations of inductive and dispersion interactions of the Li($2\,^{2}S$)-Li$^+(1\,^{1}S$) system and the Li($2\,^{2}P$)-Li$^+(1\,^{1}S$) system [Eqs.~(\ref{C623new1}) and~(\ref{C631new1})]. 

\subsection{Nonadditive coefficients: Collinear and equilateral triangle}
\label{subsec:two-specific-cases}

\begingroup
\squeezetable
\begin{table*}	
\caption{The long-range nonadditive interaction coefficients (in atomic units) of the Li($2\,^2S$)-Li($2\,^2P$)-Li$^+$($2\,^{1}S$) system
for two different types of the zeroth-order wave functions, where the three particles form an equally-spaced collinear configuration with the ion in the middle lying on the $x$-axis. The numbers in parentheses represent the computational uncertainties. }\label{TabLInon}
\begin{ruledtabular}
\begin{tabular}{lccccccc}
\multicolumn{1}{c}{Coefficients}  &  \multicolumn{1}{c}{$\Psi_{1,\bot}^{(0)}$}  & \multicolumn{1}{c}{$\Psi_{2,\bot}^{(0)}$} & \multicolumn{1}{c}{$\Psi_{1,\bot}^{(0)}$}
& \multicolumn{1}{c}{$\Psi_{2,\bot}^{(0)}$} & \multicolumn{1}{c}{$\Psi_{1,\bot}^{(0)}$}  & \multicolumn{1}{c}{$\Psi_{2,\bot}^{(0)}$}
\\
\hline
\\
&  \multicolumn{2}{c}{$^{\infty}$Li} & \multicolumn{2}{c}{$^{7}$Li} & \multicolumn{2}{c}{$^{6}$Li}
\\
\cline{2-3}  \cline{4-5}  \cline{6-7}
\\
\multicolumn{1}{c}{$C_{4,2}^{(12,23)}(1,M=0)$}  & $-$1873.904(5)  &3205.671(5) &$-$1874.274(5)  &3206.351(5)  &$-$1874.334(6)  &3206.464(5)
\\
\multicolumn{1}{c}{$C_{2,4}^{(31,12)}(1,M=0)$}   & $-$1873.904(5)  &3205.671(5) &$-$1874.274(5)   &3206.351(5)   &$-$1874.334(6)  &3206.464(5)
\\
\multicolumn{1}{c}{$C_{3,3}^{(12,23)}(1,M=0)$}  &244.58680(3)   &$-$244.58680(3)    &244.65297(5)  &$-$244.65297(5)   &244.66399(5) &$-$244.66399(5)
\\
\multicolumn{1}{c}{$C_{3,3}^{(23,31)}(1,M=0)$} &1.0592047(2)  &$-$1.0592047(2) &1.0597875(3)  &$-$1.0597875(3)  &1.0598847(2)  &$-$1.0598847(2)
\\
\multicolumn{1}{c}{$C_{3,3}^{(31,12)}(1,M=0)$} &244.58680(3)   &$-$244.58680(3)    &244.65297(5)  &$-$244.65297(5)   &244.66399(5) &$-$244.66399(5)
\\
\\
\multicolumn{1}{c}{$C_{4,2}^{(12,23)}(1,M=\pm1)$} &936.951(3)   &$-$1602.836(3) &937.136(3)   &$-$1603.176(3)   &937.167(3) &$-$1603.231(2)
\\
\multicolumn{1}{c}{$C_{2,4}^{(31,12)}(1,M=\pm1)$}  &936.951(3)    &$-$1602.836(3) &937.136(3) &$-$1603.176(3) &937.167(3)    &$-$1603.231(2)
\\
\multicolumn{1}{c}{$C_{3,3}^{(12,23)}(1,M=\pm1)$} &611.46703(6)  &$-$611.46703(6) &611.6324(1)  &$-$611.6324(1)  &611.6600(1) & $-$611.6600(1)  
\\
\multicolumn{1}{c}{$C_{3,3}^{(23,31)}(1,M=\pm1)$} &2.6480119(5) &$-$2.6480119(5) &2.649468(1) &$-$2.649468(1) &2.6497127(6) &$-$2.6497127(6)
\\
\multicolumn{1}{c}{$C_{3,3}^{(31,12)}(1,M=\pm1)$}  &611.46703(6)  &$-$611.46703(6) &611.6324(1)  &$-$611.6324(1)  &611.6600(1) & $-$611.6600(1)        \\
\end{tabular}
\end{ruledtabular}
\end{table*}
\endgroup

\begingroup
\squeezetable
\begin{table*}		
\caption{The long-range nonadditive interaction coefficients (in atomic units) of the Li($2\,^2S$)-Li($2\,^2P$)-Li$^+$($2\,^{1}S$) system
for two different types of the zeroth-order wave functions, where the three particles form an equilateral triangle. The numbers in parentheses represent the computational uncertainties. }\label{TabEQ}
\begin{ruledtabular}
\begin{tabular}{lccccccc}
\multicolumn{1}{c}{Coefficients}  &  \multicolumn{1}{c}{$\Psi_{1, \bot}^{(0)}$}  & \multicolumn{1}{c}{$\Psi_{2,\bot}^{(0)}$} & \multicolumn{1}{c}{$\Psi_{1, \bot}^{(0)}$}
& \multicolumn{1}{c}{$\Psi_{2, \bot}^{(0)}$} & \multicolumn{1}{c}{$\Psi_{1, \bot}^{(0)}$}  & \multicolumn{1}{c}{$\Psi_{2, \bot}^{(0)}$}
\\
\hline
\\
&  \multicolumn{2}{c}{$^{\infty}$Li} & \multicolumn{2}{c}{$^{7}$Li} & \multicolumn{2}{c}{$^{6}$Li}
\\
\cline{2-3}  \cline{4-5}  \cline{6-7}
\\
\multicolumn{1}{c}{$C_{4,2}^{(12,23)}(1,M=0)$}   & $-$936.951(3)  &1602.836(3) &$-$937.136(3)   &1603.176(3)  &$-$937.167(3)   &1603.231(2)
\\
\multicolumn{1}{c}{$C_{2,4}^{(31,12)}(1,M=0)$}   & $-$936.951(3)  &1602.836(3) &$-$937.136(3)   &1603.176(3)  &$-$937.167(3)   &1603.231(2)
\\
\multicolumn{1}{c}{$C_{3,3}^{(12,23)}(1,M=0)$}  & 244.58680(3)  &$-$244.58680(3)    &244.65297(5)  &$-$244.65297(5)   &244.66399(5) &$-$244.66399(5)
\\
\multicolumn{1}{c}{$C_{3,3}^{(23,31)}(1,M=0)$} &1.0592047(2)  &$-$1.0592047(2)  &1.0597875(3)  &$-$1.0597875(3)  &1.0598847(2)  &$-$1.0598847(2)
\\
\multicolumn{1}{c}{$C_{3,3}^{(31,12)}(1,M=0)$} & 244.58680(3)  &$-$244.58680(3)   &244.65297(5) &$-$244.65297(5)    &244.66399(5)  &$-$244.66399(5)
\\
\\
\multicolumn{1}{c}{$C_{4,2}^{(12,23)}(1,M=\pm1)$} & 468.476(1)  &$-$801.417(1) &468.567(2)   &$-$801.587(1)   &468.584(1)  &$-$801.616(2)
\\
\multicolumn{1}{c}{$C_{2,4}^{(31,12)}(1,M=\pm1)$}  & 468.476(1)  &$-$801.417(1) &468.567(2) &$-$801.587(1) &468.584(1)    &$-$801.616(2)
\\
\multicolumn{1}{c}{$C_{3,3}^{(12,23)}(1,M=\pm1)$}  &$-$214.01346(2) &214.01346(2)  &$-$214.07137(3) &214.07137(3)  & $-$214.08101(3)  &214.08101(3)
\\
\multicolumn{1}{c}{$C_{3,3}^{(23,31)}(1,M=\pm1)$}  &$-$0.9268041(2) & 0.9268041(2)  &$-$0.9273142(2) &0.9273142(2) &$-$0.9273991(2) &0.9273991(2)
\\
\multicolumn{1}{c}{$C_{3,3}^{(31,12)}(1,M=\pm1)$}  &$-$214.01346(2) &214.01346(2) &$-$214.07137(3)   &214.07137(3) &$-$214.08101(3)  &214.08101(3)     \\
\end{tabular}
\end{ruledtabular}
\end{table*}
\endgroup

\begin{figure} 
\begin{center}
\includegraphics[width=15cm,height=9cm]{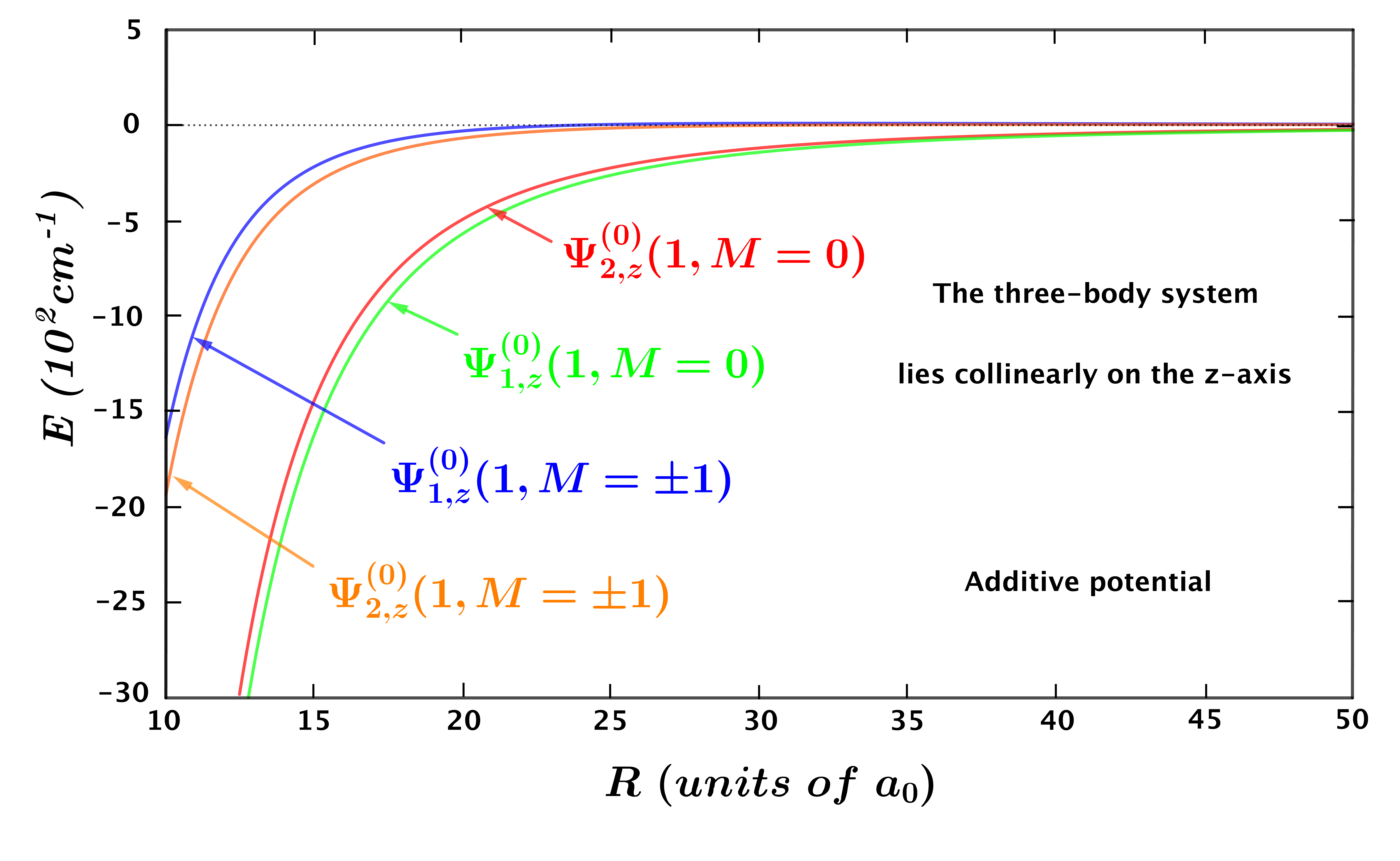}
\end{center}
\caption {\label{f1} Long-range \textit{additive} interaction potentials (in atomic units) of the $^\infty$Li($2\,^2S$)-$^\infty$Li($2\,^2P$)-$^\infty$Li$^+$($1\,^{1}S$) system for two types of the zeroth-order wave functions, where three particles lie collinearly on the $z$-axis.
For each curve labeled by a wave function, the plotted curve is the sum of $\Delta E^{(1)}$ and $\Delta E^{(2)}$.}
\label{figcolz}
\end{figure}

\begin{figure} 
\begin{center}
\includegraphics[width=15cm,height=9cm]{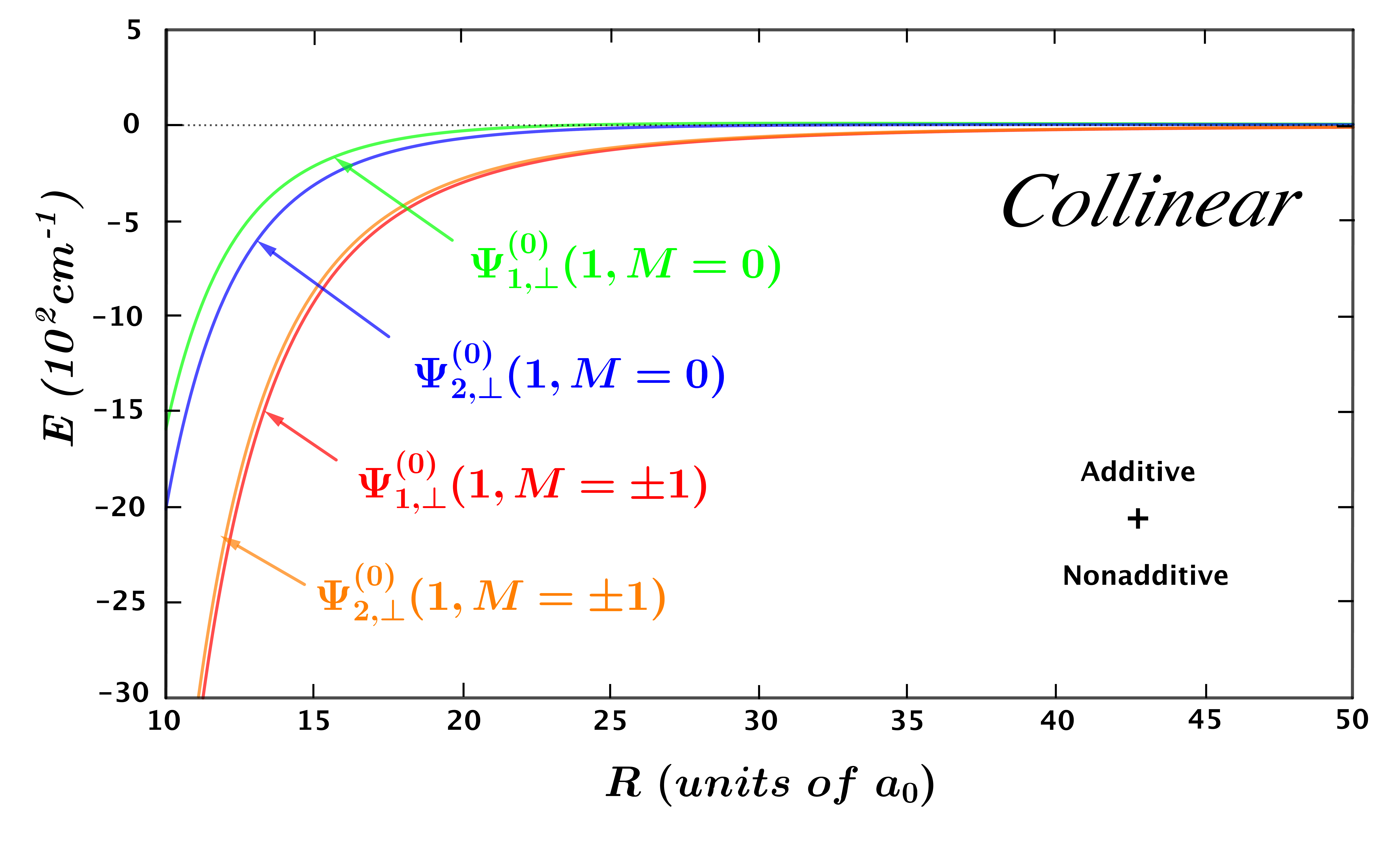}
\end{center}
\caption {Long-range interaction potentials (in atomic units) of the $^\infty$Li($2\,^2S$)-$^\infty$Li($2\,^2P$)-$^\infty$Li$^+$($1\,^{1}S$) system for two types of the zeroth-order wave functions, where three particles lie collinearly on the $x$-axis.
The plotted potentials include all electrostatic, dispersion, and induction type interactions
(additive and nonadditive)
up to $\mathcal{O}(R^{-6})$.
For each curve labeled by a wave function, the plotted curve is the sum of $\Delta E^{(1)}$ and $\Delta E^{(2)}$.}
\label{figcol}
\end{figure}

\begin{figure} 
\begin{center}
\includegraphics[width=15cm,height=9cm]{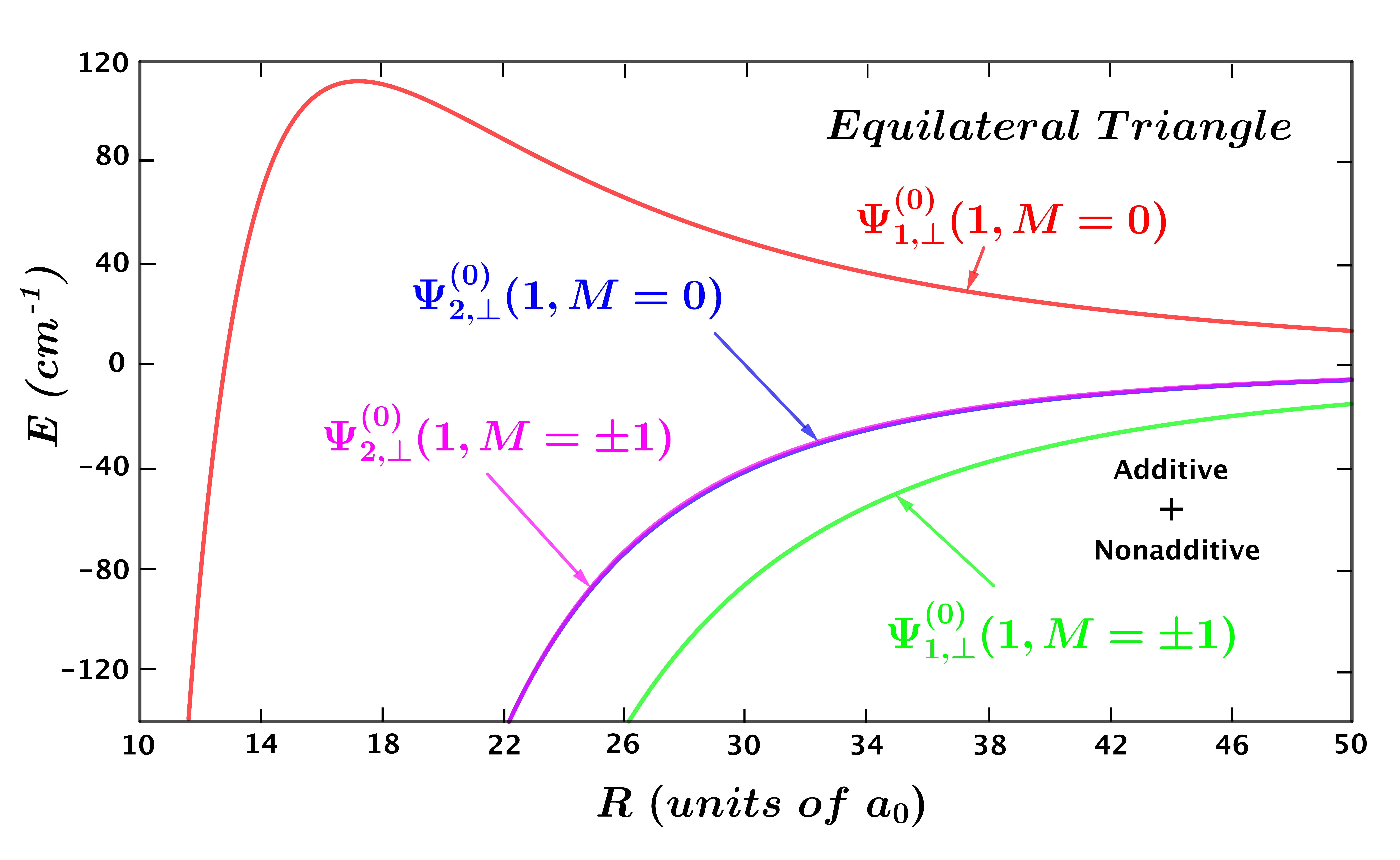}
\end{center}
\caption {Long-range interaction potentials (in atomic units) of the $^\infty$Li($2\,^2S$)-$^\infty$Li($2\,^2P$)-$^\infty$Li$^+$($1\,^{1}S$) system for two types of the zeroth-order wave functions, where three particles form an equliteral triangle on the $x$-$y$ plane.
The plotted potentials include all electrostatic, dispersion,
and induction
type interactions
(additive and nonadditive)
up to $\mathcal{O}(R^{-6})$.
For each curve labeled by a wave function, the plotted curve is the sum of $\Delta E^{(1)}$ and $\Delta E^{(2)}$.}
\label{figeqt}
\end{figure}

The nonadditive interaction coefficients of Eq.~(\ref{etotnonadd2}) show a dependence on the interior angles of the three-body system. 
It is not practical to calculate
the nonadditive coefficients for
arbitrary cases when $R_{23}=R_{31}=R$.
However, for the collinear 
and the equilateral triangle configuration,
which fortunately are probably
the most interesting configurations,
we can evaluate specific values.
In this subsection,
these coefficients are given for two geometries: an equally-spaced collinear configuration with the ion in the center ($R_{23}=R_{31}=R$),
see Table~\ref{TabLInon},
and an equilateral triangle 
configuration ($R_{23}=R_{31}=R_{12}=R$), see Table~\ref{TabEQ}. 
Different from the ground state Li$_3^+$ trimer demonstrated in Ref.~\cite{yan20}, the long-range nonadditive interactions of the current excited Li$_3^+$ trimer appear in the second-order correction, not in the third-order correction. This phenomenon is caused by the degeneracy of the three-body system introduced by the presence of the excited Li($2\,^2P$) atom.
From Tables~\ref{TabLInon} and~\ref{TabEQ}, we can find that most of the nonadditive coefficients are indeed different from each other for these two geometries, even with the same atomic states as shown in Eqs.~(\ref{delta11})  and (\ref{delta22}).
This kind of three-body effect is caused by the different interior angles of the two geometries accociated with the magnetic quantum number $M$ of the Li($2\,^{2}P$) atom, which can also be easily figured out from Eqs.~(\ref{C62331})--(\ref{C6333112}). And for the different interior angles of the geometries and for the different magnetic quantum number $M$,
these nonadditive terms can be attractive or repulsive.

Due to the induction effect of the Li$^+(1\,^{1}S$) cation, some of these nonadditive coefficients are enhanced. For example, from Table~\ref{TabLInon}, we find that the inductive nonadditive coefficients $|C_{4,2}^{(12,23)}(1,M=0)|=|C_{2,4}^{(31,12)}(1,M=0)|=1873.904(5)$ a.u. are much larger than the dispersion nonadditive one \bigg($|C_{3,3}^{(23,31)}(1,M=0)|=1.0592047(2)$ a.u. from Table~\ref{TabLInon}\bigg) and are even larger than some of the additive dispersion \bigg[$|C_{6}^{(12)}(1,M=0)|=1406.68(3)$ a.u. from Table~\ref{TabADD}\bigg], and additive inductive  ones \bigg[$|C_{6}^{(23)}(1,M=0)|=552.8371(7)$ a.u. from Table~\ref{TabADD}\bigg] at the same order.
The competition between the additive attractive and nonadditive repulsive terms of $C_6$ for particular geometries will also be discussed in the following section. These large nonadditive inductive interactions would be indispensable in constructing potential surfaces and be very useful in studies of quantum three-body effect for the excited Li$_3^+$ trimers.

\subsection{Long-range potentials: Results}

Evaluating the
additive and nonadditive
long-range potentials using the coefficients
given in Tables~\ref{TabADDzaxis}--\ref{TabEQ}, the potential
functions are displayed for two geometries:  An equally-spaced collinear configuration with $R_{23}=R_{31}=R$,
see Figs.~\ref{figcolz} and \ref{figcol}, and an equilateral triangle 
with sides of
length $R$,
see Fig.~\ref{figeqt}. We should indicate that the nonadditive interactions of the present paper are all evaluated for the geometries lying on the $x$-$y$ plane as shown in Fig.~\ref{fig:coords}. Thus for the three particles lying on the $z$-axis, only the additive potentials are shown in Fig.~\ref{figcolz} with respect to the two-body p$_1$ situation (see Fig.~\ref{fig:axis}). For the collinear configuration lying on the $x$-$y$ plane, the total additive and nonadditive potentials are displayed in Fig. \ref{figcol}. The separations between the $M=0$ and $M=\pm1$ states are mainly caused by the leading repulsive or attractive electrostatic interaction involving $C_3$ between the ion and the excited atom. In Fig.~\ref{figeqt}, we display the total long-range potentials (additive and nonadditive) for the geometry of equilateral triangle lying on the $x$-$y$ plane,  where a barrier 
of about $115\,\text{cm}^{-1}$
at internuclear distance of  $17\,a_0$ is found for the $\Psi_{1,\bot}^{(0)}(1, M=0)$ state. The barrier results
from the interplay
of the repulsive leading terms involving $C_3$ and the attractive induction interaction involving $C_4$. 
For the other states, the long-range potentials 
are attractive
at all internuclear distances.  
Note that the data
presented in this
subsection does not include exchange energies, which may contribute at these internuclear distances.
We will discuss their contributions in the following subsection.

\subsection{The strong nonadditive potentials and ``switch-off" of the additive potentials}
\label{subsec:switchoff}

\begin{figure}
\centering
\begin{minipage}[t]{0.48\textwidth}
\centering
\includegraphics[width=7.25cm]{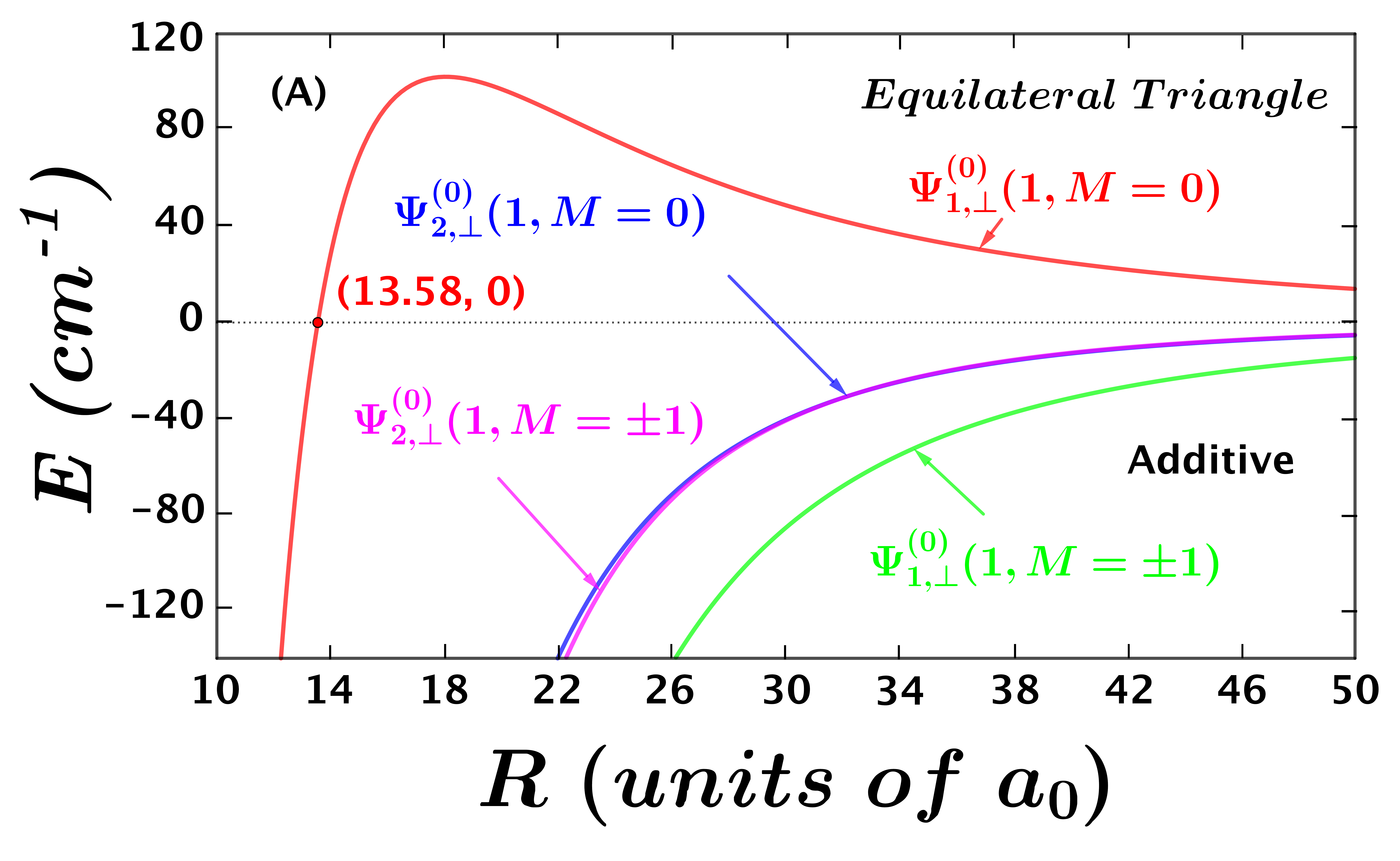}
\end{minipage}
\begin{minipage}[t]{0.48\textwidth}
\centering
\includegraphics[width=7.25cm]{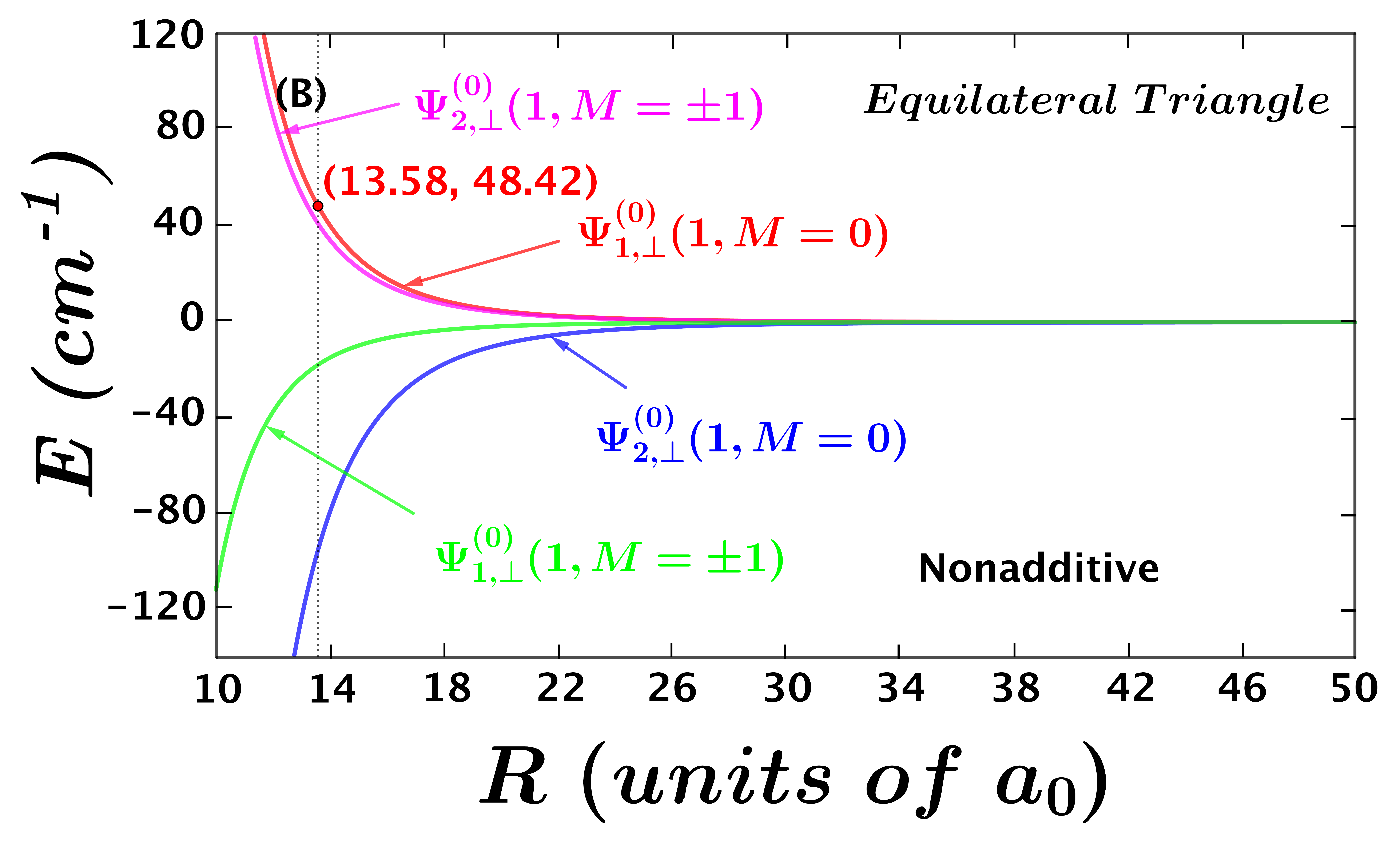}
\end{minipage}

\begin{minipage}[t]{0.48\textwidth}
\centering
\includegraphics[width=8cm]{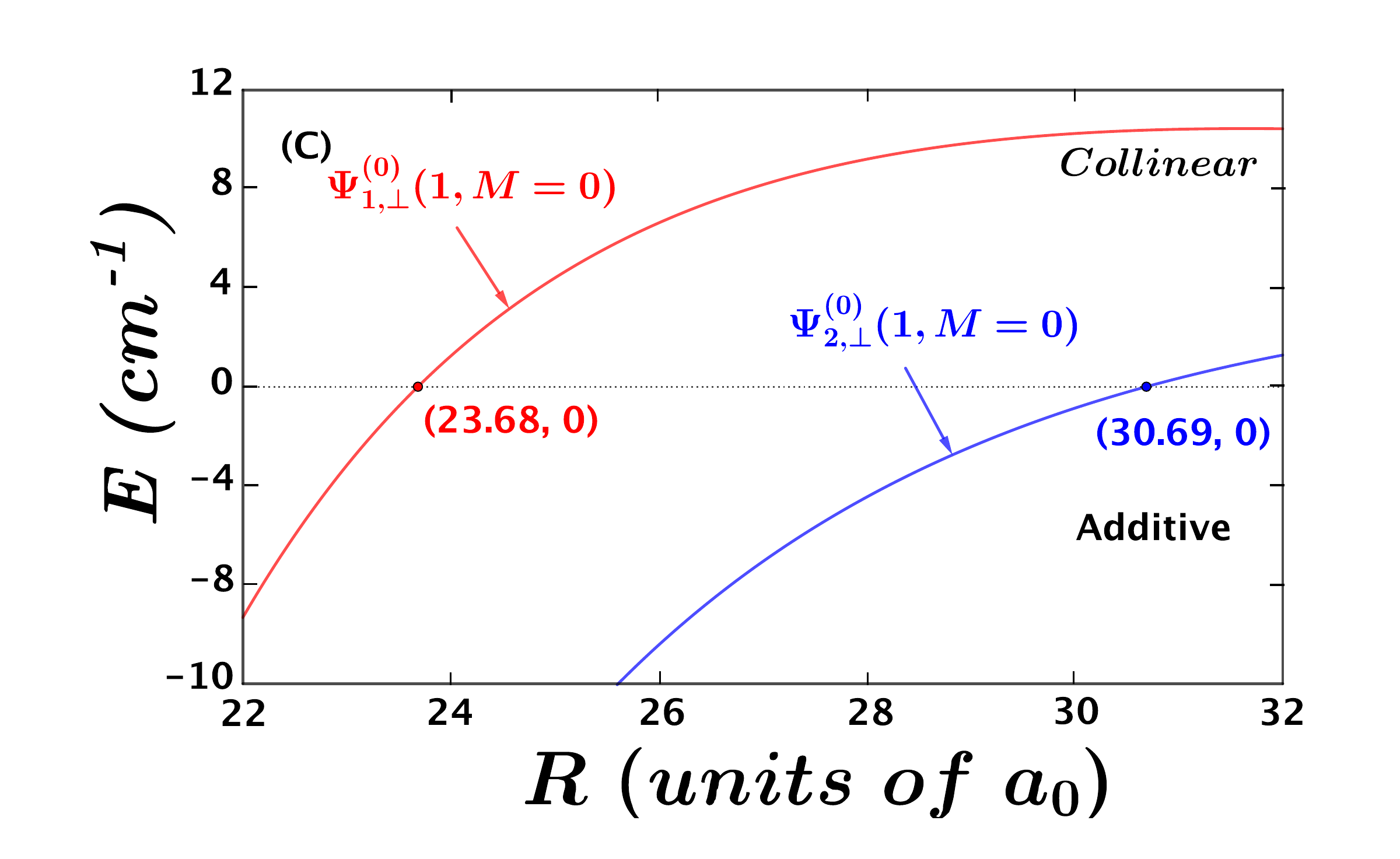}
\end{minipage}
\begin{minipage}[t]{0.48\textwidth}
\centering
\includegraphics[width=8cm]{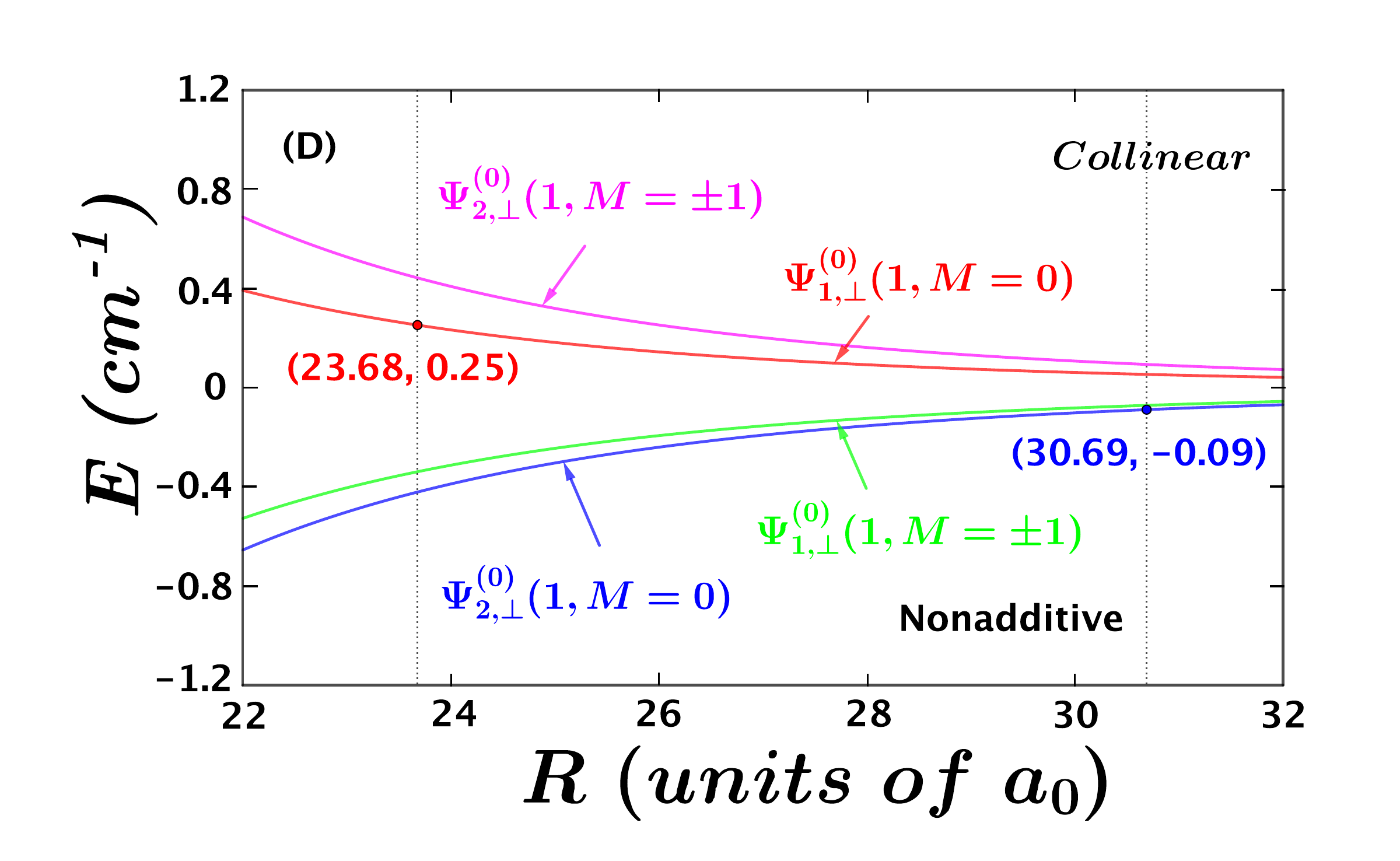}
\end{minipage}

\caption {\label{f5} 
Comparison of long-range additive potentials
[left side, (A) and (C)]
and nonadditive (collective) potentials [right side, (B) and (D)] (in atomic units)
of the $^\infty\atom(2\,^2S)$-$^\infty\atom(2\,^2P)$-$^\infty\atom^+(1\,^{1}S)$ system for two 
 types of the zeroth-order wave functions with $R_{23}=R_{31}=R$: equilateral triangle
(A) and (B), equally-spaced collinear configurations
(C) and (D). At the labeled points, the two-body additive potentials sum to zero leaving only the net nonadditive collective potentials. } 
\end{figure}

As we discussed before, the nonadditive collective effect of the three-body system is caused by its degeneracy, which is introduced by the presence of the excited Li($2\,^2P$) atom. Meanwhile, the presence of the Li$^{+}$($1\,^{1}S$) ion introduces the induction effect, which strongly 
enhances the nonadditive (collective) interaction, 
as demonstrated in Section \ref{subsec:two-specific-cases}. In this subsection, we present a graphical comparison of the additive and
nonadditive potentials for the equilateral triangle and collinear ($R_{23}=R_{31}=R$) configurations, see Fig.~\ref{f5}. 

The figure illustrates that
the nonadditive potentials are significant and can
even be stronger than the net contribution
from the additive potentials.
For example, for the equilateral triangle
configuration, the magnitude  of the additive
contribution [Fig.~\ref{f5}(A)] becomes less than the
nonadditive contribution
[Fig.~\ref{f5}(C)] around $R\sim 14\,a_0$. Indeed, we find that there are specific internuclear distances
at which the additive contributions sum to zero leaving only nonadditive contributions. Denoting these
special distances
by $\bar{R}$, for the equilateral triangle
configuration, the additive cancellation occurs at $\bar{R}=13.58\,a_0$
for the $\Psi_{1,\Delta}(1,M=0)$ state with a net
energy of $48.42\,\text{cm}^{-1}$.
For the collinear configuration the additive cancellation
occurs at $\bar{R}=23.68\,a_0$ for the $\Psi_{1,\Delta}(1,M=0)$ state with a net energy of $0.25\,\text{cm}^{-1}$ and at
$\bar{R}=30.69\,a_0$ for the $\Psi_{2,\Delta}(1,M=0)$ state
with a net energy of $-0.09\,\text{cm}^{-1}$.

We now, as promised in the Introduction, Sec.~\ref{subsec:lattice}, 
draw a comparison with trapped
cold polar molecules.
When the additive (two-body) contributions sum to zero at a distance $\bar{R}$, Eq.~(\ref{etot}) reduces to
\begin{equation}
    \Delta E(\bar{R})=\Delta E_{\text{non}}^{(2)}(\bar{R}).
\end{equation}
Comparing Eq.~(\ref{etotnonadd2}) for
$\Delta E_{\text{non}}^{(2)}$
with Eq.~(\ref{eq:latticetrimer}) for the three-body lattice interaction,
we observe that they are precisely the same form.
Since our results are specific to three-particles, the collinear case is most similar to the case of trapped polar molecules in a linear configuration,
such as shown in Fig.~1(A) of Ref.~\cite{BucMicZol07}.
Our intriguing result deserves further study. In retrospect, we can understand
the appearance of a cancellation
analogous to that found for trapped polar molecules: The anisotropy of the present system due to the $\atom (2\,^2P)$ atom in the presence
of the $\ion (1\,^1S)$ ion charge 
is physically similar to the dipole-dipole interaction in the presence of an external electric field in the optical lattice case. To gauge
precisely the physical potential energies at the special distances $\bar{R}$,
treatment of the exchange energy contributions, or equivalently
quantum-chemical calculations,
would be desirable.
However, by analogy with the $\ion(1\,^1S)$--$\atom(2\,^2P)$
results that we presented in
Sec.~\ref{subsec:PS+},
we observe that the values of $\bar{R}$
are probably sufficiently large so that it
is likely that 
only exchange energies will contribute.
Nevertheless, the present
result suggests an intriguing
similarity between 
the $\atom(2\,^2S)$-$\atom(2\,^2P)$-$\atom^+(1\,^{1}S)$
system
and the trapped cold polar molecule scenario.

\section{Conclusion}

The long-range additive and nonadditive interaction potentials for the Li($2\,^2S$)-Li($2\,^2P$)-Li$^{+}$($1\,^{1}S$) system were calculated by using degenerate perturbation theory.
 We found that all the first-order and second-order additive and nonadditive interaction coefficients show a dependence on the geometrical configurations of the system. The nonadditive interactions depend on both the atomic states and the interior angles of the configurations. 
The degeneracy of the system caused by the presence of
the $\atom(2\,^2P)$ atom leads to the three-body collective effect. The presence of the $\ion (1\,^{1}S)$ ion was found
to enhance this collective effect, which makes the three-body nonadditive collective interactions of the system  even stronger than the two-body additive interactions for some specific configurations of the three-body system.
For the two particular configurations
with $R_{23}=R_{31}=R$, the  equilateral triangle configuration and the equally-spaced collinear configuration, the interaction coefficients 
were evaluated
with highly accurate wave functions calculated variationally 
in Hylleraas coordinates.
In addition,
for the Li($2\,^2S$)-Li($2\,^2P$)-Li$^{+}$($1\,^{1}S$) system, the two-body additive interaction can be ``switched off" leaving only three-body nonadditive interactions for particular geometries, which makes this three-body system a 
prospective platform to study the quantum collective effect.
We demonstrated how two-body interaction potentials
can be extracted from our results and gave explicit expressions for the long-range potentials of the $\ion(1\,^1S)$--$\atom(2\,^2P)$
system. The present
high-precision results
can serve as benchmarks for future quantum-chemical calculations and may be of interest for constructing precise potential energy surfaces. The general formulae for $A(n_0S)$-$A(n_0'L)$-$A^{Q+}(n_0''S)$ are listed in the Supplemental Material.

\begin{acknowledgments}
This work was supported by NSERC of Canada,
and by the Strategic Priority Research Program of the Chinese Academy of Sciences under Grant No.~XDB21030300.
J.F.B.
was supported in part by the U.S. NSF through grant PHY-1521560 for the Institute of Theoretical Atomic, Molecular,
and Optical Physics at Harvard University and Smithsonian
Astrophysical Observatory.
\end{acknowledgments}

\bibliographystyle{apsrev4-2}
\bibliography{positron}

\newpage


\pagebreak

\onecolumngrid

\begin{center}
  \textbf{\large Supplemental Material: Long-range additive and nonadditive potentials in a hybrid system:
Ground state atom, excited state atom, and ion}\\[.2cm]
  Pei-Gen Yan,$^{1}$ Li-Yan Tang,$^{2}$ Zong-Chao Yan,$^{1,2}$ and James F. Babb$^3$\\[.1cm]
  {\itshape ${}^1$Department of Physics, University of New Brunswick,\\ Fredericton, New Brunswick, E3B 5A3, Canada\\ 
  ${}^2$State Key Laboratory of Magnetic Resonance and Atomic and Molecular Physics,\\ Wuhan Institute of Physics and Mathematics,\\ Innovation Academy for Precision Measurement Science and Technology, \\Chinese Academy of Sciences, Wuhan 430071, People's Republic of China\\ 
  ${}^3$ITAMP, Center for Astrophysics \textbar\ Harvard \& Smithsonian, \\MS 14, 60 Garden St., Cambridge, MA 02138, USA\\}
(Dated: \today)\\[1cm]

\end{center}


\maketitle


This Supplemental Material provides 
additional details on
the calculations.
We introduce 
elements of the
perturbative
approach in Secs.~\ref{subsec:appendix-zero}--\ref{subsec:appendix-secondorder} and
give specific
expressions
for $\ion (1\,^1S)$--$\atom (2\,^2S)$ system in Sec.~\ref{subsec:appendix-dimerion},
for $\atom (2\,^2S)$--$\atom  (2\,^2P)$ system in
Sec.~\ref{subsec:appendix-dimer},
for $\atom (2\,^2P)$--$\ion (1\,^1S)$ system in
Sec.~\ref{subsec:appendix-p+} and
for $\atom (2\,^2S)$--$\atom  (2\,^2P)$--$\ion (1\,^1S)$ system in
Sec.~\ref{subsec:appendix-trimer}.
In the present work, we take the electrostatic interaction $V_{123}$ between pairs of particles for the $A(n_0S)$
-$A(n_0'L)$-$A^{Q+}(n_0''S)$  system as a perturbation
\begin{equation}\label{v123}
H'=V_{123}=V_{12}+V_{23}+V_{31} \,,
\end{equation}
where $V_{12}$, $V_{23}$ and $V_{31}$ are the two-body mutual electrostatic interactions between atoms 1
and 2 and ion 3. For three well-separated systems, the mutual interaction energy $V_{IJ}$ can be expanded with the same method as used in Refs.~\cite{yan16, yan18, yan20},
\begin{equation}\label{e2}
 V_{IJ}=\sum_{l_Il_J}\sum_{m_Im_J}T_\text{$l_I-m_I$}(\boldsymbol{\sigma})T_\text{$l_Jm_J$}(\boldsymbol{\rho})W_{l_Il_J}^{m_I-m_J}(IJ) \,,
\end{equation}
where the multipole transition operators are
\begin{eqnarray}
 T_{l_I-m_I}(\boldsymbol{\sigma})&=&\sum_{i}Q_i\sigma ^{l_I}_{i}Y_{l_I-m_I}(\hat{\boldsymbol{\sigma}_i}) \,,  \\ \label{e3a}
T_{l_Jm_J}(\boldsymbol{\rho})&=&\sum_{j}q_j\rho ^{l_J}_{j}Y_{l_Jm_J}(\hat{\boldsymbol{\rho}_j}) \,,  \label{e3b}
\end{eqnarray}
 where $Q_i$ and $q_j$,
respectively, are the charges of the $i$-th and $j$-th sub-particles of the atoms $I$ and $J$. 
The geometry factor is
\begin{eqnarray}\label{e4}
W_{l_Il_J}^{m_I-m_J}(IJ)&=&\frac{4\pi(-1)^{l_J}}{R_{IJ}^{l_I+l_J+1}}\frac{(l_I+l_J-m_I+m_J)!(l_I,l_J)^{-1/2}}{[(l_I+m_I)!
(l_I-m_I)!(l_J+m_J)!(l_J-m_J)!]^{1/2}}
P_{l_I+l_J}^{m_I-m_J}(\cos\theta_{IJ})\nonumber\\
&\times & \exp[{i(m_I-m_J)\Phi_{IJ}}] \,,
\end{eqnarray}
where ${\bf R}_{IJ}={\bf R}_J-{\bf R}_I$ is the relative position vector from particle $I$ to particle $J$, the notation
$(l_I,l_J,\ldots)=(2l_I+1)(2l_J+1)\ldots$, and $P_{l_I+l_J}^{m_I-m_J}(\cos\theta_{IJ})$ is the associated Legendre function with $\theta_{IJ}$ representing the angle between ${\bf R}_{IJ}$ and the $z$-axis. If we now choose the $z$ axis to be normal to the plane of the three particles, ~\textit{i. e.}, $\theta_{12}=\theta_{23}=\theta_{31}=\pi/2$, the associated Legendre functions can be  simplified as
\begin{eqnarray}
P_{l}^{m}(0)&=&\frac{1}{2^{l+1}}[1+(-1)^{l+m}](-1)^{\frac{l+m}{2}}(l+m)!
\bigg[\bigg(\frac{l+m}{2}\bigg)!\bigg]^{-1}\bigg[\bigg(\frac{l-m}{2}\bigg)!\bigg]^{-1} \,.
\end{eqnarray}
$\Phi_{IJ}$ denotes the angle between ${\bf R}_{IJ}$ and the $x$-axis. It shows the dependence of the mutual dipole-dipole interaction between two atoms on the orientation of the interacting dipoles relative to the line connecting them \cite{axilrod43}. The expressions for $V_{JK}$ and $V_{KI}$ are similar to $V_{IJ}$. For simplicity, in this work, we transform all $\Phi_{IJ}$
into interior angles ($\alpha$, $\beta$, $\gamma$) of the triangle formed by the three lithium nuclei with the same method
as used in Ref.~\cite{yan16}.

\subsection{The zeroth-order wave function}
\label{subsec:appendix-zero}

According to degenerate perturbation theory, the zeroth-order wave function of the unperturbed system $A(n_0S)$
-$A(n_0'L)$-$A^{Q+}(n_0''S)$ can be written as
\begin{eqnarray}
\ket{\Psi^{(0)}}=a\ket{\phi_1}+b\ket{\phi_2} \,,
\end{eqnarray}
where the two orthonormalized degenerate eigenvectors of the unperturbed Hamiltonian with the energy eigenvalue $E_{n_0n_0'n_0''}^{(0)}=E_{n_0S}^{(0)}+E_{n_0'L}^{(0)}+E_{n_0''S}^{(0)}$ can be writen as,
\begin{eqnarray}
\ket{\phi_1}&=&\ket{{n_0'}L;{n_0}0;{n_0''}0}\,, \label{e6a_a} \\
\ket{\phi_2}&=&\ket{{n_0}0;{n_0'}L;{n_0''}0}\,. \label{e6b_b}
\end{eqnarray}

The expansion coefficients $a$, $b$ are determined by diagonalizing the perturbation in the basis set \{$\phi_1$, $\phi_2$\}, which depend on the geometrical configuration formed by the three particles. In the following, we show that all the long-range interaction coefficients would contain $a$, $b$ or one of them. This
leads to the dependence of these coefficients on the configurations of the three particles. Thus the zeroth-order energy correction is obtained by the perturbation matrix with respect to $\{\phi_1, \phi_2\}$
\begin{equation}\label{ehp}
{H}^{\prime}=
\left(
  \begin{array}{ccc}
    \Delta_{11} & \Delta_{12} \\
    \Delta_{12}^* & \Delta_{22} \\
  \end{array}
\right)\,,
\end{equation}
where
\begin{eqnarray}\label{Delta_11}
\Delta_{11}&=& \langle\phi_1|V_{123}|\phi_1\rangle =\langle\phi_1|V_{12}|\phi_1\rangle
+\langle\phi_1|V_{23}|\phi_1\rangle
+\langle\phi_1|V_{31}|\phi_1\rangle
\nonumber \\
&=&  \sum_{l_1}
\frac{ Q (-1)^{L-M} }{R_{31}^{l_{1}+1}}
 \sqrt{\frac{{4\pi}}{2l_1+1}} 
P_{l_{1}}(0)
\left(
 \begin{array}{ccc}
    L & l_{1} & L \\
    -M & 0 &  M \\
  \end{array}
  \right)
 \langle{n'_0}L\|T_{l_{1}}\|{n'_0}L \rangle \,,
\end{eqnarray}
\begin{eqnarray}\label{Delta_12}
\Delta_{12}&=& \langle\phi_1|V_{123}|\phi_2\rangle =\langle\phi_1|V_{12}|\phi_2\rangle
+\langle\phi_1|V_{23}|\phi_2\rangle
+\langle\phi_1|V_{31}|\phi_2\rangle
\nonumber \\
&=&
\frac{4\pi}{R_{12}^{2L+1}}\frac{(-1)^{L+M}(2L)!P_{2L}{(0)}}{(2L+1)^2(L-M)!(L+M)!}  |\langle{n_0}0\|T_L
\|{n_0'}L\rangle|^2 \,,
\end{eqnarray}
\begin{eqnarray}\label{Delta_22}
\Delta_{22}&=& \langle\phi_2|V_{123}|\phi_2\rangle =\langle\phi_2|V_{12}|\phi_2\rangle
+\langle\phi_2|V_{23}|\phi_2\rangle
+\langle\phi_2|V_{31}|\phi_2\rangle
\nonumber \\
&=& \sum_{l_2}
\frac{ Q  (-1)^{L-M} }{R_{23}^{l_{2}+1}}
\sqrt{\frac{{4\pi}}{2l_2+1}} 
P_{l_{2}}(0)
\left(
 \begin{array}{ccc}
    L & l_{2} & L \\
    -M & 0 &  M \\
  \end{array}
  \right)
 \langle{n'_0}L\|T_{l_{2}}\|{n'_0}L \rangle \,.
\end{eqnarray}
We solve this eigenvalue problem to get the eigenvalues and corresponding eigenfunctions.

\subsection{The first-order energy correction}
\label{subsec:appendix-firstorder}

According to perturbation theory, the first-order energy correction is
\begin{eqnarray}
&&\Delta E^{(1)}=\langle \Psi^{(0)}|V_{123}|\Psi^{(0)}\rangle \nonumber \\
&&=|a|^{2}\langle\phi_1|V_{123}|\phi_1\rangle+|b|^{2}\langle\phi_2|V_{123}|\phi_2\rangle+(a^{*}b+b^{*}a)\langle\phi_1|V_{123}|\phi_2\rangle \nonumber \\
&&=|a|^{2} \sum_{l_1} \frac{ Q (-1)^{L-M}  }{R_{31}^{l_{1}+1}} \sqrt{\frac{{4\pi}}{2l_1+1}}  P_{l_{1}}({0} )
\left(
 \begin{array}{ccc}
    L & l_{1} & L \\
    -M & 0 &  M \\
  \end{array}
  \right)
 \langle {n'_0}L\|T_{l_{1}}\|{n'_0}L \rangle
\nonumber\\
&&+|b|^{2} \sum_{l_2} \frac{ Q  (-1)^{L-M} }{R_{23}^{l_{2}+1}}  \sqrt{\frac{{4\pi}}{2l_2+1}}  P_{l_{2}}(0)
\left(
 \begin{array}{ccc}
    L & l_{2} & L \\
    -M & 0 &  M \\
  \end{array}
  \right)
 \langle{n'_0}L\|T_{l_{2}}\|{n'_0}L \rangle
\nonumber\\
&&+(a^{*}b+b^*a)\frac{4\pi}{R_{12}^{2L+1}}\frac{(-1)^{L+M}(2L)!P_{2L}{( 0 )}}{(2L+1)^2(L-M)!(L+M)!}  |\langle{n_0}0\|T_L
\|{n_0'}L\rangle|^2 \,.
\end{eqnarray}

\subsection{The second-order energy correction}
\label{subsec:appendix-secondorder}

The second-order energy correction is given by
\begin{eqnarray}\label{E2}
\Delta E^{(2)}&=&-\sum_{n_sn_tn_u}\sum_{L_sL_tL_u}\sum_{M_sM_tM_u}
\frac{|\langle\Psi^{(0)}|V_{123}|{n_s}L_s;{n_t}L_t;{n_u}L_u \rangle |^{2}}{E_{n_sL_s;n_tL_t;n_uL_u}-E_{n_0S;n_0'L;n_0''S}^{(0)}} \nonumber \\
&=&V_{12}^{(2)}+V_{23}^{(2)}+V_{31}^{(2)}+V_{12,23}^{(2)}+V_{23,31}^{(2)}+V_{31,12}^{(2)}\,,
\end{eqnarray}
where $ |{n_s}L_s;{n_t}L_t;
{n_u}L_u \rangle$ is an intermediate state of the system with the energy eigenvalue $E_{n_sL_s;n_tL_t;n_uL_u}=E_{n_sL_s}+E_{n_tL_t}+E_{n_uL_u}$. It is noted that the above summations should exclude terms
with $E_{n_sL_s;n_tL_t;n_uL_u}=E_{n_0S;n_0'L;n_0''S}^{(0)}$. The three additive interaction terms, denoted by $V_{12}^{(2)}$, $V_{23}^{(2)}$, $V_{31}^{(2)}$ become, respectively,

\begin{eqnarray}\label{V12-12b}
 V_{12}^{(2)}  &=& -|a|^2\sum_{n_sn_t}\sum_{L_sL_t l_1l_{1}^{\prime}}\sum_{M_sM_t m_{1}} \frac{16\pi^2}{R_{12}^{2L_{t}+l_1+l_1^{\prime}+2}}\left(
  \begin{array}{ccc}
    L & l_1 & L_{s}\\
    -M & m_1 & M_{s}\\
  \end{array}
\right)
\left(
  \begin{array}{ccc}
    L & l_{1}^{\prime} & L_{s}\\
    -M & m_{1} & M_{s}\\
  \end{array}
\right) \nonumber \\
&\times&
\frac{P_{L_{t}+l_1}^{M_{t}-m_1}(0)P_{L_{t}+l_1^{\prime}}^{M_{t}-m_1}(0)(L_{t}+l_1-M_{t}+m_1)!(L_{t}+l_1^{\prime}-M_{t}+m_1)!(L_{t},L_{t})^{-1}(l_1,l_1^{\prime})^{-1/2}}
{(L_{t}+M_{t})!(L_{t}-M_{t})![(l_1+m_1)!(l_1-m_1)!(l_1^{\prime}+m_1)!(l_1^{\prime}-m_1)!]^{1/2}}   \nonumber \\
&\times&
\frac{\langle{n_0'}L\|T_{l_1}\|{n_s}L_s\rangle^*
\langle{n_0'}L\|T_{l_{1}^{\prime}}\|{n_s}L_s\rangle
|\langle{n_0}0\|T_{L_t}\|{n_t}L_t\rangle|^{2}}
{E_{n_sL_s}+E_{n_tL_t}-E_{n_0S}^{(0)}-E_{n_0'L}^{(0)}}
 \nonumber \\
&-&|b|^2\sum_{n_sn_t}\sum_{L_sL_tl_2l_{2}^{\prime}}\sum_{M_sM_t m_{2}}\frac{16\pi^2}
{R_{12}^{2L_{s}+l_2+l_2^{\prime}+2}}\left(
  \begin{array}{ccc}
    L & l_2 & L_{t}\\
    -M & m_2 & M_{t}\\
  \end{array}
\right)
\left(
  \begin{array}{ccc}
    L & l_{2}^{\prime} & L_{t}\\
    -M & m_{2} & M_{t}\\
  \end{array}
\right)  \nonumber \\
&\times&
\frac{P_{L_{s}+l_2}^{M_{s}-m_2}(0) P_{L_{s}+l_2^{\prime}}^{M_{s}-m_2}(0)(L_{s}+l_2-M_{s}+m_2)!(L_{s}+l_2^{\prime}-M_{s}+m_2)!(L_{s},L_{s})^{-1}(l_2,l_2^{\prime})^{-1/2}}
{(L_{s}+M_{s})!(L_{s}-M_{s})![(l_2+m_2)!(l_2-m_2)!(l_2^{\prime}+m_2)!(l_2^{\prime}-m_2)!]^{1/2}}  \nonumber \\
&\times&
\frac{|\langle{n_0}0\|T_{L_s}\|{n_s}L_s\rangle|^{2}
\langle{n_0'}L\|T_{l_2}\|{n_t}L_t\rangle^*
\langle{n_0'}L\|T_{l_{2}^{\prime}}\|{n_t}L_t\rangle}
{E_{n_sL_s}+E_{n_tL_t}-E_{n_0S}^{(0)}-E_{n_0'L}^{(0)}}\nonumber  \\
&-& a^{*}b\sum_{n_sn_t}\sum_{L_sL_t l_1l_{2}^{\prime}}\sum_{M_sM_t m_{1}m_{2}^{\prime}}
\frac{16\pi^2(-1)^{L_{s}+l_2^{\prime}-M_{s}-M_{t}}}{R_{12}^{l_1+L_{s}+L_{t}+l_2^{\prime}+2}}\left(
  \begin{array}{ccc}
    L & l_1 & L_{s}\\
    -M & -m_1 & M_{s}\\
  \end{array}
\right)
\left(
  \begin{array}{ccc}
    L & l_{2}^{\prime} & L_{t}\\
    -M & m_{2}^{\prime} & M_{t}\\
  \end{array}
\right)\nonumber  \\
&\times&\frac{P_{L_{t}+l_{1}}^{M_{t}+m_{1}}(0)P_{L_{s}+l_{2}^{\prime}}^{M_{s}-m_{2}^{\prime}}(0) (L_{t}+l_{1}-M_{t}-m_{1})!(L_{s}+l_2^{\prime}-M_{s}+m_2^{\prime})!{(L_{s},L_{t})^{-1}(l_1,l_2^{\prime})^{-1/2}}}
{[(L_{s}+M_{s})!(L_{s}-M_{s})!(L_{t}+M_{t})!(L_{t}-M_{t})!(l_{1}+m_{1})!(l_{1}-m_{1})!(l_2^{\prime}+m_2^{\prime})!
(l_2^{\prime}-m_2^{\prime})!]^{1/2}} \nonumber \\
&\times&
\frac{\langle{n_0'}L\|T_{l_1}\|{n_s}L_s\rangle^*
\langle{n_0}0\|T_{L_t}\|{n_{t}}L_t\rangle^*
\langle{n_0}0\|T_{L_s}\|{n_s}L_s\rangle
\langle{n_0'}L\|T_{l_{2}^{\prime}}\|{n_t}L_t\rangle}
{E_{n_sL_s}+E_{n_tL_t}-E_{n_0S}^{(0)}-E_{n_0'L}^{(0)}}
 \nonumber \\
&-&b^{*}a\sum_{n_sn_t}\sum_{L_sL_t l_1^{\prime}l_{2}}\sum_{M_sM_t m_{1}^{\prime}m_2} \frac{16\pi^2(-1)^{L_{s}+l_2-M_{s}-M_{t}}}{R_{12}^{l_1^{\prime}
+L_{t}+L_{s}+l_2+2}}\left(
  \begin{array}{ccc}
    L & l_1^{\prime} & L_{s}\\
    -M & -m_1^{\prime} & M_{s}\\
  \end{array}
\right)
\left(
  \begin{array}{ccc}
    L & l_{2} & L_{t}\\
    -M & m_{2} & M_{t}\\
  \end{array}
\right) \nonumber \\
&\times& \frac{P_{L_{s}+l_{2}}^{M_{s}-m_{2}}(0)P_{L_{t}+l_{1}^{\prime}}^{M_{t}+m_{1}^{\prime}}(0)
(L_{t}+l_{1}^{\prime}-M_{t}-m_{1}^{\prime})!(L_{s}+l_2-M_{s}+m_2)!{(L_{s},L_{t})^{-1}(l_1^{\prime},l_2)^{-1/2}}}
{[(L_{s}+M_{s})!
(L_{s}-M_{s})!(L_{t}+M_{t})!(L_{t}-M_{t})!(l_{1}^{\prime}+m_{1}^{\prime})!(l_{1}^{\prime}-m_{1}^{\prime})!(l_2+m_2)!(l_2-m_2)!]^{1/2}}
\nonumber \\
&\times&
\frac{\langle{n_0}0\|T_{L_{s}}\|{n_s}L_s\rangle^*
\langle{n_0'}L\|T_{l_{2}}\|{n_t}L_t\rangle^*\langle{n_0'}L\|T_{l_{1}^{\prime}}\|{n_s}L_s\rangle\langle{n_0}0\|T_{L_t}\|{n_t}L_t\rangle
}
{E_{n_sL_s}+E_{n_tL_t}-E_{n_0S}^{(0)}-E_{n_0'L}^{(0)}}\nonumber \\
&=&-\bigg\{|a|^2\sum_{n_sn_t}\sum_{L_sL_t l_1l_{1}^{\prime}} \frac{F_1(n_s,n_t,L_s,L_t;l_1,l_1';L,M)}{R_{12}^{2L_{t}+l_1+l_1^{\prime}+2}}
+|b|^2\sum_{n_sn_t}\sum_{L_sL_tl_2l_{2}^{\prime}}
\frac{F_1(n_t,n_s,L_t,L_s;l_2,l_2';L,M)}
{R_{12}^{2L_{s}+l_2+l_2^{\prime}+2}} \nonumber  \\
&+& a^{*}b\sum_{n_sn_t}\sum_{L_sL_t l_1l_{2}^{\prime}}
\frac{F_3(n_s,n_t,L_s,L_t;l_1,l_2';L,M)}{R_{12}^{L_{s}+L_{t}+l_1+l_2^{\prime}+2}} + b^{*}a\sum_{n_sn_t}\sum_{L_sL_t l_1^{\prime}l_{2}}
\frac{ F_3^*(n_s,n_t,L_s,L_t;l_1',l_2;L,M)}{R_{12}^{L_{s}+L_{t}+l_1^{\prime}
+l_2+2}}\bigg\} \,, \nonumber  \\
\end{eqnarray}

\begin{eqnarray}\label{V23-23b}
V_{23}^{(2)}&=&-|a|^2\sum_{n_tn_u}\sum_{L_tL_u}\sum_{M_tM_u}\frac{16\pi^2}{R_{23}^{2L_{t}+2L_{u}+2}}
\frac{[P_{L_{t}+L_{u}}^{M_{t}+M_{u}}(0)(L_{t}+L_{u}-M_{t}-M_{u})!]^2(L_{t},L_{u})^{-2}}{(L_{t}+M_{t})!(L_{t}-M_{t})!(L_{u}+M_{u})!
(L_{u}-M_{u})!} \nonumber \\
&\times&
\frac{|\langle{n_0}0\|T_{L_{t}}\|{n_t}L_t\rangle|^{2}
|\langle{n_0''}0\|T_{L_u}\|{n_u}L_u\rangle|^{2}}
{E_{n_tL_t}+E_{n_uL_u}-E_{n_0S}^{(0)}-E_{n_0''S}^{(0)}} \nonumber \\
&-&|b|^2\sum_{n_tn_u}\sum_{L_tL_u l_2l_{2}^{\prime}}\sum_{M_tM_u m_2}
\frac{16\pi^2}{R_{23}^{2L_{u}+l_2+l_2^{\prime}+2}}\left(
  \begin{array}{ccc}
    L & l_2 & L_{t}\\
    -M & m_2 & M_{t}\\
  \end{array}
\right)
\left(
  \begin{array}{ccc}
    L & l_{2}^{\prime} & L_{t}\\
    -M & m_{2} & M_{t}\\
  \end{array}
\right)  \nonumber \\
&\times&\frac{ P_{L_{u}+l_2}^{M_{u}-m_2}(0)P_{{L_{u}+l_2^\prime}}^{M_{u}-m_2}(0)  (L_{u}+l_2-M_{u}+m_2)!(L_{u}+l_2^{\prime}-M_{u}+m_2)!(L_{u},L_{u})^{-1}(l_2,l_2^{\prime})^{-1/2}}{(L_{u}+M_{u})!
(L_{u}-M_{u})![(l_2+m_2)!(l_2-m_2)!(l_2^{\prime}+m_2)!(l_2^{\prime}-m_2)!]^{1/2}} \nonumber \\
&\times&
\frac{\langle{n_0'}L\|T_{l_2}\|{n_t}L_t\rangle^*
\langle{n_0'}L\|T_{l_2'}\|{n_t}L_t\rangle
|\langle{n_0''}0\|T_{L_{u}}\|{n_u}L_u\rangle|^{2}}
{E_{n_tL_t}+E_{n_uL_u}-E_{n_0''S}^{(0)}-E_{n_0'L}^{(0)}} \nonumber \\
&=&-\bigg\{|a|^2\sum_{n_tn_u}\sum_{L_tL_u}
\frac{F_2(n_t,n_u,L_t,L_u)}{R_{23}^{2L_{t}+2L_u+2}}
+|b|^2\sum_{n_tn_u}\sum_{L_tL_u l_2l_{2}^{\prime}}
\frac{F_1(n_t,n_u,L_t,L_u;l_2,l_2';L,M)}{R_{23}^{2L_{u}+l_2+l_2^{\prime}+2}}
\bigg\}  \,, \nonumber  \\
\end{eqnarray}

\begin{eqnarray}\label{V31-31b}
V_{31}^{(2)}&=&-|a|^2 \sum_{n_sn_u}\sum_{L_sL_u l_1l_{1}^{\prime}}\sum_{M_sM_u m_{1}}
\frac{16\pi^2}{R_{31}^{2L_{u}+l_1+l_1^{\prime}+2}}\left(
  \begin{array}{ccc}
    L & l_1 & L_{s}\\
    -M & m_1 & M_{s}\\
  \end{array}
\right)
\left(
  \begin{array}{ccc}
    L & l_{1}^{\prime} & L_{s}\\
    -M & m_{1} & M_{s}\\
  \end{array}
\right)\nonumber \\
&\times&
 \frac{P_{L_{u}+l_1}^{M_{u}-m_1}(0)P_{L_{u}+l_1^{\prime}}^{M_{u}-m_1}(0)(L_{u}+l_1-M_{u}+m_1)!
 (L_{u}+l_1^{\prime}-M_{u}+m_1)!(L_{u},L_{u})^{-1}(l_1,l_1^{\prime})^{-1/2}}{(L_{u}+M_{u})!
(L_{u}-M_{u})![(l_1+m_1)!(l_1-m_1)!(l_1^{\prime}+m_1)!(l_1^{\prime}-m_1)!]^{1/2}}\nonumber \\
&\times&
\frac{\langle{n_0'}L\|T_{l_1}\|{n_s}L_s\rangle^*
\langle{n_0'}L\|T_{l_{1}^{\prime}}\|{n_s}L_s \rangle
|\langle{n_0''}0\|T_{L_{u}}\|{n_u}L_u\rangle|^{2}}
{E_{n_sL_s}+E_{n_uL_u}-E_{n_0''S}-E_{n_0'L}}
\nonumber \\
&-&|b|^2\sum_{n_sn_u}\sum_{L_sL_u}\sum_{M_sM_u} \frac{16\pi^2}{R_{31}^{2L_{s}+2L_{u}+2}} \frac{[P_{L_{u}+L_{s}}^{M_{u}+M_{s}}(0)(L_{u}+L_{s}-M_{u}-M_{s})!]^2(L_{u},L_{s})^{-2}}{(L_{u}+M_{u})!(L_{u}-M_{u})!
(L_{s}+M_{s})!(L_{s}-M_{s})!} \nonumber \\
&\times&
\frac{|\langle{n_0}0\|T_{L_{s}}\|{n_s}L_s\rangle|^{2}
|\langle{n_0''}0\|T_{L_{u}}\|{n_u}L_u\rangle|^{2}}
{E_{n_sL_s}+E_{n_uL_u}-E_{n_0S}-E_{n_0''S}}  \nonumber \\
&=&-\bigg\{
|a|^2 \sum_{n_sn_u}\sum_{L_sL_u l_1l_{1}^{\prime}}
\frac{F_1(n_s,n_u,L_s,L_u;l_1,l_1';L,M)}{R_{31}^{2L_{u}+l_1+l_1^{\prime}+2}}
+ |b|^2\sum_{n_sn_u}\sum_{L_sL_u} \frac{F_2(n_s,n_u,L_s,L_u)}{R_{31}^{2L_{s}+2L_{u}+2}}
\bigg\} \,. \nonumber  \\
\end{eqnarray}

The three nonadditive interaction terms, denoted by  $V_{12,23}^{(2)}$, $V_{23,31}^{(2)}$, $V_{31,12}^{(2)}$ become, respectively,

\begin{eqnarray}\label{d12}
 V_{12,23}^{(2)}
&=&-|a|^2 \sum_{n_tL_tM_t}\sum_{l_1}
\frac{8 \sqrt{\pi^3} Q (-1)^{L_{t}-M_{t}+L-M}}{R_{12}^{l_1+L_{t}+1}R_{23}^{L_{t}+1}}
\left(
  \begin{array}{ccc}
    L & l_{1} & L\\
    -M & 0 & M\\
  \end{array}
\right)
cos(M_{t}\beta) \nonumber\\
&\times&  \frac{ P_{l_1+L_{t}}^{M_{t}}(0)P_{L_{t}}^{M_{t}}(0) (l_1+L_{t}-M_{t})!(L_{t})^{-2}(l_1)^{-1/2}}{
(l_1)!(L_{t}+M_{t})!}
\nonumber\\
&\times &
\frac{\langle{n'_0}L\|T_{l_{1}}\|{n'_0}L \rangle
|\langle{n_0}0\|T_{L_t}\|{n_t}L_t\rangle|^2}{E_{n_tL_t}-E_{n_0S}}
\nonumber\\
\nonumber\\
&-& \sum_{n_tL_tM_t}\sum_{l_2'm_2'}
\frac{8\sqrt{\pi^3}Q}{R_{12}^{L+L_{t}+1}R_{23}^{l_2'+1}}  \left(
  \begin{array}{ccc}
    L & l_{2}' & L_{t}\\
    -M & -m_{2}' & M_{t}\\
  \end{array}
\right) \nonumber \\
&\times& \{a^*b\exp[{-i(m_2')\beta}]+ b^*a\exp[{i(m_2')\beta}] \}
\nonumber\\
&\times& \frac{P_{L+L_{t}}^{-M+M_{t}}(0)P_{l_2'}^{m_2'}(0)(L+L_{t}+M-M_{t})!(l_2'-m_2')!(l_2')^{-1/2}(L,L_{t})^{-1}}{[(L+M)!
(L-M)!(L_{t}+M_{t})!(L_{t}-M_{t})!(l_2'+m_2')!
(l_2'-m_2')!]^{1/2}}
\nonumber\\
&\times& \frac{ \langle {n_0}0 \|T_L
\|{n_0'}L\rangle \langle{n_0}0\|T_{L_t}\|{n_t}L_t\rangle^*
\langle{n_0'}L\|T_{l_2'}\|{n_t}L_t\rangle}{E_{n_tL_t}-E_{n_0'L}}
\nonumber\\
\nonumber\\
&-&\sum_{n_tL_tM_t}\sum_{l_2m_2}
\frac{8 \sqrt{\pi^3}Q(-1)^{L}}{R_{12}^{L+l_2+1}R_{23}^{L_{t}+1}}\left(
  \begin{array}{ccc}
    L & l_2 & L_{t}\\
    -M & m_{2} & M_{t}\\
  \end{array}
\right)\nonumber\\
&\times&
\{a^*b \exp[{i(M_{t})\beta}]+b^*a \exp[{-i(M_{t})\beta}]\}
\nonumber\\
&\times&
\frac{P_{L+l_2}^{M-m_2}(0) P_{L_{t}}^{M_{t}}(0)(L+l_2-M+m_2)!(L_{t}-M_{t})!(l_2)^{-1/2}(L_{t},L)^{-1}}{[(L+M)!
(L-M)!(l_2+m_2)!(l_2-m_2)!(L_{t}+M_{t})!
(L_{t}-M_{t})!]^{1/2}}
\nonumber\\
&\times& \frac{ \langle {n_0}0 \|T_L
\|{n_0'}L\rangle^*
\langle{n_0'}L\|T_{l_2}\|{n_t}L_t\rangle^* \langle{n_0}0\|T_{L_t}\|{n_t}L_t\rangle}{E_{n_tL_t}-E_{n_0S}}
\nonumber\\
\nonumber\\
&=&-\bigg\{ |a|^2\sum_{n_tL_tM_t}\sum_{l_1} \frac{F_5(n_t,L_t,M_t;l_1;L,M;Q) cos(M_t\beta)}{R_{12}^{l_1+L_{t}+1}R_{23}^{L_{t}+1}}
\nonumber\\
&+&\sum_{n_tL_tM_t} \sum_{l_2'm_2'} \frac{\{a^*b\exp[{-i(m_2')\beta}]+ b^*a\exp[{i(m_2')\beta}]\}F_6(n_t,L_t,M_t;l_2',m_{2}';L,M;Q)}{R_{12}^{L+L_{t}+1}R_{23}^{l_2'+1}}
\nonumber\\
&+&\sum_{n_tL_tM_t}\sum_{l_2} \frac{\{a^*b \exp[{i(M_{t})\beta}]+b^*a \exp[{-i(M_{t})\beta}]\}F_7(n_t,L_t,M_t;l_2;L,M;Q)}{R_{12}^{L+l_2+1}R_{23}^{L_{t}+1}}
\bigg\}\,,\nonumber\\
\end{eqnarray}

\begin{eqnarray}\label{V23-31b}
V_{23,31}^{(2)}&=&-\sum_{n_uL_uM_u} \frac{16\pi^2(-1)^{L_{u}+L+M_{u}-M}}{R_{23}^{L_{u}+L+1}R_{31}^{L_{u}+L+1}} \frac{[P_{L_u+L}^{M_u-M}(0)(L_u +L-M_u+M)!(L_{u},L)^{-1}]^2 }{(L_{u}+M_{u})!(L_{u}-M_{u})!(L+M)!(L-M)!} \nonumber \\
&\times& \{(a^*b) \exp[i(M_{u}-M)\gamma]+(b^*a) \exp[-i(M_{u}-M)\gamma]\} \nonumber \\
&\times&
\frac{|\langle{n_0'}L\|T_{L}\|{n_0}0\rangle|^2|\langle{n_0''}0\|T_{L_{u}}\|{n_u}L_u\rangle|^{2}}
{E_{n_uL_u}-E_{n_0''S}+E_{n_0S}-E_{n_0'L}} \nonumber \\
&-&\sum_{n_uL_uM_u} \frac{16\pi^2(-1)^{L_{u}+L+M_{u}+M}}{R_{23}^{L_{u}+L+1}R_{31}^{L_{u}+L+1}} \frac{[P_{L_u+L}^{M_u+M}(0)(L_u +L-M_u-M)!(L_{u},L)^{-1}]^2 }{(L_{u}+M_{u})!(L_{u}-M_{u})!(L+M)!(L-M)!} \nonumber \\
&\times& \{(a^*b) \exp[-i(M_{u}+M)\gamma]+(b^*a) \exp[i(M_{u}+M)\gamma]\} \nonumber \\
&\times&
\frac{|\langle{n_0}0\|T_{L}\|{n_0'}L\rangle|^2|\langle{n_0''}0\|T_{L_{u}}\|{n_u}L_u\rangle|^{2}}
{E_{n_uL_u}+E_{n_0'L}-E_{n_0S}-E_{n_0''S}} \nonumber \\
&=& - \sum_{n_uL_uM_u} \bigg\{ \{(a^*b)\exp[i(M_{u}-M)\gamma]\} +  \{(b^*a)\exp[-i(M_{u}-M)\gamma]\} \bigg\}\frac{F_4(n_u,L_u,M_u;L,M)}{R_{23}^{L_{u}+L+1}R_{31}^{L_{u}+L+1}} \,, \nonumber \\
\end{eqnarray}

\begin{eqnarray}\label{d13}
 V_{31,12}^{(2)} &=& - |b|^2 \sum_{n_sL_sM_s}\sum_{l_2}
\frac{8\sqrt{\pi^3}Q(-1)^{L-M+L_{s}-M_{s}}}{R_{31}^{L_{s}+1}R_{12}^{L_{s}+l_2+1}}\left(
  \begin{array}{ccc}
    L & l_2 & L \\
    -M & 0  & M  \\
  \end{array}
\right) cos({M_{s} \alpha})
\nonumber\\
&\times &
\frac{P_{L_{s}+l_2}^{M_{s}}(0) P_{L_{s}}^{M_{s}}(0)(L_{s}+l_2-M_{s})!(L_{s})^{-2}(l_2)^{-1/2}}{(L_{s}+M_{s})!(l_2)!}
\nonumber\\
&\times &
\frac{\langle{n'_0}L\|T_{l_{2}}\|{n'_0}L \rangle^*
|\langle{n_0}0\|T_{L_s}\|{n_s}L_s\rangle|^2}{E_{n_sL_s}-E_{n_0S}}
\nonumber\\
\nonumber\\
&-& \sum_{n_sL_sM_s}\sum_{l_1m_1}
\frac{8\sqrt{\pi^3}Q(-1)^{L}}{R_{12}^{l_1+L+1}R_{31}^{L_{s}+1}} \left(
  \begin{array}{ccc}
    L & l_{1} & L_{s}\\
    -M & -m_{1} & M_{s}\\
  \end{array}
\right)\nonumber\\
&\times&
 \{a^*b\exp[{i(M_{s})\alpha}] + b^*a\exp[{-i(M_{s})\alpha}]\}\nonumber\\
&\times&\frac{ P_{l_1+L}^{m_1+M}(0) P_{L_{s}}^{M_{s}}(0)(l_1+L-m_1-M)!(L_{s}-M_{s})!(L_{s},L)^{-1}(l_1)^{-1/2}}{[(l_1+m_1)!
(l_1-m_1)!(L+M)!(L-M)!(L_{s}+M_{s})!(L_{s}-M_{s})!]^{1/2}}  \nonumber\\
\nonumber\\
&\times& \frac{ \langle {n_0}0 \|T_L
\|{n_0'}L\rangle^*
\langle{n_0'}L\|T_{l_1}\|{n_s}L_s\rangle^* \langle{n_0}0\|T_{L_s}\|{n_s}L_s\rangle}{E_{n_sL_s}-E_{n_0S}}
\nonumber\\
\nonumber\\
&-& \sum_{n_sL_sM_s}\sum_{l_1'm_1'}
\frac{8\sqrt{\pi^3}Q}{R_{12}^{L_{s}+L+1}R_{31}^{l_1'+1}} \left(
  \begin{array}{ccc}
    L & l_{1}' & L_{s}\\
    -M & -m_{1}' & M_{s}\\
  \end{array}
\right)
\nonumber\\
&\times&
\{a^*b\exp[{i(m_1')\alpha}]+b^*a\exp[{-i(m_1')\alpha}]\}
\nonumber\\
&\times&
  \frac{P_{L_{s}+L}^{m_{s}-M}(0)P_{l_1'}^{m_1'}(0) (L_{s}+L-M_{s}+M)!(l_1'-m_1')!(l_1')^{-1/2}(L_{s},L)^{-1}}{[(L_{s}+M_{s})!
(L_{s}-M_{s})!(L+M)!(L-M)!(l_1'+m_1')!(l_1'-m_1')!]^{1/2}}
 \nonumber\\
&\times& \frac{ \langle {n_0}0 \|T_{L_s}
\|{n_s}L_s\rangle^* \langle{n_0}0\|T_{L}\|{n_0'}L\rangle
\langle{n_0'}L\|T_{l_1'}\|{n_s}L_s\rangle}{E_{n_sL_s}-E_{n_0'L}}
\nonumber\\
\nonumber\\
&=&-\bigg\{ |b|^2\sum_{n_sL_sM_s}\sum_{l_2} \frac{F_5(n_s,L_s,M_s;l_2;L,M;Q) cos({M_{s} \alpha})}{R_{31}^{L_{s}+1}R_{12}^{L_{s}+l_2+1}}
\nonumber\\
&+&\sum_{n_sL_sM_s}\sum_{l_1} \frac{\{a^*b \exp[{i(M_{s})\alpha}]+b^*a \exp[{-i(M_{s})\alpha}]\}F_7(n_s,L_s,M_s;l_1;L,M;Q)}{R_{12}^{l_1+L+1}R_{31}^{L_{s}+1}}
\nonumber\\
&+&\sum_{n_sL_sM_s} \sum_{l_1'm_1'} \frac{\{a^*b\exp[{i(m_1')\alpha}]+ b^*a\exp[{-i(m_1')\alpha}]\}F_6(n_s,L_s,M_s;l_1',m_{1}';L,M;Q)}{R_{12}^{L_{s}+L+1}R_{31}^{l_1'+1}}
\bigg\}\,. \nonumber\\
\end{eqnarray}
In the above Eqs.~(\ref{V12-12b})-(\ref{d13}),  the $F_1$, $F_2$, and $F_3$ functions and the corresponding $G_1(L_i,L_j,\ell_k,\ell_k';L,M)$, $G_2(L_i,L_j,\ell_{k_1},\ell_{k_2}';L,M)$, and $G_3(L_i,L_j)$ are defined in Ref. \cite{yan16}. The $F_4$, $F_5$, $F_6$, and $F_7$ functions and the corresponding $G_4(L_i,M_i;L,M)$, $G_5(L_i,M_i;\ell_{k_1};L,M;Q)$, $G_6(L_i,M_i;\ell_{k_1},m_{k_1};L,M;Q)$ and $G_7(L_i,M_i;\ell_{k_1};L,M;Q)$ are are defined by
\begin{eqnarray}\label{AF1}
F_1(n_s,n_t,L_s,L_t;l_1,l_1';L,M)  &=& G_1(L_s,L_t,l_1,l_1';L,M)
\langle{n_0'}L\|T_{l_1}\|{n_s}L_s\rangle^* 
\nonumber \\
&\times&\frac{
\langle{n_0'}L\|T_{l_{1}^{\prime}}\|{n_s}L_s\rangle
|\langle{n_0 ( n_0'' )0 }\|T_{L_t}\|{n_t}L_t\rangle|^{2}}
{E_{n_sL_s}+E_{n_tL_t}-E_{n_0 ( n_0'' ) S}^{(0)}-E_{n_0'L}^{(0)}} \,,
\end{eqnarray}
\begin{eqnarray}\label{AF2}
F_2(n_s,n_t,L_s,L_t;l_1,l_2';L,M)  &=&  (-1)^{L_{s}+l_2^{\prime}}G_2(L_s,L_t,l_1,l_2';L,M)  \nonumber \\
&\times&
\langle{n_0 (n_0'')}0\|T_{L_s}\|{n_s}L_s\rangle
\langle{n_0'}L\|T_{l_{2}^{\prime}}\|{n_t}L_t\rangle
 \nonumber \\
&\times&
\frac{\langle{n_0'}L\|T_{l_1}\|{n_s}L_s\rangle ^*
\langle{n_0 (n_0'')}0\|T_{L_t}\|{n_{t}}L_t\rangle ^*
}
{E_{n_sL_s}+E_{n_tL_t}-E_{n_0 (n_0'') S}^{(0)}-E_{n_0'L}^{(0)}} \,,
\end{eqnarray}
\begin{eqnarray}\label{AF3}
F_3(n_s,n_t,L_s,L_t)  &=&  G_3(L_s,L_t)
\frac{|\langle{n_0}0\|T_{L_{t}}\|{n_t}L_t\rangle|^{2}
|\langle{n_0''}0\|T_{L_u}\|{n_u}L_u\rangle|^{2}}
{E_{n_tL_t}+E_{n_uL_u}-E_{n_0S}^{(0)}-E_{n_0''S}^{(0)}}  \,,
\end{eqnarray}
\begin{eqnarray}\label{AF4}
F_4(n_u,L_u,M_u;L,M) &=& (-1)^{L_{u}+L+M_u+M} G_4(L_u,M_u;L,M)
\nonumber \\
&\times&\bigg[
\frac{|\langle{n_0'}L\|T_{L}\|{n_0}0\rangle|^2|\langle{n_0''}0\|T_{L_{u}}\|{n_u}L_u\rangle|^{2}}
{E_{n_uL_u}-E_{n_0''S}+E_{n_0S}-E_{n_0'L}}
\nonumber \\
&+&
\frac{|\langle{n_0}0\|T_{L}\|{n_0'}L\rangle|^2|\langle{n_0''}0\|T_{L_{u}}\|{n_u}L_u\rangle|^{2}}
{E_{n_uL_u}+E_{n_0'L}-E_{n_0S}-E_{n_0''S}}\bigg] \,,
\end{eqnarray}
\begin{eqnarray}\label{AF5}
F_5(n_t,L_t,M_t;l_1;L,M;Q) &=& (-1)^{L_{t}+L+M_t+M} G_5(L_t,M_t;l_1;L,M;Q)
\nonumber \\
&\times&
\frac{\langle{n_0'}L\|T_{l_1}\|{n_0'}L\rangle|\langle{n_0}0\|T_{L_{t}}\|{n_t}L_t\rangle|^{2}}{E_{n_tL_t}-E_{n_0S}}
 \,,
\end{eqnarray}
\begin{eqnarray}\label{AF6}
F_6(n_t,L_t,M_t;l_2',m_{2}';L,M;Q) &=&  G_6(L_t,M_t;l_2',m_{2}';L,M;Q)
\nonumber \\
&\times&
\frac{ \langle {n_0}0 \|T_L
\|{n_0'}L\rangle  \langle{n_0}0\|T_{L_t}\|{n_t}L_t\rangle^*
\langle{n_0'}L\|T_{l_2'}\|{n_t}L_t\rangle}{E_{n_tL_t}-E_{n_0'L}}
 \,,
 \end{eqnarray}
\begin{eqnarray}\label{AF7}
F_7(n_t,L_t,M_t;l_2;L,M;Q) &=& (-1)^{L} G_7(L_t,M_t;l_2;L,M;Q)
\nonumber \\
&\times&
\frac{ \langle {n_0}0 \|T_L \|{n_0'}L\rangle^*  \langle{n_0'}L\|T_{l_2}\|{n_t}L_t\rangle^* \langle{n_0}0\|T_{L_t}\|{n_t}L_t\rangle}{E_{n_tL_t}-E_{n_0S}}
 \,,
\end{eqnarray}
\begin{eqnarray}\label{G1}
G_1(L_i,L_j,\ell_k,\ell_k';L,M) 
&=&  \frac{16\pi^2(\ell_k,\ell_k^{\prime})^{-1/2}}{(2L_{j}+1)^2}\sum_{M_i M_j m_k} \left(
  \begin{array}{ccc}
    L & \ell_k & L_i\\
    -M & m_k & M_i\\
  \end{array}
\right)
\left(
  \begin{array}{ccc}
    L & \ell_{k}^{\prime} & L_i\\
    -M & m_{k} & M_i\\
  \end{array}
\right)  \nonumber \\
&\times&
\frac{(L_j+\ell_k-M_j+m_k)!(L_j+\ell_k^{\prime}-M_j+m_k)!P_{L_j+\ell_k}^{M_j-m_k}(0)P_{L_j+\ell_k^{\prime}}^{M_j-m_k}(0)}
{(L_j+M_j)!(L_j-M_j)![(\ell_k+m_k)!(\ell_k-m_k)!(\ell_k^{\prime}+m_k)!(\ell_k^{\prime}-m_k)!]^{1/2}} \,, \nonumber \\ \\
G_2(L_i,L_j,\ell_{k},\ell_{k}';L,M) &=&   \frac{16\pi^2(\ell_{k},\ell_{k}^{\prime})^{-1/2}}{(2L_i+1)(2L_j+1)}\sum_{M_iM_j m_{k}m_{k}^{\prime}}
\left(
  \begin{array}{ccc}
    L & \ell_{k} & L_{i}\\
    -M & -m_{k} & M_{i}\\
  \end{array}
\right)
\left(
  \begin{array}{ccc}
    L & \ell_{k}^{\prime} & L_{j}\\
    -M & m_{k}^{\prime} & M_{j}\\
  \end{array}
\right)   \nonumber \\
&\times& \frac{(-1)^{M_{i}+M_{j}}P_{L_{j}+\ell_{k}}^{M_{j}+m_{k}}(0)P_{L_{i}+\ell_{k}^{\prime}}^{M_{i}-m_{k}^{\prime}}(0)}
{[(L_{i}+M_{i})!(L_{i}-M_{i})!(L_{j}+M_{j})!(L_{j}-M_{j})!]^{1/2}} \nonumber  \\
&\times&\frac{ (L_{j}+\ell_{k}-M_{j}-m_{k})!(L_{i}+\ell_{k}^{\prime}-M_{i}+m_{k}^{\prime})!}
{[(\ell_{k}+m_{k})!(\ell_{k}-m_{k})!(\ell_{k}^{\prime}+m_k^{\prime})!
(\ell_{k}^{\prime}-m_k^{\prime})!]^{1/2}} \, ,  \\
G_3(L_i,L_j) &=&  16\pi^2(L_i,L_j)^{-2}\sum_{M_iM_j}
\frac{[P_{L_i+L_j}^{M_i+M_j}(0)(L_i+L_j-M_i-M_j)!]^2}{(L_i+M_i)!
(L_i-M_i)!(L_j+M_j)!(L_j-M_j)!} \,,  \nonumber  \\ \\
G_4(L_i,M_i;L,M) &=&  16\pi^2\frac{[P_{L_i+L}^{M_i-M}(0)(L_i+L-M_i+M)!(L_{i},L)^{-1}]^2 }{(L_{i}+M_{i})!(L_{i}-M_{i})!(L+M)!(L-M)!}  \,,
 \\
G_5(L_i,M_i;\ell_{k};L,M;Q)&=& \frac{8 \sqrt{\pi^3} Q  P_{\ell_{k}+L_{i}}^{M_{i}}(0)P_{L_{i}}^{M_{i}}(0) (\ell_{k}+L_{i}-M_{i})! }{(2L_i+1)^2\sqrt{2\ell_{k}+1}(l_1)!(L_{i}+M_{i})!}
\left(
  \begin{array}{ccc}
    L & \ell_{k} & L\\
    -M & 0 & M\\
  \end{array}
\right) \,,
 \\
G_6(L_i,M_i;\ell_{k},m_{k};L,M;Q)&=&
\frac{8\sqrt{\pi^3}Q(\ell_{k})^{-1/2}}{(2L+1)(2L_i+1)} \left(
  \begin{array}{ccc}
    L & \ell_{k} & L_{i}\\
    -M & -m_{k} & M_{i}\\
  \end{array}
\right) \nonumber \\
&\times& \frac{P_{L+L_{i}}^{-M+M_{i}}(0)P_{\ell_{k}}^{m_{k}}(0)(L+L_{i}+M-M_{i})!(\ell_{k}-m_{k})!}{[(L+M)!
(L-M)!(L_{i}+M_{i})!(L_{i}-M_{i})!(\ell_{k}+m_{k})!
(\ell_{k}-m_{k})!]^{1/2}}  \,,
\nonumber \\ \\
G_7(L_i,M_i;\ell_{k};L,M;Q)&=&
\frac{8\sqrt{\pi^3}Q(\ell_{k})^{-1/2}}{(2L+1)(2L_i+1)} \sum_{m_{k}} \left(
  \begin{array}{ccc}
    L & \ell_{k} & L_{i}\\
    -M & m_{k} & M_{i}\\
  \end{array}
\right) \nonumber \\
&\times& \frac{P_{L+\ell_{k}}^{M-m_{k}}(0)P_{L_i}^{M_{i}}(0)(L+\ell_{k}-M+m_{k})!(L_{i}-M_{i})!}{[(L+M)!
(L-M)!(L_{i}+M_{i})!(L_{i}-M_{i})!(\ell_{k}+m_{k})!
(\ell_{k}-m_{k})!]^{1/2}} \,,
\nonumber\\
\end{eqnarray}
where $n_0$ and $n''_0$, respectively,  are the principal quantum numbers for $A(n_0 S)$ and $A^{Q+}(n_0'' S)$.

For the specific Li($2\,^{2}S$)-Li($2\,^{2}P$)-Li$^+(1\,^{1}S$) system, the second-order energy correction is simplified as,

\begin{eqnarray}
\Delta E^{(2)}
&=&-\sum_{ n_2\geq 2, n_3\geq 3 }\bigg(\frac{C_{2n_3}^{(12)}(L,M)}{R_{12}^{2n_3}}+\frac{C_{2n_2}^{(23)}(L,M)}{R_{23}^{2n_2}}+\frac{C_{2n_2}^{(31)}(L,M)}{R_{31}^{2n_2}}\bigg)\nonumber \\
&-&\sum_{\substack{ n_1\ne n_2 \\n_1\geq 1,n_2\geq 2,n_3\geq 3 \\ n_1+n_2+2=2n_3}}\bigg(\frac{C_{n_2+1,n_1+1}^{(12,23)}(L,M)}
{R_{12}^{n_2+1}R_{23}^{n_1+1}}+\frac{C_{n_1+1,n_2+1}^{(31,12)}(L,M)}{R_{31}^{n_1+1}R_{12}^{n_2+1}}\bigg)
\nonumber \\
&-&\sum_{\substack{n_3\geq 3  }}\bigg(\frac{C_{n_3,n_3}^{(12,23)}(L,M)}{R_{12}^{n_3}R_{23}^{n_3}}+\frac{C_{n_3,n_3}^{(23,31)}(L,M)}{R_{23}^{n_3}R_{31}^{n_3}}+\frac{C_{n_3,n_3}^{(31,12)}(L,M)}{R_{31}^{n_3}R_{12}^{n_3}}\bigg)
\,, 
\end{eqnarray}
where $C_{2n_3}^{(12)}(L,M)$, $C_{2n_2}^{(23)}(L,M)$, $C_{2n_3}^{(31)}(L,M)$, $C_{n_3,n_3}^{(23,31)}(L,M)$, $C_{n_2+1,n_1+1}^{(12,23)}(L,M)$, and $C_{n_1+1,n_2+1}^{(31,12)}(L,M)$ are the additive and nonadditive dispersion coefficients. These coefficients can be expressed as
\begin{eqnarray}
C_{2n_3}^{(12)}(L,M)
&=&|a|^2\sum_{n_sn_t}\sum_{\substack{L_sL_t l_1l_{1}^{\prime} \\ 2L_{t}+l_1+l_1^{\prime}+2=2n_3}}
F_1(n_s,n_t,L_s,L_t;l_1,l_1';L,M) \nonumber \\
&+&|b|^2\sum_{n_sn_t}\sum_{\substack{L_sL_tl_2l_{2}^{\prime} \\ 2L_{s}+l_2+l_2^{\prime}+2=2n_3}}
F_1(n_t,n_s,L_t,L_s;l_2,l_2';L,M) \nonumber  \\
&+& a^{*}b\sum_{n_sn_t}\sum_{\substack{L_sL_t l_1l_{2}^{\prime} \\ L_{s}+L_{t}+l_1+l_2^{\prime}+2=2n_3}}
F_3(n_s,n_t,L_s,L_t;l_1,l_2';L,M) \nonumber \\
&+& b^{*}a\sum_{n_sn_t}\sum_{\substack{L_sL_t l_1^{\prime}l_{2} \\ L_{s}+L_{t}+l_1^{\prime}
+l_2+2=2n_3}}F_3^*(n_s,n_t,L_s,L_t;l_1',l_2;L,M) \,,
\end{eqnarray}
\begin{eqnarray}
C_{2n_2}^{(23)}(L,M)&=&|a|^2\sum_{n_tn_u}\sum_{\substack{L_tL_u  \\ 2L_{t}+2L_u+2 =2n_2}} F_2(n_t,n_u,L_t,L_u)
 \nonumber \\
&+&|b|^2\sum_{n_tn_u}\sum_{\substack{L_tL_u l_2l_{2}^{\prime} \\ 2L_{u}+l_2+l_2^{\prime}+2 =2n_2}} F_1(n_t,n_u,L_t,L_u;l_2,l_2';L,M)
 \,,
\end{eqnarray}
\begin{eqnarray}
C_{2n_2}^{(31)}(L,M)&=&|a|^2 \sum_{n_sn_u}\sum_{\substack{L_sL_u l_1l_{1}^{\prime} \\ 2L_{u}+l_1+l_1^{\prime}+2 =2n_2}}
F_1(n_s,n_u,L_s,L_u;l_1,l_1';L,M)
\nonumber \\
&+& |b|^2\sum_{n_sn_u}\sum_{\substack{L_sL_u  \\ 2L_{s}+2L_{u}+2=2n_2}} F_2(n_s,n_u,L_s,L_u)  \,,
\end{eqnarray}
\begin{eqnarray}
C_{n_2+1,n_1+1}^{(12,23)}(L,M) &=& \sum_{\substack{L_tl_2 \\ l_2+L+1=n_2+1 \\ L_{t}+1 = n_1+1 }} \sum_{\substack{L'_tl_2'l_1 \\ L+L'_t+1=n_2+1 \\ l_2'+1 = n_1+1 }}
\mathbb{P}_1(a,b,\beta,Q,L_t,l_2,L'_t,l'_2,l_1,L,M )\,,
\end{eqnarray}
\begin{eqnarray}
C_{n_1+1,n_2+1}^{(31,12)}(L,M) &=& \sum_{\substack{L_sl_1l_2 \\ l_1+L+1=n_2+1 \\ L_{s}+1 = n_1+1 }}
\sum_{\substack{L'_sl_1' \\ L'_s+L+1=n_2+1 \\ l_1'+1 = n_1+1 }}
 \mathbb{P}_1(b,a,\alpha,Q,L_s,l_1,L'_s,l'_1,l_2,L,M ) \,,
\end{eqnarray}
\begin{eqnarray}
C_{n_3,n_3}^{(12,23)}(L,M) &=& \sum_{\substack{\\L_t l_2\\ l_2+L+1=n_3 \\ L_{t}+1 = n_3   }} \sum_{\substack{\\L'_t l_2' \\ L+L'_t+1=n_3 \\ l_2'+1 = n_3 }} \mathbb{P}_2(a,b,\beta,Q,L_t,l_2,L'_t,l'_2,L,M ) \,,
\end{eqnarray}
\begin{eqnarray}
C_{n_3,n_3}^{(23,31)}(L,M)=\sum_{\substack{L_u \\ L_u+L+1=n_3}}\mathbb{P}_3(a,b,\gamma,L_u,L,M)  \,,
\end{eqnarray}
\begin{eqnarray}
C_{n_3,n_3}^{(31,12)}(L,M) &=& \sum_{\substack{L_sl_1 \\ l_1+L+1=n_3 \\ L_{s}+1 = n_3 }}
\sum_{\substack{L'_sl_1' \\ L'_s+L+1=n_3 \\ l_1'+1 = n_3 }}\mathbb{P}_2(b,a,\alpha,Q,L_s,l_1,L'_s,l'_1,L,M )
 \,,
\end{eqnarray}
where the $\mathbb{P}_n$ functions are defined as
\begin{eqnarray}\label{eq:p1}
 \mathbb{P}_1(a,b,\beta,Q,L_t,l_2,L'_t,l'_2,l_1,L,M )
&=& |a|^2 \sum_{\substack{n_tM_t}}  F_5(n_t,L_t,M_t;l_1;L,M;Q) cos(M_t\beta)
\nonumber\\
&+&\sum_{\substack{n'_tM'_tm_2'}}  \{a^*b\exp[{-i(m_2')\beta}]+ c.c.\}F_6(n'_t,L'_t,M'_t;l_2',m_{2}';L,M;Q)
\nonumber\\
&+&  \sum_{\substack{n_tM_t }}  \{a^*b \exp[{i(M_{t})\beta}]+c.c.\}F_7(n_t,L_t,M_t;l_2;L,M;Q)
 \,, \nonumber  \\
\end{eqnarray}
\begin{eqnarray}\label{eq:p2}
 \mathbb{P}_2(a,b,\beta,Q,L_t,l_2,L'_t,l'_2,L,M )
&=& \sum_{\substack{n'_tM'_tm_2' }}  \{a^*b\exp[{-i(m_2')\beta}]+ c.c.\}F_6(n'_t,L'_t,M'_t;l_2',m_{2}';L,M;Q)
\nonumber\\
&+& \sum_{\substack{n_tM_t }}  \{a^*b \exp[{i(M_{t})\beta}]+c.c.\}F_7(n_t,L_t,M_t;l_2;L,M;Q)
 \,, \nonumber  \\
\end{eqnarray}
\begin{eqnarray}\label{eq:p3}
\mathbb{P}_3(a,b,\gamma,L_u,L,M)
= \sum_{n_uM_u} \bigg\{(a^*b)\exp[i(M_u-M)\gamma] + c.c. \bigg\}   F_4(n_u,L_u,M_u;L,M)   \,, \nonumber  \\
\end{eqnarray}
and where $c.c.$ indicates the complex conjugate of the preceding term. 

\subsection{Long-range interaction coefficients for the Li($2\,^{2}S$)-Li$^+(1\,^{1}S$) system}
\label{subsec:appendix-dimerion}

The induction and dispersion coefficients $C_{4,\text{ind}}^{(S-S^+)}$, $C_{6,\text{ind}}^{(S-S^+)}$, and $C_{6,\text{disp}}^{(S-S^+)}$ for the Li($2\,^{2}S$)-Li$^+(1\,^{1}S$) system can be written as
\begin{eqnarray}\label{c423s}
C_{4,\text{ind}}^{(S-S^+)}=\sum_{n_tn_u}  F_2(n_t,n_u,1,0) \, ,
\end{eqnarray}
\begin{eqnarray}\label{c623s}
C_{6,\text{ind}}^{(S-S^+)}=\sum_{n_tn_u}   F_2(n_t,n_u,2,0)  \, ,
\end{eqnarray}
\begin{eqnarray}\label{c623d}
C_{6,\text{disp}}^{(S-S^+)}= \sum_{n_tn_u}F_2(n_t,n_u,1,1)  \, .
\end{eqnarray}

\subsection{Long-range interaction coefficients for the Li($2\,^{2}S$)-Li$(2\,^{2}P$) system}
\label{subsec:appendix-dimer}

The dipolar and dispersion interaction coefficients $C_{3,\text{dip}}^{(S-P)}(M)$ and $C_{6,\text{disp}}^{(S-P)}(M)$ for the Li($2\,^{2}S$)-Li($2\,^{2}P$) system can be written as

\begin{eqnarray}\label{c312new}
C_{3,\text{dip}}^{(S-P)}(M)&=& (a^{*}b+b^{*}a) \frac{4\pi(-1)^{1+M}}{9(1-M)!(1+M)!}
|\langle{n_0}0\|T_1
\|{n_0}1\rangle|^2
 \,, \label{C3anew}
\end{eqnarray}
\begin{eqnarray}\label{c612SP}
C_{6,\text{disp}}^{(S-P)}(M)=\sum_{n_sn_tL_s} F_1(n_s,n_t,L_s,1;1,1;1,M) \, .
\end{eqnarray}

\subsection{Long-range interaction coefficients for the Li$(2\,^{2}P$)-Li$^+$($1\,^{1}S$) system}
\label{subsec:appendix-p+}

The electrostatic, dispersion and induction interaction coefficients for the Li($2\,^{2}P$)-Li$^+(1\,^{1}S$) system can be written as
\begin{eqnarray}\label{c312}
 C_{3,\text{elst}}^{(P-S^+)}(M)= Q (-1)^{1+M} \sqrt{\frac{\pi}{5}}
\left(
  \begin{array}{ccc}
    1 & 2 & 1\\
    -M & 0 & M\\
  \end{array}
\right)
\langle{n_0}1\|T_2
\|{n_0}1\rangle\ \, ,
\end{eqnarray}
\begin{eqnarray}\label{c423i}
C_{4,\text{ind}}^{(P-S^+)}(M)=\sum_{n_tn_u L_t} F_1(n_t,n_u,L_t,0;1,1;1,M) \, ,
\end{eqnarray}
\begin{eqnarray}\label{c623disp}
C_{6,\text{disp}}^{(P-S^+)}(M)=\sum_{n_tn_uL_t}  F_1(n_t,n_u,L_t,1;1,1;1,M) \, ,
\end{eqnarray}
and
\begin{eqnarray}\label{c623i}
C_{6,\text{ind}}^{(P-S^+)}(M)= &\sum_{n_tn_uL_t}& \bigg \{  F_1(n_t,n_u,L_t,0;2,2;1,M)  + F_1(n_t,n_u,L_t,0;1,3;1,M)  \nonumber \\ &&+ F_1(n_t,n_u,L_t,0;3,1;1,M)  \bigg \} \, ,
\end{eqnarray}

\subsection{Long-range interaction coefficients for the Li($2\,^{2}S$)-Li$(2\,^{2}P$)-Li$^+(1\,^{1}S$) system}
\label{subsec:appendix-trimer}

The additive interaction coefficients for the Li($2\,^{2}S$)-Li($2\,^{2}P$)-Li$^+(1\,^{1}S$) system can be written as
\begin{eqnarray}
C_{4}^{(23)}(1,M)&=& |a|^2 \mathbb{T}_1 + |b|^2\mathbb{T}_3(M)
\,, \label{C423addSM}
\end{eqnarray}
\begin{eqnarray}
C_{4}^{(31)}(1,M)&=& |a|^2 \mathbb{T}_3(M) + |b|^2 \mathbb{T}_1
\,, \label{C431addSM}
\end{eqnarray}
\begin{eqnarray}
C_{6}^{(12)}(1,M)&=& |a|^2 \mathbb{T}_4(M) + |b|^2\mathbb{T}_4(M)
\,, \label{C612addSM}
\end{eqnarray}
\begin{eqnarray}
C_{6}^{(23)}(1,M)= |a|^2 \mathbb{T}_2
+ |b|^2 \mathbb{T}_5(M)
\,, \label{C623addSM}
\end{eqnarray}
\begin{eqnarray}
C_{6}^{(31)}(1,M)= |a|^2\mathbb{T}_5(M)+ |b|^2 \mathbb{T}_2
\,, \label{C631addSM}
\end{eqnarray}
where
\begin{eqnarray}
\mathbb{T}_1&=&\sum_{n_tn_u}  F_3(n_t,n_u,1,0) \, ,
\end{eqnarray}
\begin{eqnarray}
\mathbb{T}_2&=&\sum_{n_tn_u}  \bigg \{ F_3(n_t,n_u,2,0) + F_3(n_t,n_u,1,1) \bigg \} \, ,
\end{eqnarray}
\begin{eqnarray}
\mathbb{T}_3(M)&=&\sum_{n_tn_u L_t} F_1(n_t,n_u,L_t,0;1,1;1,M) \, ,
\end{eqnarray}
\begin{eqnarray}
\mathbb{T}_4(M)&=&\sum_{n_sn_tL_s} F_1(n_s,n_t,L_s,1;1,1;1,M) \, ,
\end{eqnarray}
\begin{eqnarray}
\mathbb{T}_5(M)=\sum_{n_tn_uL_t}& \bigg \{& F_1(n_t,n_u,L_t,1;1,1;1,M) + F_1(n_t,n_u,L_t,0;2,2;1,M) \nonumber \\ & + & F_1(n_t,n_u,L_t,0;1,3;1,M)  + F_1(n_t,n_u,L_t,0;3,1;1,M)  \bigg \} \, . \label{T5SM}
\end{eqnarray}
The nonadditive interaction coefficients for the Li($2\,^{2}S$)-Li($2\,^{2}P$)-Li$^+(1\,^{1}S$) system are given by 
\begin{eqnarray}
C_{4,2}^{(12,23)}(1,M)  =\mathbb{P}_1(a,b,\beta,Q,1,2,2,1,2,1,M) \,, \label{C6421223SM}
\end{eqnarray}
\begin{eqnarray}
C_{2,4}^{(31,12)}(1,M) = \mathbb{P}_1(b,a,\alpha,Q,1,2,2,1,2,1,M )
\,, \label{C6243112SM}
\end{eqnarray}
\begin{eqnarray}
C_{3,3}^{(12,23)}(1,M) = \mathbb{P}_2(a,b,\beta,Q,2,1,1,2,1,M )
\,, \label{C6331223SM}
\end{eqnarray}
\begin{eqnarray}
C_{3,3}^{(23,31)}(1,M)= \mathbb{P}_3(a,b,\gamma,1,1,M ) \,,   \label{C62331SM}
\end{eqnarray}
and
\begin{eqnarray}
C_{3,3}^{(31,12)}(1,M) = \mathbb{P}_2(b,a,\alpha,Q,2,1,1,2,1,M )
\,. \label{C6333112SM}
\end{eqnarray}

\clearpage

\end{document}